\documentclass[%
    aps,
    prd,
    twocolumn,
    reprint,
    10pt,
    superscriptaddress,
    longbibliography,
    amsfonts,
    amssymb,
    amsmath,
    preprintnumbers,
    nobibnotes,
    nofootinbib,
    noshowpacs,
    noshowkeyws,
    tightenlines,
    floats,
    notitlepage,
    final,
    letterpaper,
    oneside,
    eqsecnum,
    balancelastpage,
    flushbottom,
    noraggedfooter,
    ]{revtex4-2}
\pdfoutput=1
\usepackage[utf8]{inputenc}
\usepackage[T1]{fontenc}
\usepackage[british]{babel}
\usepackage[british,cleanlook,printdayon]{isodate}
\usepackage[Symbolsmallscale]{upgreek}
\usepackage{mathtools}
\usepackage{isomath}
\usepackage[protrusion,tracking,kerning,spacing,final,babel]{microtype}
\usepackage{physics}
\usepackage[dvipsnames]{xcolor}
\usepackage[final]{hyperref}
\hypersetup{%
    colorlinks=true,
    citecolor=MidnightBlue,
    linkcolor=MidnightBlue,
    urlcolor=MidnightBlue
    }%
\usepackage{cleveref}
\usepackage{wasysym}

\newcommand{\rme}{\mathrm{e}}
\newcommand{\rmi}{\mathrm{i}}
\newcommand{\asc}{\textit{\ascnode}}
\newcommand{\eps}{\varepsilon}
\newcommand{\obs}{\mathrm{obs}}

\begin{document}
\title{Detecting stochastic gravitational waves with binary resonance}

\author{Diego~Blas}
\affiliation{Grup  de  F\'isica  Te\`orica,  Departament  de  F\'isica,  Universitat  Aut\`onoma  de  Barcelona,  08193  Bellaterra, Spain,}
\affiliation{Institut de Fisica d’Altes Energies (IFAE), The Barcelona Institute of Science and Technology, Campus UAB, 08193 Bellaterra, Spain}
\affiliation{Theoretical Particle Physics and Cosmology Group, Physics Department, King's College London, University of London, Strand, London WC2R 2LS, United Kingdom}

\author{Alexander~C.~Jenkins}
\email{alex.jenkins@ucl.ac.uk}
\altaffiliation{Corresponding author. Present address: Department of Physics and Astronomy, University College London, London WC1E 6BT, United Kingdom}
\affiliation{Theoretical Particle Physics and Cosmology Group, Physics Department, King's College London, University of London, Strand, London WC2R 2LS, United Kingdom}

\date{\today}
\preprint{KCL-PH-TH/2021-34}

\begin{abstract}
    LIGO and Virgo have initiated the era of gravitational-wave (GW) astronomy; but in order to fully explore GW frequency spectrum, we must turn our attention to innovative techniques for GW detection.
    One such approach is to use binary systems as dynamical GW detectors by studying the subtle perturbations to their orbits caused by impinging GWs.
    We present a powerful new formalism for calculating the orbital evolution of a generic binary coupled to a stochastic background of GWs, deriving from first principles a secularly-averaged Fokker-Planck equation which fully characterises the statistical evolution of all six of the binary's orbital elements.
    We also develop practical tools for numerically integrating this equation, and derive the necessary statistical formalism to search for GWs in observational data from binary pulsars and laser-ranging experiments.
\end{abstract}

\maketitle
\tableofcontents

\section{Introduction}

\begin{figure*}[t!]
    \includegraphics[width=\textwidth]{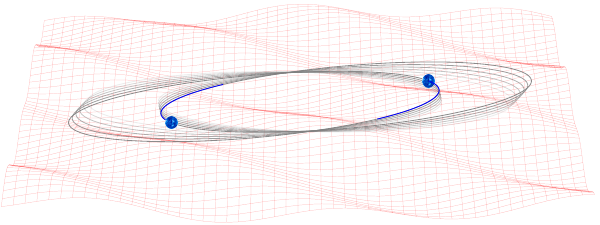}
    \caption{%
    Cartoon illustration of the binary resonance effect. Stochastic fluctuations in the background spacetime geometry due to incoming GWs perturb the trajectories of two orbiting masses, causing cumulative changes to their orbital elements.}
    \label{fig:cartoon}
\end{figure*}

We have entered the era of gravitational-wave (GW) astronomy.
The advanced laser interferometers LIGO~\cite{TheLIGOScientific:2014jea}, Virgo~\cite{TheVirgo:2014hva}, and KAGRA~\cite{Akutsu:2018axf} now form a global GW observatory network~\cite{KAGRA:2013rdx}, which has already proved its capabilities by detecting dozens of GW signals in the 10--1000~Hz band from coalescing compact binaries~\cite{LIGOScientific:2020ibl}.
Concurrently, GW searches in the nHz band with pulsar timing arrays (PTAs)~\cite{Kramer:2013kea,McLaughlin:2013ira,Hobbs:2013aka,IPTA:2013lea} are, after more than a decade of continuous operation, widely expected to make their first detections in the coming few years, with the NANOGrav Collaboration having possibly already seen the first hints of a GW detection in their 12.5-yr dataset~\cite{Arzoumanian:2020vkk}.
By the mid-2030s, these experiments are expected to be superseded by third generation interferometers such as Einstein Telescope~\cite{Punturo:2010zz} and PTA campaigns by next-generation radio telescopes such as the Square Kilometre Array~\cite{Janssen:2014dka}, as well as the space-based interferometer LISA~\cite{Amaro-Seoane:2017drz}, which will survey the as-yet-unexplored mHz frequency band.
Further proposals to cover the intermediate frequency band between LISA and LIGO/Virgo/KAGRA include the atom interferometers AION~\cite{Badurina:2019hst} and MAGIS~\cite{Abe:2021ksx}.

These experiments each hold enormous scientific potential, and together will probe a dizzying range of exotic astrophysical and cosmological phenomena throughout the history of the Universe.
However, practical and technical limitations restrict the sensitivity of each experiment to a narrow frequency band, leaving broad swathes of the GW frequency spectrum essentially unexplored.
These gaps in the GW spectrum could contain signals which are inaccessible to any current or planned GW observatory; perhaps the clearest example is the GW signal expected from a cosmological first-order phase transition~\cite{Kamionkowski:1993fg,Caprini:2015zlo,Caprini:2019egz}, which is sharply peaked and could, for a broad range of the physical parameter space, be missed by all of the GW experiments listed above.
There is also the distinct possibility of such unexplored frequency bands containing unexpected signals, with the potential for exciting discoveries beyond just those models that have been proposed in the literature.

This problem motivates us to explore alternatives that can bridge these gaps in the GW spectrum.
One possibility is to study the interaction between GWs and astronomical binary systems.
Rather than searching for oscillations in the proper distance between the test masses in an interferometer (as in LIGO/Virgo/KAGRA, etc.), or between pulsars and the Earth (as in PTAs), we consider here the GW-induced oscillations between two freely-falling masses in a gravitationally-bound orbit.
While these oscillations are extremely challenging to observe directly for any realistic binary, they can leave lasting imprints on the binary's orbit, particularly if they occur at an integer multiple of the binary's orbital frequency, as this causes the perturbations to be resonantly amplified.
For long-duration GW signals, these imprints accumulate over time, eventually giving rise to observable deviations which can be used to infer the GW amplitude, turning the binary into a dynamical GW detector (as illustrated in Fig.~\ref{fig:cartoon}).
This idea has a long history~\cite{Bertotti:1973clo,Misner:1974qy,Rudenko:1975tes,Mashhoon:1978res,Turner:1979inf,Futamase:1979gu,Mashhoon:1981cos,Linet:1982xc,Nelson:1982pk,Chicone:1995nn,Chicone:1996yc,Chicone:1996pm,Chicone:1999sg,Iorio:2011eu,Iorio:2021cxt}, and has been used to search for GWs with the orbit of the binary pulsar B1913+16~\cite{Hui:2012yp}.\footnote{%
    Similar ideas have also been used to search for orbital changes induced by ultralight dark matter~\cite{Blas:2016ddr,LopezNacir:2018epg,Armaleo:2019gil,Desjacques:2020fdi,Blas:2019hxz}.}
Nonetheless, this binary GW resonance effect has received relatively little attention in the GW community, and, we feel, has not yet been exploited to its full potential.

In this paper we develop, from first principles, a new formalism for calculating the GW-induced evolution of a generic binary system.
Motivated by the requirements for the GW signal to be persistent (so that the orbital deviations accumulate over time) and broadband (so that the frequency content of the signal overlaps with the narrow resonant frequencies bands of the binary), we focus on the \emph{stochastic GW background} (SGWB)~\cite{Allen:1996vm,Maggiore:1999vm,Regimbau:2011rp,Romano:2016dpx,Christensen:2018iqi,Caprini:2018mtu}: a pseudo-random signal formed from the incoherent superposition of many independent GW sources throughout cosmic history.
The stochastic nature of this signal means that we cannot deterministically calculate the evolution of a given binary; instead we treat each of the binary's orbital elements as a time-dependent random variable, and study the statistics of the orbital perturbations over time.
Our key result is a secularly-averaged \emph{Fokker-Planck equation} (FPE) which fully specifies the evolution of the probability distribution for all six orbital elements.
By comparing solutions of this FPE with high-precision orbital data from various binary systems, one can place stringent new constraints on the SGWB at frequencies that are inaccessible to all other current and future GW observatories, as we demonstrate in a companion paper~\cite{Blas:2021mqw}.

The remainder of this paper is structured as follows.
In Sec.~\ref{sec:dynamics} we give a brief, self-contained overview of Keplerian orbits, orbital perturbations, and the formalism of osculating orbital elements.
In Sec.~\ref{sec:gw-resonance} we specialise this formalism to perturbations from the SGWB, and develop a FPE for the orbital elements, expressing the coefficients of the equation in terms of GW transfer functions.
In Sec.~\ref{sec:KM} we derive these coefficients explicitly as functions of the orbital elements, and briefly discuss their properties.
In Sec.~\ref{sec:circular} we obtain some exact late-time results for the case where the binary's eccentricity is held fixed at zero, which greatly simplifies the FPE.
In Sec.~\ref{sec:full-fpe} we tackle the more complicated general-eccentricity case, consider practical approaches for solving the FPE on observational timescales, and present some example results for the Hulse-Taylor binary pulsar B1913+16.
In Sec.~\ref{sec:observations} we develop the necessary statistical formalism to search for SGWB-induced orbital perturbations in observational data, discussing how to compute upper limits and sensitivity forecasts, and how to apply these tools to pulsar timing and laser-ranging experiments.
We summarise our results in Sec.~\ref{sec:summary}.
The Appendices give various technical details for the derivation of the FPE coefficients.
We use units where $c=1$ throughout, but keep $G\ne1$.

\section{Binary dynamics}
\label{sec:dynamics}

\begin{figure}[t!]
    \includegraphics[width=0.48\textwidth]{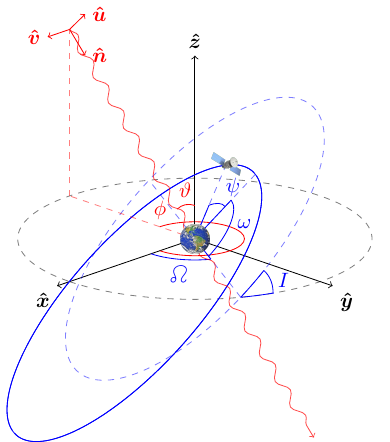}
    \caption{%
    Schematic diagram of a Keplerian orbit.
    The plane of the orbit (shown in blue) is defined relative to the fixed reference frame $(\vu*x,\vu*y,\vu*z)$ (shown in black) by the inclination $I$ and the longitude of ascending node $\asc$, while the orientation of the orbit within this plane is specified by the argument of pericentre $\omega$.
    The true anomaly $\psi$ acts as an angular coordinate for the position of the orbit within the orbital plane, measured relative to the pericentre.
    The polarisation tensors describing the effects of an incoming plane GW are described in terms of the basis $(\vu*n,\vu*u,\vu*v)$ (shown in red), which is related to the reference frame by the angles $\vartheta$ and $\phi$.
    (Note that the wavelength of the GW is not shown to scale here---our analysis assumes a wavelength much larger than the size of the orbit.)}
    \label{fig:orbital-elements}
\end{figure}

In this section we introduce the machinery of osculating orbital elements.
We start by recalling the basic properties of Keplerian orbits (i.e., orbits of two point masses interacting only through Newtonian gravity), before introducing the equations of motion (EoM) for the osculating elements, and discussing the most important contribution to these EoM from relativistic effects.
We also introduce alternative sets of orbital elements that are useful in cases where the eccentricity or inclination of the orbit are small.

\subsection{Keplerian orbits}

We start with a Keplerian binary, working in cylindrical coordinates $(\vu*r,\vu*\theta,\vu*\ell)$ in a frame with the centre of mass fixed at the origin.
We also introduce a fixed Cartesian reference frame $(\vu*x,\vu*y,\vu*z)$, such the line of sight of the observer is in the positive $\vu*z$ direction.
The unperturbed EoM for the separation vector $\vb*r$ is
    \begin{equation}
    \label{eq:Keplerian-EoM}
        \ddot{\vb*r}+\frac{GM}{r^2}\vu*r=\vb*0,
    \end{equation}
    where $r\equiv|\vb*r|$ is the radial separation, $\vu*r\equiv\vb*r/r$ is the radial unit vector, and $M\equiv m_1+m_2$ is the total mass.
We define the energy and angular momentum of the binary (in units of the reduced mass $\mu\equiv m_1m_2/M$) by
    \begin{align}
    \begin{split}
        \mathcal{E}&=\frac{1}{2}\dot{\vb*r}\vdot\dot{\vb*r}-\frac{GM}{r},\\
        \vb*\ell&=\vb*r\cp\dot{\vb*r}=r^2\dot{\theta}\,\vu*\ell.
    \end{split}
    \end{align}
For the total angular momentum we write $\ell\equiv|\vb*\ell|=r^2\dot{\theta}$.
These are all conserved, since
    \begin{align}
    \begin{split}
    \label{eq:e-l-conservation}
        \dot{\mathcal{E}}&=\dot{\vb*r}\vdot\qty(\ddot{\vb*r}+\frac{GM}{r^2}\vu*r)=0,\\
        \dot{\vb*\ell}&=\vb*r\cp\ddot{\vb*r}=\vb*r\cp\qty(-\frac{GM}{r^2}\vu*r)=\vb*0,
    \end{split}
    \end{align}
    where we used $\dot{r}=\dot{\vb*r}\vdot\vu*r$.
The fact that $\dot{\vb*\ell}=\vb*0$ means that the binary orbit is confined to a fixed 2D plane, which is specified with respect to the $(\vu*x,\vu*y,\vu*z)$ reference frame by two angles: the inclination $I$, which is the angle between the binary's angular momentum vector $\vb*\ell$ and the observer's line of sight $\vu*z$, and the longitude of ascending node $\asc$, which is the angle between $\vu*x$ and the point where the orbit passes through the reference plane with positive velocity in the $\vu*z$ direction.
(These angles are illustrated in Fig.~\ref{fig:orbital-elements}.)

We can find the shape of the orbit in this plane by integrating Eq.~\eqref{eq:Keplerian-EoM}, yielding a family of elliptical solutions
    \begin{equation}
    \label{eq:Kepler-solutions}
        r=\frac{\ell^2/(GM)}{1+e\cos\psi},\quad\mathcal{E}=-\frac{GM}{2a},\quad\ell=\sqrt{GMa(1-e^2)},
    \end{equation}
    with the shape of the ellipse described by its semi-major axis $a$ and eccentricity $e$.
Here we have introduced the true anomaly $\psi$ as the angular position of the orbit within the orbital plane, defined such that the pericentre (minimum separation) occurs at $\psi=0$.
The point at which this occurs is defined by the argument of pericentre $\omega\equiv\theta-\psi$, which is measured relative to the ascending node.
Since $\omega$ is constant, the orbit is closed and the motion of the binary is periodic in time, with period $P$ related to the semi-major axis by Kepler's third law,
    \begin{equation}
    \label{eq:kepler-iii}
        \frac{GM}{a^3}=\qty(\frac{2\uppi}{P})^2.
    \end{equation}
(In what follows, we work entirely in terms of the period rather than the semi-major axis, as the former is more closely linked to the resonant frequencies of the orbit.)

The five constants $(P,e,I,\asc,\omega)$ are almost enough information to specify a particular Keplerian orbit; all that remains is to specify the time at which the binary is at pericentre, $t_0$.
In practice, it is more convenient to replace $t_0$ with the compensated mean anomaly,
    \begin{equation}
    \label{eq:epsilon}
        \eps\equiv\frac{2\uppi}{P}(t-t_0)-\int_0^t\dd{t'}\frac{2\uppi}{P(t')},
    \end{equation}
    as this is more well-behaved when the orbit is perturbed~\cite{Brouwer:1961mcm}.
Note that in the absence of perturbations this reduces to $\eps=-2\uppi t_0/P$.
We call the set $(P,e,I,\asc,\omega,\eps)$ the \emph{orbital elements} of a binary.

\subsection{Perturbations and osculating orbits}
\label{sec:perturbing-force}

Now, following Refs.~\cite{Burns:1976cm,Murray:2000ssd,Hui:2012yp,Blas:2016ddr}, suppose the binary is acted upon by some small perturbing force
    \begin{equation}
    \label{eq:perturbing-force}
        \updelta\ddot{\vb*r}=r(\mathcal{F}_r\vu*r+\mathcal{F}_\theta\vu*\theta+\mathcal{F}_\ell\vu*\ell),
    \end{equation}
    so that the EoM is instead
    \begin{equation}
        \ddot{\vb*r}+\frac{GM}{r^2}\vu*r=\updelta\ddot{\vb*r}.
    \end{equation}
[The factor of $r$ in Eq.~\eqref{eq:perturbing-force} is included for later convenience, as the GW perturbations we consider are proportional to the orbital separation.]
Inserting this into Eq.~\eqref{eq:e-l-conservation} gives
    \begin{align}
    \begin{split}
    \label{eq:e-l-evolution}
        \dot{\mathcal{E}}&=\dot{\vb*r}\vdot\updelta\ddot{\vb*r}=r\dot{r}\mathcal{F}_r+r^2\dot{\theta}\mathcal{F}_\theta,\\
        \dot{\vb*\ell}&=\vb*r\cp\updelta\ddot{\vb*r}=r^2\mathcal{F}_\theta\vu*\ell-r^2\mathcal{F}_\ell\vu*\theta,
    \end{split}
    \end{align}
    so the binary's energy and angular momentum are no longer constant.
As a result, the binary is no longer described by a fixed set of orbital elements.
However, the orbit is still tangent to some Keplerian ellipse at each moment in time.
We therefore define $(P,e,I,\asc,\omega,\eps)$ as functions of time which track the evolution of this tangent ellipse; these are called the \emph{osculating} orbital elements.

A full derivation of the EoM for the osculating elements is given in, e.g., Refs.~\cite{Burns:1976cm,Murray:2000ssd}; here we simply quote the resulting set of equations,
    \begin{align}
    \begin{split}
    \label{eq:Xdot}
        \dot{P}&=\frac{3P^2\gamma}{2\uppi}\qty[\frac{e\sin\psi \mathcal{F}_r}{1+e\cos\psi}+\mathcal{F}_\theta],\\
        \dot{e}&=\frac{\dot{P}\gamma^2}{3Pe}-\frac{P\gamma^5\mathcal{F}_\theta}{2\uppi e(1+e\cos\psi)^2},\\
        \dot{I}&=\frac{P\gamma^3\cos\theta \mathcal{F}_\ell}{2\uppi(1+e\cos\psi)^2},\\
        \dot{\asc}&=\frac{\tan\theta}{\sin I}\dot{I},\\
        \dot{\omega}&=\frac{P\gamma^3}{2\uppi e}\qty[\frac{(2+e\cos\psi)\sin\psi \mathcal{F}_\theta}{(1+e\cos\psi)^2}-\frac{\cos\psi \mathcal{F}_r}{1+e\cos\psi}]-\cos I\dot{\asc},\\
        \dot{\eps}&=-\frac{P\gamma^4\mathcal{F}_r}{\uppi(1+e\cos\psi)^2}-\gamma(\cos I\dot{\asc}+\dot{\omega}),
    \end{split}
    \end{align}
    where we have defined the dimensionless angular momentum,
    \begin{equation}
        \gamma\equiv\sqrt{1-e^2}=\frac{\ell}{\sqrt{GMa}}.
    \end{equation}
Note that the size and shape of the orbit (determined by $P$ and $e$) is only affected by forces within the plane (i.e., $\mathcal{F}_r$ and $\mathcal{F}_\theta$), while the plane of the orbit (determined by $I$ and $\asc$) is only affected by forces normal to the plane (i.e., $\mathcal{F}_\ell$), and the radial and angular phases of the orbit (determined by $\omega$ and $\eps$) are affected by both.

\subsection{Secular evolution due to relativistic effects}
\label{sec:relativistic-drift}

In Section~\ref{sec:gw-resonance}, we use Eq.~\eqref{eq:Xdot} to calculate the perturbations to the osculating orbital elements caused by resonance with the SGWB.
However, we can use the same set of equations to calculate the perturbations caused by relativistic corrections to the equations of motion, which are particularly important for binaries with short periods.
Following Ref.~\cite{Poisson:2014gr}, we write the relativistic force components to leading post-Newtonian (PN) order as
    \begin{align}
    \begin{split}
    \label{eq:relativistic-drift}
        \mathcal{F}_r&=\qty(\frac{2\uppi}{P})^2\frac{v_P^2}{\gamma^8}(1+e\cos\psi)^3\bigg[3-\eta-e^2(1+3\eta)\\
        &\qquad\qquad\qquad\quad+e(2-4\eta)\cos\psi+e^2\frac{8-\eta}{2}\sin^2\psi\bigg],\\
        \mathcal{F}_\theta&=\qty(\frac{2\uppi}{P})^2\frac{2ev_P^2}{\gamma^8}\sin\psi(1+e\cos\psi)^4(2-\eta),\\
        \mathcal{F}_\ell&=0,
    \end{split}
    \end{align}
    where
    \begin{equation}
        v_P\equiv\qty(\frac{2\uppi GM}{P})^{1/3},
    \end{equation}
    is the binary's rms velocity, and
    \begin{equation}
        \eta\equiv\frac{\mu}{M}=\frac{m_1m_2}{(m_1+m_2)^2},
    \end{equation}
    is the dimensionless mass ratio.
Inserting these expressions into Eq.~\eqref{eq:Xdot}, we average over the orbit to find the secular perturbations, using the fact that $\dot{\psi}=\ell/r^2$ for Keplerian orbits to write
    \begin{equation}
        \dot{X}_\mathrm{sec}\equiv\int_{t_0}^{t_0+P}\frac{\dd{t}}{P}\dot{X}=\int_0^{2\uppi}\frac{\dd{\psi}}{2\uppi}\frac{\gamma^3\dot{X}}{(1+e\cos\psi)^2},
    \end{equation}
    where $X\in(P,e,I,\asc,\omega,\eps)$.
The only non-vanishing perturbations to the osculating elements in this case are then
    \begin{equation}
    \label{eq:relativistic-drift-conservative}
        \dot{\omega}_\mathrm{sec}=\frac{6\uppi v_P^2}{P\gamma^2},\qquad\dot{\eps}_\mathrm{sec}=\frac{2\uppi v_P^2}{P}\qty(6-7\eta-\frac{15-9\eta}{\gamma}),
    \end{equation}
    where the first equality is the famous perihelion precession.
These are both 1PN corrections (i.e., order $v_P^2$).

Note that the above implies that the binary's period and eccentricity are conserved at 1PN order.
However, since these are typically the most precisely-measured orbital elements, it is worth including their leading-order relativistic evolution, even if this is at higher PN order and is thus much smaller than the terms in Eq.~\eqref{eq:relativistic-drift-conservative} for most binaries.
This leading-order evolution of the period and eccentricity is due to GW radiation reaction, and can be calculated by applying simple energy-balance arguments to the unperturbed Keplerian orbit~\cite{Peters:1963ux}, giving
    \begin{align}
    \begin{split}
    \label{eq:relativistic-drift-radiative}
        \dot{P}_\mathrm{sec}&=-\frac{192\uppi\eta v_P^5}{5\gamma^7}\qty(1+\tfrac{73}{24}e^2+\tfrac{37}{96}e^4),\\
        \dot{e}_\mathrm{sec}&=-\frac{608\uppi\eta v_P^5}{15P\gamma^5}\qty(e+\tfrac{121}{304}e^3).
    \end{split}
    \end{align}
We see that these are both 2.5PN effects (i.e., order $v_P^5$).
The inclination and longitude of ascending node are both conserved at this PN order.

\subsection{Small-eccentricity and small-inclination orbits}
\label{sec:small-e-I}

For binaries where the eccentricity is small ($e\lesssim10^{-3}$) the argument of pericentre $\omega$ becomes ill-defined, and this in turn means that the compensated mean anomaly $\eps$ becomes ill-defined, as this is defined relative to the time at which the binary is at pericentre, $t_0$.
These issues can be resolved by defining
    \begin{equation}
    \label{eq:small-e-I}
        \zeta\equiv e\sin\omega,\qquad\kappa\equiv e\cos\omega,\qquad\xi\equiv\omega+\eps.
    \end{equation}
The first two quantities here are sometimes called the ``Laplace-Lagrange eccentric parameters'', while the latter is the compensated mean argument.
We can then describe the orbit of a near-circular binary in terms of the alternative set of osculating elements $(P,\zeta,\kappa,I,\asc,\xi)$~\cite{Murray:2000ssd,Lange:2001rn}.
These evolve according to
    \begin{align}
    \begin{split}
        \dot{\zeta}&=\dot{e}\sin\omega+\dot{\omega}e\cos\omega,\\
        \dot{\kappa}&=\dot{e}\cos\omega-\dot{\omega}e\sin\omega,\\
        \dot{\xi}&=\dot{\omega}+\dot{\eps},
    \end{split}
    \end{align}
    with $\dot{e}$, $\dot{\asc}$, $\dot{\omega}$, and $\dot{\eps}$ given by Eq.~\eqref{eq:Xdot}.
For the relativistic perturbations~\eqref{eq:relativistic-drift}, we thus have
    \begin{align}
    \begin{split}
        \dot{\zeta}_\mathrm{sec}&=\frac{6\uppi v_P^2}{P}\qty(\kappa-\tfrac{304}{45}\eta v_P^3\zeta),\\
        \dot{\kappa}_\mathrm{sec}&=-\frac{6\uppi v_P^2}{P}\qty(\zeta+\tfrac{304}{45}\eta v_P^3\kappa),\\
        \dot{\xi}_\mathrm{sec}&=-\frac{4\uppi v_P^2}{P}(3-\eta),
    \end{split}
    \end{align}
    where we have neglected $\order*{e^2}$ terms.

Similarly, $\asc$ is ill-defined for orbits with very small inclination, so in this case we define
    \begin{equation}
        p\equiv I\sin\asc,\qquad q\equiv I\cos\asc,\qquad\lambda=\asc+\xi,
    \end{equation}
    and describe the orbit using $(P,\zeta,\kappa,p,q,\lambda)$.

\section{Resonant gravitational-wave perturbations}
\label{sec:gw-resonance}

In this section we calculate the evolution of the osculating orbital elements of a binary system due to resonance with the SGWB.
We start by specifying the perturbing force associated with an incoming plane GW in the limit where the wavelength is much larger than the size of the orbit.
This allows us to write down a Langevin equation describing individual random realisations of the stochastic evolution of the osculating elements.
Using the statistical properties of the SGWB, we then derive a secularly-averaged FPE which describes the evolution of the full statistical distribution of the orbital elements over timescales much longer than the binary period.

\subsection{Coupling to the gravitational-wave polarisation modes}

The response of a binary to an impinging plane GW can be expressed in the proper detector frame (i.e., using Fermi normal coordinates to construct a freely-falling frame with the binary's centre of mass fixed at the origin, in which coordinate distances correspond to proper distances to a good approximation, and in which we can treat the GW as a perturbing Newtonian force) as~\cite{Misner:1974qy,Maggiore:1900zz}
    \begin{equation}
    \label{eq:gw-force}
        \updelta\ddot{r}^i=\frac{1}{2}\ddot{h}_{ij}r^j,
    \end{equation}
    so that the resulting evolution of the binary is described in terms of the perturbing force terms
    \begin{equation}
        \mathcal{F}_r=\frac{1}{2}\ddot{h}_{ij}\hat{r}^i\hat{r}^j,\quad \mathcal{F}_\theta=\frac{1}{2}\ddot{h}_{ij}\hat{r}^i\hat{\theta}^j,\quad \mathcal{F}_\ell=\frac{1}{2}\ddot{h}_{ij}\hat{r}^i\hat{\ell}^j,
    \end{equation}
    where $h_{ij}(t)$ is the transverse-traceless part of the metric perturbation at the position of the binary's centre of mass.

We decompose the GW strain in terms of plane waves of each polarisation arriving from each direction on the sky,
    \begin{equation}
        h_{ij}(t)=\int_{S^2}\dd[2]{\vu*n}e_{ij}^A(\vu*n)h_A(t,\vu*n),
    \end{equation}
    where $A=+,\times$ are the two GW polarisations, and summation over the repeated polarisation index is implied.
We define the standard polarisation tensors
    \begin{equation}
        e_{ij}^+=\hat{u}^i\hat{u}^j-\hat{v}^i\hat{v}^j,\qquad e_{ij}^\times=\hat{u}^i\hat{v}^i+\hat{v}^i\hat{u}^j,
    \end{equation}
    with $\vu*u,\vu*v$ unit vectors that are orthogonal to the GW propagation direction $\vu*n$ and to each other, as illustrated in Fig.~\ref{fig:orbital-elements}.
We therefore write
    \begin{equation}
        \label{eq:force-ddot-h}
        \mathcal{F}_\alpha=\frac{1}{2}\int_{S^2}\dd[2]{\vu*n}e_{ij}^A\hat{r}^i\hat{\alpha}^j\ddot{h}_A,
    \end{equation}
    where $\alpha$ runs over the cylindrical coordinates $(r,\theta,\ell)$.

Note that Eq.~\eqref{eq:gw-force} is only correct in the limit where the GW wavelength $\lambda$ is much larger than the size of the binary's orbit, $\lambda\gg a$.
Using Kepler's third law~\eqref{eq:kepler-iii}, this condition can be rewritten as
    \begin{equation}
        fP\ll1/v_P,
    \end{equation}
    where $f$ is the GW frequency.
Since we are interested in GW frequencies which are harmonics of the binary period ($f=n/P$ for some $n\in\mathbb{Z}_+$), this tells us that the analysis below, based on Eq.~\eqref{eq:gw-force}, is only valid for harmonics that satisfy $n\ll 1/v_P$.
This is not an impediment, since in the cases we are interested in the binary is sub-relativistic, $v_P\ll1$, and the strongest contribution typically comes from the lowest few harmonics anyway.

\subsection{Langevin formulation}

The stochastic evolution of the binary is described by the evolution equations derived in Sec.~\ref{sec:perturbing-force}, with the perturbing force terms given by Eq.~\eqref{eq:force-ddot-h}.
All this can be rewritten as a coupled set of nonlinear Langevin equations,
    \begin{equation}
    \label{eq:langevin}
        \dot{X}_i(\vb*X,t)=V_i(\vb*X)+\Gamma_i(\vb*X,t),
    \end{equation}
    where $X_i$ runs over the set of orbital elements [either $(P,e,I,\asc,\omega,\eps)$, or the small-eccentricity and/or small-inclination alternatives described in Sec.~\ref{sec:small-e-I}], $V$ is the deterministic drift term due to the relativistic effects described in Sec.~\ref{sec:relativistic-drift}, and $\Gamma$ is the stochastic diffusion term due to resonance with the SGWB.
The latter can be written as
    \begin{equation}
    \label{eq:Gamma-def}
        \Gamma_i(\vb*X,t)=\int_{S^2}\dd[2]{\vu*n}T_i^A(X,t,\vu*n)\ddot{h}_A(t,\vu*n),
    \end{equation}
where the $T_i^A$ are transfer functions describing the coupling between the SGWB strain and the orbital elements.
For example, using Eqs.~\eqref{eq:Xdot} and~\eqref{eq:force-ddot-h}, the transfer functions for the period $P$ are
    \begin{equation}
        T_{P}^A=\frac{3P^2\gamma}{4\uppi}\qty(\frac{e\sin\psi}{1+e\cos\psi}\hat{r}^i+\hat{\theta}^i)\hat{r}^je^A_{ij}.
    \end{equation}
(The full set of transfer functions are written out in Appendix~\ref{sec:transfer} in terms of their Fourier components.)

The explicit time-dependence in the transfer functions is via the true anomaly $\psi(t)$, due to the variation of the binary's response over the course of each orbit.
These variations occur only on timescales less than or equal to the orbital period $P$, which is much shorter than the timescales over which the orbital elements evolve, $X/\dot{X}\gg P$.
The latter are \emph{secular} effects, whereas the former are \emph{resonant} ones, which contribute only at integer multiples of the fundamental frequency $1/P$.
This property of the resonant spectrum, combined with the separation of scales between the resonant and secular evolution, allows us to approximate the transfer functions as Fourier series with period $P$,
    \begin{equation}
    \label{eq:transfer-function-sum}
        T^A_i(\vb*X,t,\vu*n)=\sum_{n=-\infty}^{+\infty}\rme^{-2\uppi\rmi nt/P}T^A_{i,n}(\vb*X,\vu*n).
    \end{equation}

In Sec.~\ref{sec:fokker-planck}, we recast Eq.~\eqref{eq:langevin} as a FPE, allowing us to calculate the ensemble-averaged properties of the binary's orbital evolution.
However, in order to do so, we must first describe the statistical properties of the SGWB strain, which we do in Sec.~\ref{sec:sgwb} below.

\subsection{Statistical properties of the stochastic background}
\label{sec:sgwb}

The stochastic GW strain components $h_A$ are time-dependent random variables whose statistical properties are usually specified in terms of their Fourier transforms,
    \begin{equation}
        \tilde{h}_A(f,\vu*n)=\int_{-\infty}^{+\infty}\dd{t}\rme^{2\uppi\rmi ft}h_A(t,\vu*n).
    \end{equation}
Since the SGWB is generated by the superposition of a large number of statistically independent sources, it is usually assumed that the SGWB strain is Gaussian by the central limit theorem.
This implies that the statistics of the strain Fourier components are fully specified by their first two moments.
With the further standard assumptions that the SGWB is isotropic, stationary, unpolarised, and has zero phase correlation between different sky locations, these moments can be written as
    \begin{align}
    \begin{split}
        \ev*{\tilde{h}_A(f,\vu*n)}&=0,\\
        \ev*{\tilde{h}_A(f,\vu*n)\tilde{h}^*_{A'}(f',\vu*n')}&=\frac{3H_0^2\Omega(f)}{32\uppi^3|f|^3}\delta_{AA'}\delta(f-f')\delta(\vu*n,\vu*n'),
    \end{split}
    \end{align}
    with $\Omega(f)$ the one-sided SGWB energy density spectrum,
    \begin{equation}
        \Omega(f)\equiv\frac{1}{\rho_\mathrm{crit}}\dv{\rho_\mathrm{gw}}{(\ln f)},
    \end{equation}
    which is normalised with respect to the cosmological critical density $\rho_\mathrm{crit}\equiv3H_0^2/(8\uppi G)$, where $H_0$ is the Hubble constant.

We are interested in the statistics of the second time derivative of the strain, which is related to the Fourier components by
    \begin{align}
    \begin{split}
        \ddot{h}_A&=\dv[2]{}{t}\int_{-\infty}^{+\infty}\dd{f}\rme^{-2\uppi\rmi ft}\tilde{h}_A\\
        &=-4\uppi^2\int_{-\infty}^{+\infty}\dd{f}\rme^{-2\uppi\rmi ft}f^2\tilde{h}_A.
    \end{split}
    \end{align}
The corresponding moments are therefore
    \begin{align}
    \begin{split}
        \label{eq:ddot-h-moments}
        \ev*{\ddot{h}_A(t,\vu*n)}&=0,\\
        \ev*{\ddot{h}_A(t,\vu*n)\ddot{h}_{A'}(t',\vu*n')}&=3\uppi H_0^2\delta_{AA'}\delta(\vu*n,\vu*n')\\
        &\times\int_0^\infty\dd{f}\cos[2\uppi f(t-t')]f\Omega(f).
    \end{split}
    \end{align}

\subsection{From the Langevin equation to the Fokker-Planck equation}
\label{sec:fokker-planck}

Solving the Langevin equation~\eqref{eq:langevin} gives individual random trajectories of the binary through parameter space, corresponding to different random realisations of the SGWB.
However, we are more interested in the ensemble of all possible random trajectories, which we describe in terms of the time-dependent distribution function (DF) for the orbital elements, $W(\vb*X,t)$.
This is defined such that the probability of the orbital elements $\vb*X$ belonging to any region $\mathcal{X}$ of parameter space at time $t$ is given by the corresponding integral over the DF,
    \begin{equation}
        \Pr(\vb*X\in\mathcal{X}|t)=\int_{\mathcal{X}}\dd{\vb*X}W(\vb*X,t).
    \end{equation}
This DF can either be interpreted as the probability density function for the stochastic orbital elements of an individual binary, or as the cumulative distribution for the orbital elements of a population of multiple binaries.

Formally, we can write down the time evolution of the DF in terms of the Kramers-Moyal (KM) forward expansion~\cite{Risken:1989fpe,Gardiner:2004hsm},
    \begin{equation}
    \label{eq:KM-forward}
        \pdv{W}{t}=\sum_{n=1}^\infty(-)^n\frac{\partial^n}{\partial{X_{i_1}}\partial{X_{i_2}}\cdots\partial{X_{i_n}}}\qty(D^{(n)}_{i_1i_2\cdots i_n}W),
    \end{equation}
    where repeated indices are summed over.
This is determined by the KM coefficients,
    \begin{equation}
        D^{(n)}_{i_1i_2\cdots i_n}(\vb*X,t)\equiv\lim_{\tau\to0}\frac{1}{\tau n!}\ev{\prod_{j=1}^n\qty[X_{i_j}(t+\tau)-X_{i_j}(t)]},
    \end{equation}
    with angle brackets indicating an ensemble average under the distribution $W$ at time $t$.
Since we take the SGWB as Gaussian, the KM coefficients for orders $n\ge3$ all vanish,\footnote{%
One could derive this explicitly by using the approach described later in this Section to show that $D^{(3)}_{ijk}=0$, as the Pawula theorem~\cite{Pawula:1967zz,Risken:1989fpe} then implies that all higher-order coefficients $n>3$ must vanish in order to guarantee that the DF is normalised.}
leaving just the first two coefficients,
    \begin{align}
    \begin{split}
        D^{(1)}_i&=\lim_{\tau\to0}\frac{1}{\tau}\ev{X_i(t+\tau)-X_i(t)},\\
        D^{(2)}_{ij}&=\lim_{\tau\to0}\frac{1}{2\tau}\ev{[X_i(t+\tau)-X_i(t)][X_j(t+\tau)-X_j(t)]},
    \end{split}
    \end{align}
    which we call the \emph{drift vector} and the \emph{diffusion matrix}, respectively.
The KM forward expansion~\eqref{eq:KM-forward} then becomes the \emph{Fokker-Planck equation} (FPE),
    \begin{equation}
    \label{eq:fpe}
        \pdv{W}{t}=-\partial_i(D^{(1)}_iW)+\partial_i\partial_j(D^{(2)}_{ij}W),
    \end{equation}
    with $\partial_i\equiv\pdv*{}{X_i}$.

We can calculate the KM coefficients by directly integrating the Langevin equation~\eqref{eq:langevin}, from some initial time $t$ where the orbital elements are ``sharp'' (i.e., known exactly rather than randomly distributed), $X_i(t)\equiv x_i$, over some small time interval $\tau$,
    \begin{equation}
    \label{eq:direct-integration}
        X_i(t+\tau)-x_i=\int^{t+\tau}_t\dd{t'}[V_i(\vb*X(t'))+\Gamma_i(\vb*X(t'),t')].
    \end{equation}
(This derivation closely follows that in Sec.~3.3.2 of Ref.~\cite{Risken:1989fpe}.)
Both terms under the integral on the RHS are random, due to the random spread in the orbital elements for all times $t'>t$.
However, we can express these in terms of the sharp values $x_i$ by Taylor expanding,
    \begin{align}
    \begin{split}
    \label{eq:taylor-expansion}
        V_i(\vb*X(t'))&=V_i(\vb*x)+\partial_jV_i(\vb*x)[X_j(t')-x_j]+\cdots,\\
        \Gamma_i(\vb*X(t'),t')&=\Gamma_i(\vb*x,t')+\partial_j\Gamma_i(\vb*x,t')[X_j(t')-x_j]+\cdots.
    \end{split}
    \end{align}
Wherever $X_i(t')-x_i$ appears on the RHS of Eq.~\eqref{eq:taylor-expansion} we can insert Eq.~\eqref{eq:direct-integration} and iterate, such that the only orbital elements that appear are the sharp initial values $x_i$.
Then the only randomness is through the stochastic term $\Gamma_i$, whose statistics are completely specified in terms of the SGWB moments~\eqref{eq:ddot-h-moments}.

The first few terms in the iterative expansion are
    \begin{align}
    \begin{split}
        X_i&(t+\tau)-x_i\\
        &=\int_t^{t+\tau}\dd{t'}V_i(\vb*x)\\
        &+\int_t^{t+\tau}\dd{t'}\partial_jV_i(\vb*x)\int_t^{t'}\dd{t''}[V_j(\vb*x)+\Gamma_j(\vb*x,t'')]\\
        &+\cdots\\
        &+\int_t^{t+\tau}\dd{t'}\Gamma_i(\vb*x,t')\\
        &+\int_t^{t+\tau}\dd{t'}\partial_j\Gamma_i(\vb*x,t')\int_t^{t'}\dd{t''}[V_j(\vb*x)+\Gamma_j(\vb*x,t'')]\\
        &+\cdots.
    \end{split}
    \end{align}
By virtue of the time-independence of the sharp drift term $V_i(\vb*x)$, this immediately simplifies to
    \begin{align}
    \begin{split}
    \label{eq:simplified-expansion}
        &X_i(t+\tau)-x_i\\
        &=\tau V_i(\vb*x)+\frac{1}{2}\tau^2V_j(\vb*x)\partial_jV_i(\vb*x)\\
        &+\partial_jV_i(\vb*x)\int_t^{t+\tau}\dd{t'}\int_t^{t'}\dd{t''}\Gamma_j(\vb*x,t'')+\cdots\\
        &+\int_t^{t+\tau}\dd{t'}\Gamma_i(\vb*x,t')+V_j(x)\int_t^{t+\tau}\dd{t'}(t'-t)\partial_j\Gamma_i(\vb*x,t')\\
        &+\int_t^{t+\tau}\dd{t'}\partial_j\Gamma_i(\vb*x,t')\int_t^{t'}\dd{t''}\Gamma_j(\vb*x,t'')+\cdots.
    \end{split}
    \end{align}
Taking the first moment of Eq.~\eqref{eq:simplified-expansion}, all terms linear in $\Gamma_i$ vanish, and we are left with
    \begin{align}
    \begin{split}
        &\ev{X_i(t+\tau)-x_i}\\
        &=\tau V_i(\vb*x)+\frac{1}{2}\tau^2V_j(\vb*x)\partial_jV_i(\vb*x)+\cdots\\
        &\quad+\int_t^{t+\tau}\dd{t'}\int_t^{t'}\dd{t''}\ev{\Gamma_j(\vb*x,t'')\partial_j\Gamma_i(\vb*x,t')}+\cdots,
    \end{split}
    \end{align}
so that the drift vector is given by
    \begin{equation}
    \label{eq:KM1}
        D^{(1)}_i=V_i+\lim_{\tau\to0}\frac{1}{\tau}\int_t^{t+\tau}\dd{t'}\int_t^{t'}\dd{t''}\ev{\Gamma_j(\vb*x,t'')\partial_j\Gamma_i(\vb*x,t')}.
    \end{equation}
Similarly, when taking the second moment of Eq.~\eqref{eq:simplified-expansion} and taking the $\tau\to0$ limit, the only term that survives for the diffusion matrix is
    \begin{equation}
    \label{eq:KM2}
        D^{(2)}_{ij}=\lim_{\tau\to0}\frac{1}{2\tau}\int_t^{t+\tau}\dd{t'}\int_t^{t+\tau}\dd{t''}\ev{\Gamma_i(\vb*x,t')\Gamma_j(\vb*x,t'')}.
    \end{equation}
By calculating the KM coefficients from Eqs.~\eqref{eq:KM1} and~\eqref{eq:KM2}, we can then obtain the full DF for the orbital elements by integrating the FPE~\eqref{eq:fpe}.

\subsection{Secular drift and diffusion}

We rewrite the integrands of Eqs.~\eqref{eq:KM1} and~\eqref{eq:KM2} in terms of the GW strain and the binary transfer functions using Eq.~\eqref{eq:Gamma-def}.
We can then use Eq.~\eqref{eq:ddot-h-moments} to specify the SGWB strain covariance, and insert Eq.~\eqref{eq:transfer-function-sum} to form a sum over harmonics of the binary's orbital frequency, $f=n/P$.
The resulting expressions are
    \begin{align}
    \begin{split}
    \label{eq:gamma-2nd-moments}
        &\ev{\Gamma_i(\vb*x,t')\Gamma_j(\vb*x,t'')}\\
        &\quad=3\uppi H_0^2\sum_{n=-\infty}^{+\infty}\sum_{m=-\infty}^{+\infty}\int_{S^2}\dd[2]{\vu*n}T^{A*}_{i,n}T^A_{j,m}\\
        &\qquad\times\int_0^\infty\dd{f}\rme^{2\uppi\rmi(nt'-mt'')/P}\cos[2\uppi f(t'-t'')]f\Omega(f),\\
        &\ev{\Gamma_j(\vb*x,t'')\partial_j\Gamma_i(\vb*x,t')}\\
        &\quad=3\uppi H_0^2\sum_{n=-\infty}^{+\infty}\sum_{m=-\infty}^{+\infty}\int_{S^2}\dd[2]{\vu*n}T^A_{j,m}\partial_jT^{A*}_{i,n}\\
        &\qquad\times\int_0^\infty\dd{f}\rme^{2\uppi\rmi(nt'-mt'')/P}\cos[2\uppi f(t'-t'')]f\Omega(f).
    \end{split}
    \end{align}
In order to derive the corresponding KM coefficients, we must evaluate two oscillatory time integrals,
    \begin{align}
    \begin{split}
    \label{eq:time-integrals}
        \int_t^{t+\tau}\dd{t'}\int_t^{t+\tau}\dd{t''}&\rme^{-2\uppi\rmi(f-n/P)t'}\rme^{2\uppi\rmi(f-m/P)t''},\\
        \int_t^{t+\tau}\dd{t'}\int_t^{t'}\dd{t''}&\rme^{-2\uppi\rmi(f-n/P)t'}\rme^{2\uppi\rmi(f-m/P)t''}.
    \end{split}
    \end{align}
To do so, we recall that the timescales $\lesssim P$ associated with the binary's resonant frequencies are much shorter than the secular timescales $X/\dot{X}$ we are interested in.\footnote{%
    It would also be interesting to study the dynamics on sub-orbital timescales, as was done in Refs.~\cite{Rozner:2019gba,Desjacques:2020fdi} for the case of ultralight dark matter.
    We leave this for future work.}
Even though we will later take the limit $\tau\to0$, all we really require to derive a FPE valid on secular timescales $\sim X/\dot{X}$ is that $\tau\ll X/\dot{X}$, and since we have $P\ll X/\dot{X}$, we can consistently also demand that $\tau\gg P$.
In this limit, the first integral in Eq.~\eqref{eq:time-integrals} is approximated by
    \begin{align}
    \begin{split}
        \int_t^{t+\tau}\dd{t'}\int_t^{t+\tau}\dd{t''}&\rme^{-2\uppi\rmi(f-n/P)t'}\rme^{2\uppi\rmi(f-m/P)t''}\\
        &\qquad\simeq\tau\delta_{mn}\delta(f-n/P).
    \end{split}
    \end{align}
For the second integral, notice that we can write
    \begin{align}
    \begin{split}
        &\tau\delta_{mn}\delta(f-n/P)\\
        &\simeq\int_t^{t+\tau}\dd{t'}\int_t^{t+\tau}\dd{t''}\rme^{-2\uppi\rmi(f-n/P)t'}\rme^{2\uppi\rmi(f-m/P)t''}\\
        &=\int_t^{t+\tau}\dd{t'}\int_t^{t'}\dd{t''}\rme^{-2\uppi\rmi(f-n/P)t'}\rme^{2\uppi\rmi(f-m/P)t''}\\
        &\quad+\int_t^{t+\tau}\dd{t'}\int_{t'}^{t+\tau}\dd{t''}\rme^{-2\uppi\rmi(f-n/P)t'}\rme^{2\uppi\rmi(f-m/P)t''},
    \end{split}
    \end{align}
and that in the limit $\tau\gg P$ the latter two terms are approximately equal to each other, so that
    \begin{align}
    \begin{split}
        \int_t^{t+\tau}\dd{t'}\int_t^{t'}\dd{t''}&\rme^{-2\uppi\rmi(f-n/P)t'}\rme^{2\uppi\rmi(f-m/P)t''}\\
        &\qquad\simeq\frac{1}{2}\tau\delta_{mn}\delta(f-n/P).
    \end{split}
    \end{align}
Plugging this back in to Eqs.~\eqref{eq:KM1},~\eqref{eq:KM2}, and~\eqref{eq:gamma-2nd-moments}, we are able to write the KM coefficients directly in terms of the GW transfer functions,
    \begin{align}
    \begin{split}
    \label{eq:km-final}
        D^{(1)}_i&=V_i+\frac{3\uppi}{2}H_0^2\sum_{n=1}^\infty\int_{S^2}\dd[2]{\vu*n}\frac{n\Omega_n}{P}\Re\qty(T^A_{j,n}\partial_jT^{A*}_{i,n}),\\
        D^{(2)}_{ij}&=\frac{3\uppi}{2}H_0^2\sum_{n=1}^\infty\int_{S^2}\dd[2]{\vu*n}\frac{n\Omega_n}{P}\Re\qty(T^{A*}_{i,n}T^A_{j,n}),
    \end{split}
    \end{align}
    where we have used $T_{i,-n}=T^*_{i,n}$, and where $\Omega_n\equiv\Omega(n/P)$ is the SGWB energy density at the binary's $n^\mathrm{th}$ harmonic frequency.
(We always take only the real part when defining the KM coefficients, so for brevity we will leave this implicit from now on.)
Equation~\eqref{eq:km-final} describes the secular drift and diffusion of the orbital elements over timescales much longer than the orbital period $P$.

Note that the drift vector includes not just the expected deterministic drift $V_i$, but also a stochastic term.
As we show below, this ``noise-induced drift'' leads to a net evolution of the \emph{mean values} of the orbital elements, and not just their variance.
This is somewhat counter-intuitive, and justifies the careful derivation presented in this Section; otherwise, it would be tempting to assume that the SGWB affects only the variance of the orbital elements (as was assumed in Ref.~\cite{Hui:2012yp}, for example).
One can understand this drift as being the result of a ``diffusion gradient'' due to the nonlinear coupling between the SGWB and the binary: the orbital elements change in response to the GW strain, thereby changing the values of the transfer functions and modifying the response to the strain, with the resulting feedback loop creating a preferred direction in parameter space for the orbital elements to evolve towards.
The partial derivative acting on the transfer function in Eq.~\eqref{eq:km-final} shows that this term would vanish in the linear case (i.e., where the transfer functions are constant), and therefore that this is a purely nonlinear effect.

The stochastic drift is also a result of our treatment of the SGWB strain, as we have started with a finite correlation time and taken the limit where this vanishes, rather than specifying zero correlation time from the start.
(Physically, the SGWB strain at time $t_1$ cannot be uncorrelated with that at $t_2$ for arbitrarily small $|t_1-t_2|$, as this would require $\Omega(f)$ to remain finite as $f\to\infty$, which would make the total GW energy density diverge.
However, the time over which the strain is correlated is still much shorter than the timescale over which the orbital elements evolve.)
This is the Stratonovich prescription~\cite{Stratonovich:1968mp}, which is appropriate for most physical applications, rather than the more formal It\^o prescription~\cite{Ito:1944si}.
For further discussion of the two prescriptions see, e.g., Ref.~\cite{Risken:1989fpe}.

\section{Calculating the Kramers-Moyal coefficients}
\label{sec:KM}

\begin{figure*}[p!]
    \includegraphics[width=0.4965\textwidth]{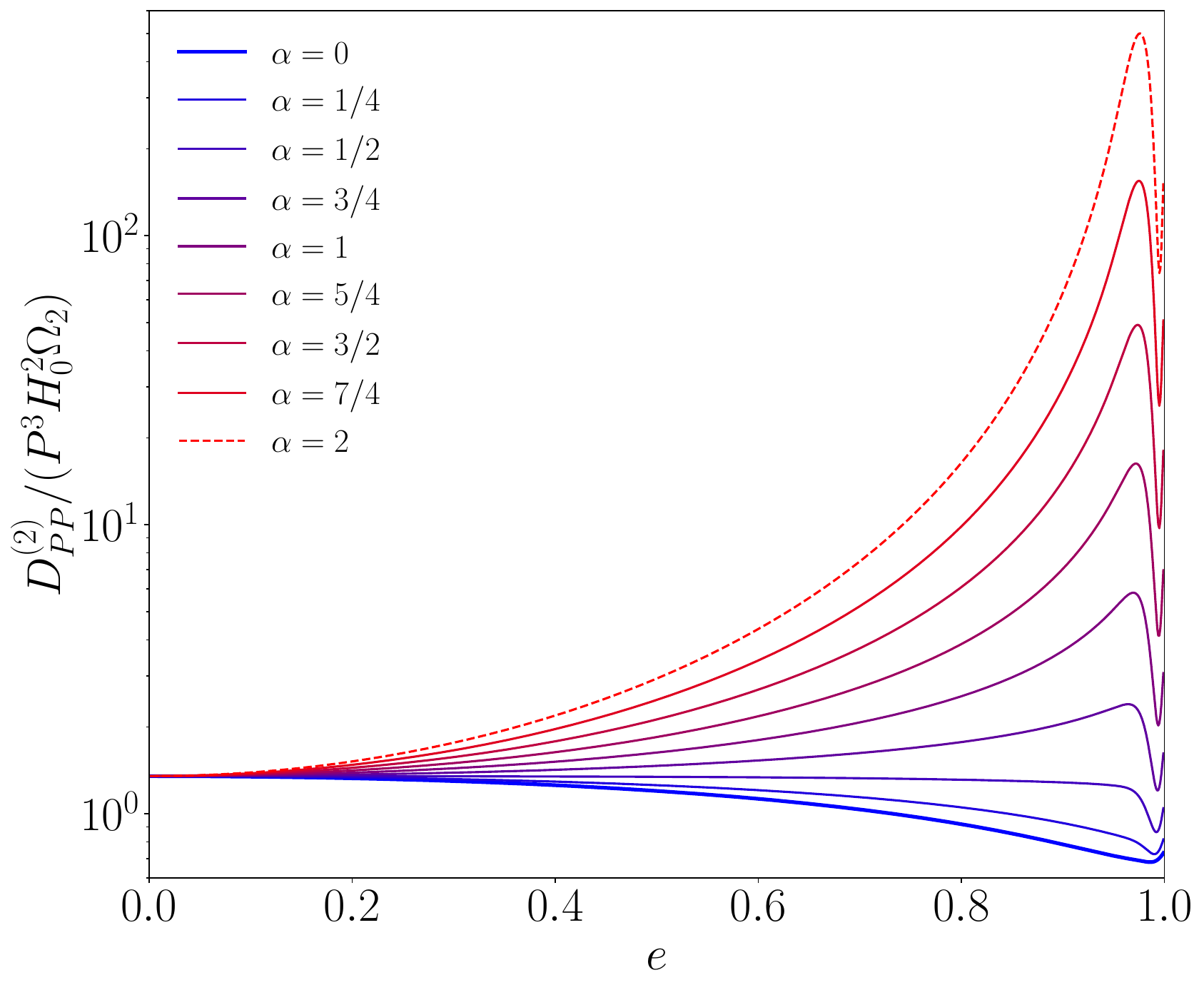}
    \includegraphics[width=0.4965\textwidth]{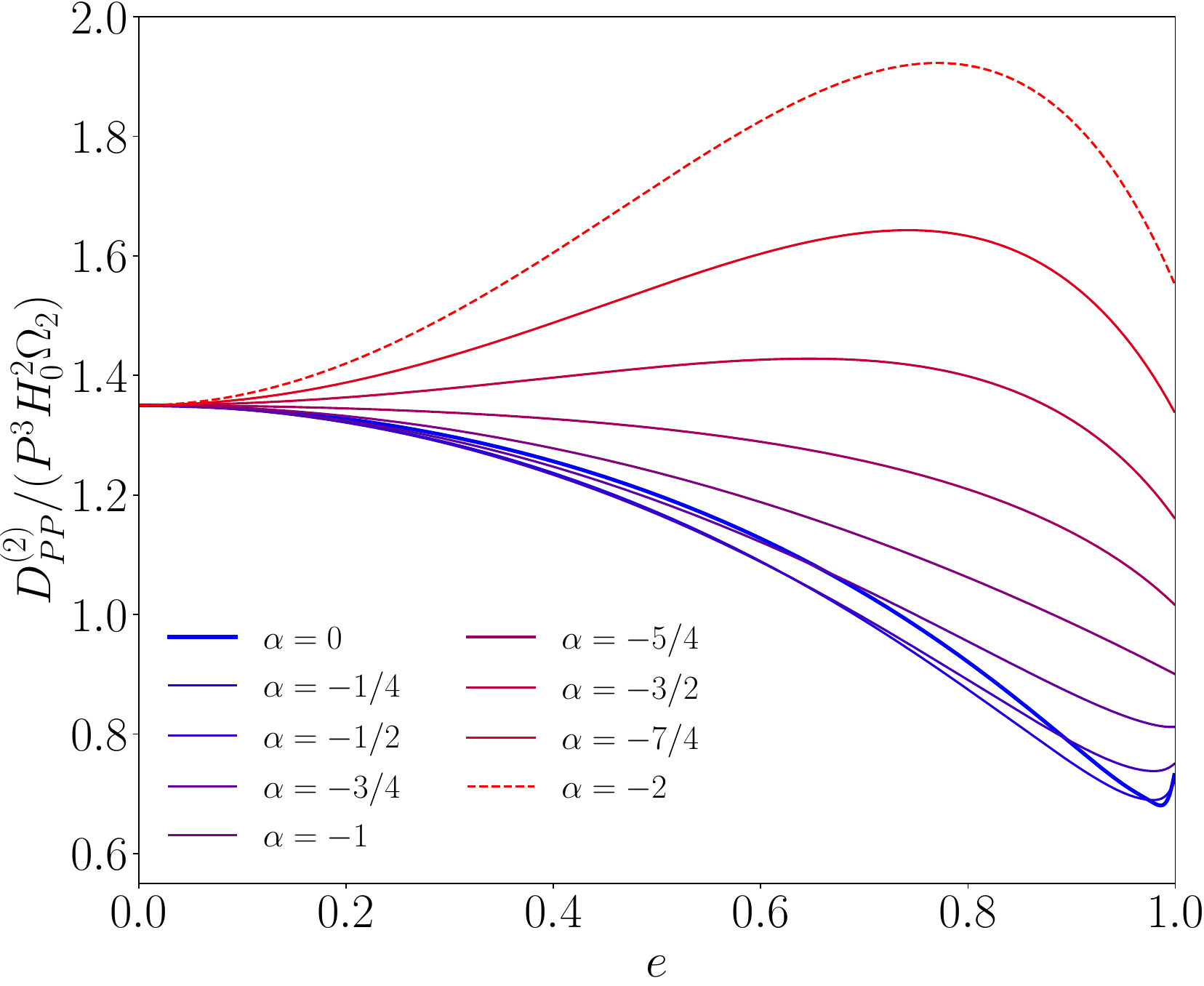}
    \includegraphics[width=0.4965\textwidth]{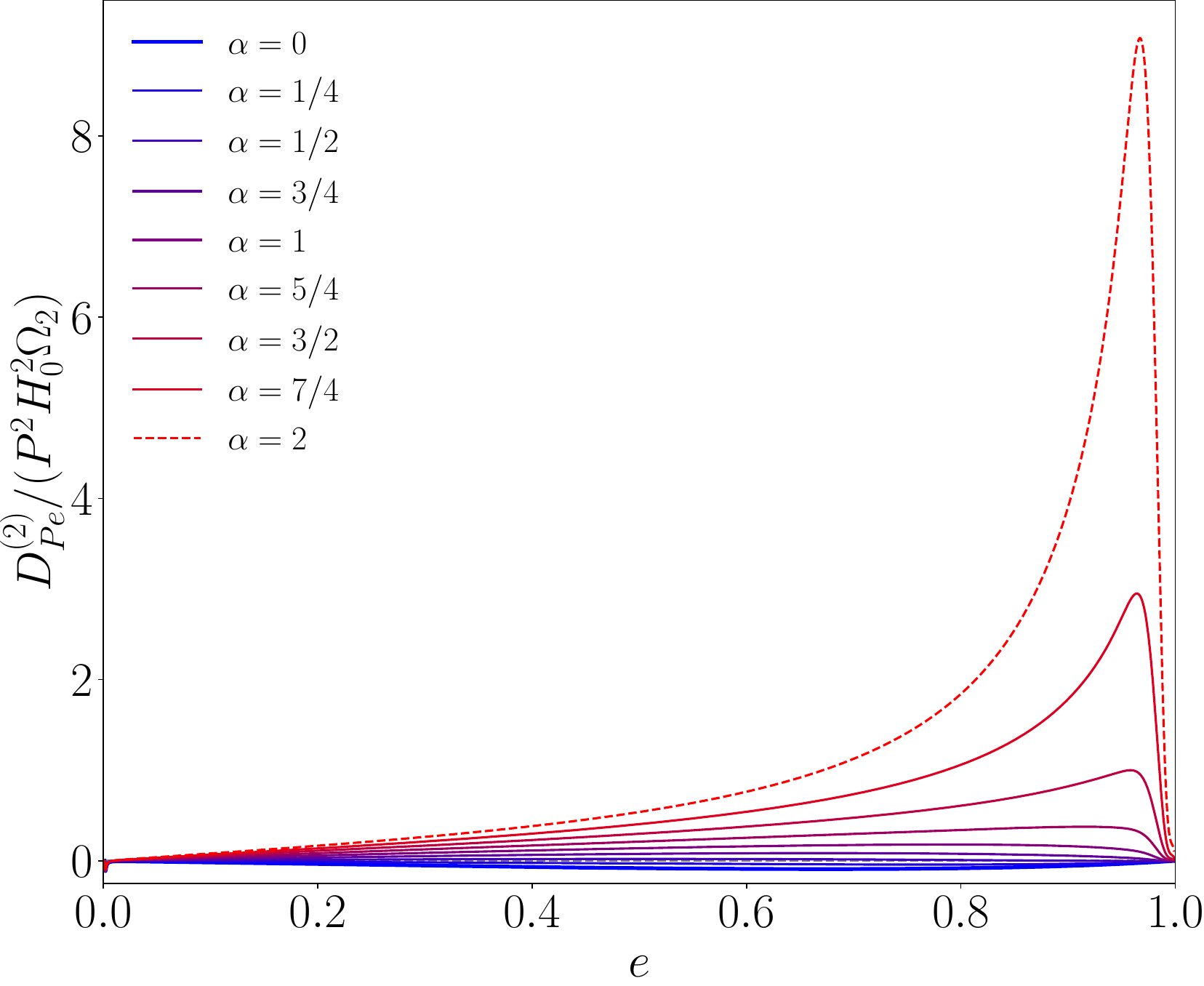}
    \includegraphics[width=0.4965\textwidth]{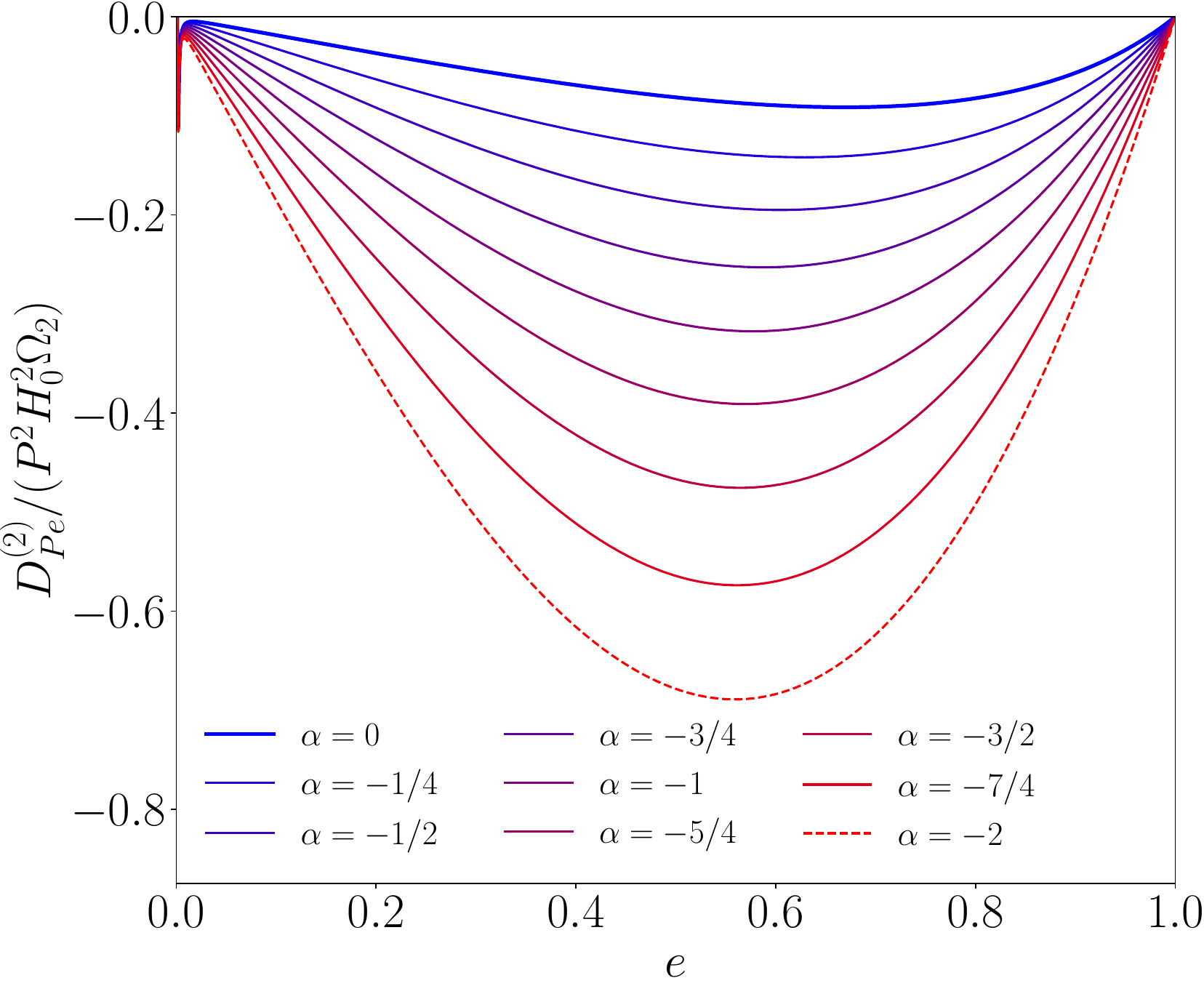}
    \includegraphics[width=0.4965\textwidth]{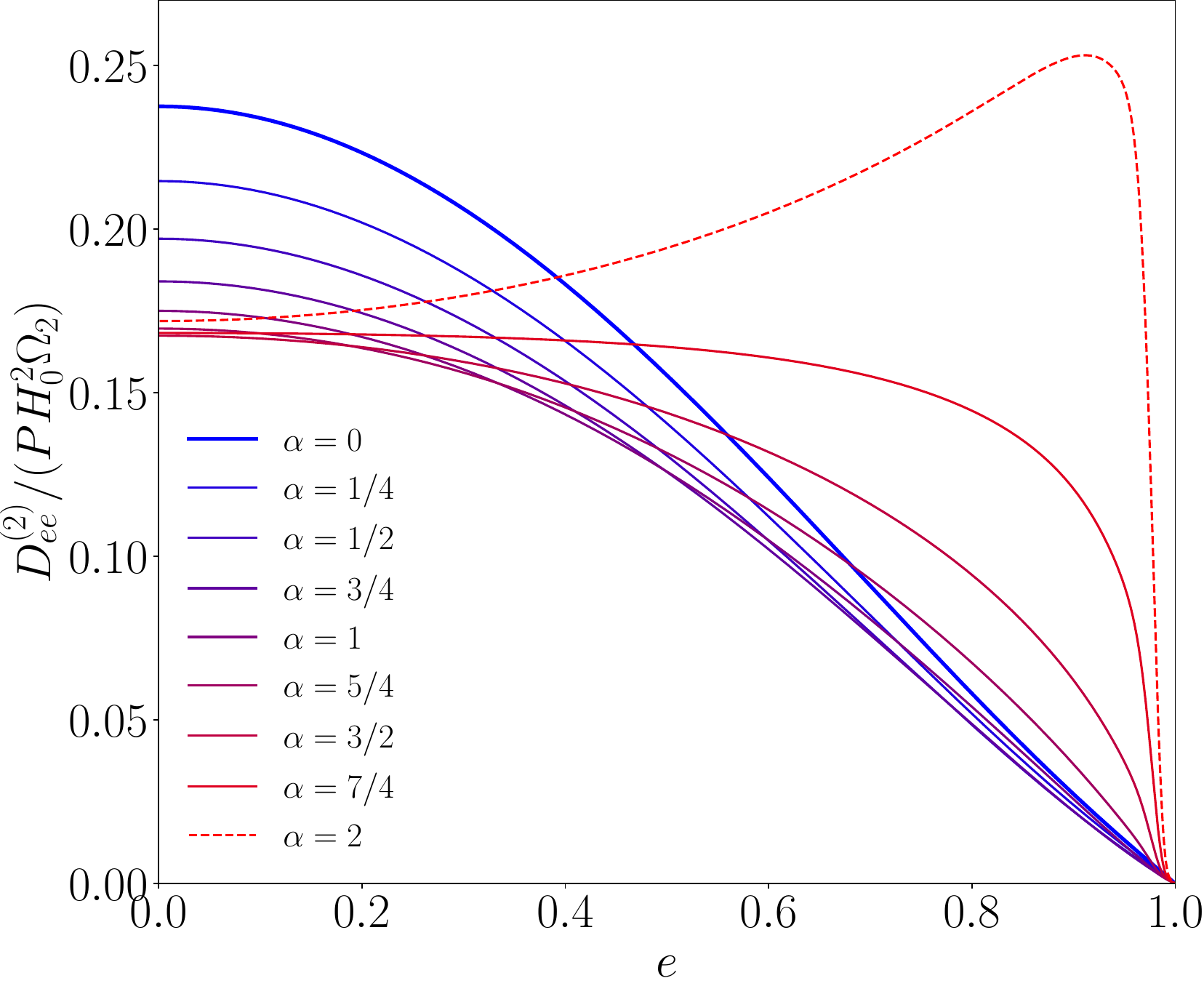}
    \includegraphics[width=0.4965\textwidth]{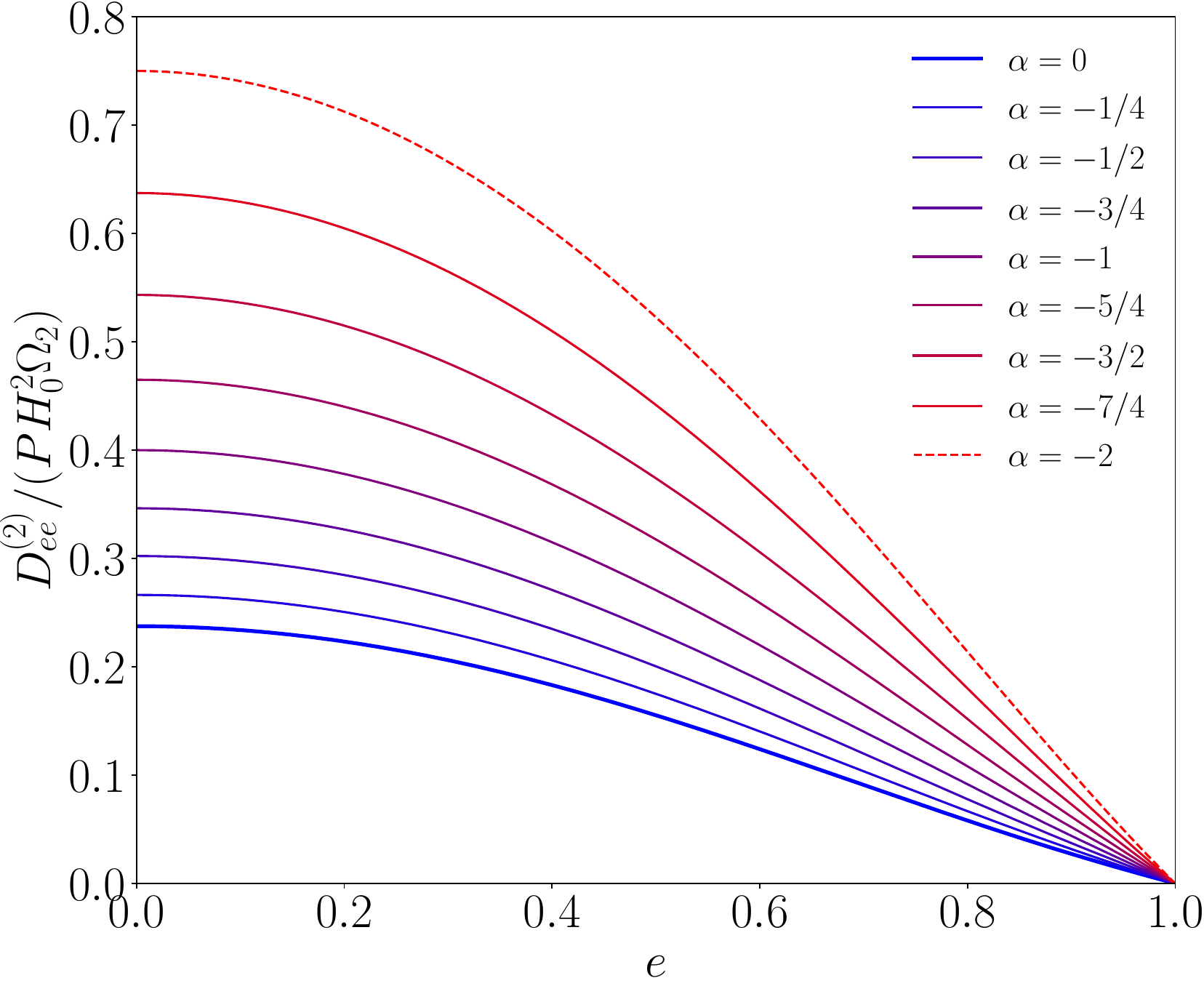}
    \caption{%
    The secular diffusion coefficients $D^{(2)}_{PP}$, $D^{(2)}_{Pe}$, and $D^{(2)}_{ee}$ (first, second, and third rows, respectively) as functions of eccentricity, for various power-law SGWB spectra, $\Omega(f)\sim f^\alpha$.
    The left column shows positive power-law indices, $\alpha=0,\dots,2$, while the right column shows negative indices, $\alpha=0,\dots,-2$.}
    \label{fig:km2-ecc}
\end{figure*}

In this section we use Eq.~\eqref{eq:km-final} to derive explicit expressions for the KM coefficients as functions of the binary orbital elements and the SGWB spectrum.
We start with the most general expressions, before specialising to the cases of small eccentricity and small inclination.

\subsection{General orbits}

We present here the secular KM coefficients for general eccentricity $e\in(0,1)$.
The details of this calculation are lengthy and unimportant for the final results, so we quote only the final expressions here, with some further details given in the Appendices.
In particular, Appendix~\ref{sec:polarisation-tensor-projections} derives the necessary projections of the GW polarisation tensors onto the binary's cylindrical coordinate basis; Appendix~\ref{sec:transfer} uses these projections to write down the transfer functions for all six orbital elements $(P,e,I,\asc,\omega,\eps)$ in terms of functions of the eccentricity called \emph{Hansen coefficients}, which we introduce below; Appendix~\ref{sec:reference-frame} derives the equations describing how each of the KM coefficients transforms under a change of the reference frame, allowing us to select a particular frame which simplifies the calculations; Appendix~\ref{sec:KM-primed-frame} presents the values of the KM coefficients in this frame in terms of the Hansen coefficients; and Appendix~\ref{sec:hansen} writes out all of the necessary Hansen coefficients explicitly as functions of eccentricity.

\begin{widetext}
Putting all of these ingredients together, we find that the secular drift vector $D^{(1)}_{i}$ is given in terms of the Hansen coefficients $(C^{lm}_n,S^{lm}_n,E^{lm}_n)$ by
    \begin{align}
    \begin{split}
    \label{eq:ecc-drift}
        D^{(1)}_P&=V_P+\frac{9P^2\gamma^2}{4}\sum_{n=1}^\infty nH_0^2\Omega_n\bigg[\qty|E^{02}_n+\frac{e}{2}(E^{11}_n-E^{13}_n)|^2-\frac{1+4e^2}{15}(S^{11}_n)^2-\frac{e\gamma^2}{15}{S^{11}_n}'S^{11}_n\\
        &\quad+\frac{\gamma^2}{10e}\qty(E^{02}_n+\frac{e}{2}(E^{11}_n-E^{13}_n))\qty(3E^{11}_n+E^{13}_n+eE^{20}_n+4E^{21}_n+4eE^{22}_n-4E^{23}_n-eE^{24}_n+2{E^{02}_n}'+e({E^{11}_n}'-{E^{13}_n}'))^*\\
        &\quad-\frac{\gamma^4}{10e}E^{22}_n\qty(E^{11}_n-E^{13}_n+2{E^{02}_n}'+e({E^{11}_n}'-{E^{13}_n}'))^*\bigg],\\
        D^{(1)}_e&=V_e-\frac{P\gamma^6}{20}\sum_{n=1}^\infty nH_0^2\Omega_n\bigg[{S^{11}_n}'S^{11}_n+\frac{3}{e^3}\qty(E^{02}_n+\frac{e}{2}(E^{11}_n-E^{13}_n)-\gamma^2E^{22}_n)\\
        &\quad\times\qty(E^{02}_n-E^{22}_n-e(E^{11}_n+E^{13}_n+2E^{21}_n-2E^{23}_n+{E^{02}_n}'-\gamma^2{E^{22}_n}')-\frac{e^2}{2}(E^{20}_n+10E^{22}_n-E^{24}_n+{E^{11}_n}'-{E^{13}_n}'))^*\bigg],\\
        D^{(1)}_I&=\frac{\sin2I}{2}D^{(2)}_{\asc\asc},\qquad\qquad D^{(1)}_\asc=-2\cot ID^{(2)}_{I\asc},\\
        D^{(1)}_\omega&=V_\omega+\frac{3P\gamma^6}{40}\sin2\omega\frac{2-\sin^2I}{\sin^2I}\sum_{n=1}^\infty nH_0^2\Omega_nE^{20}_n(E^{22}_n)^*,\qquad\qquad D^{(1)}_\eps=V_\eps,
    \end{split}
    \end{align}
    while the secular diffusion matrix $D^{(2)}_{ij}$ is given by
        \begin{figure*}[t!]
            \includegraphics[width=0.497\textwidth]{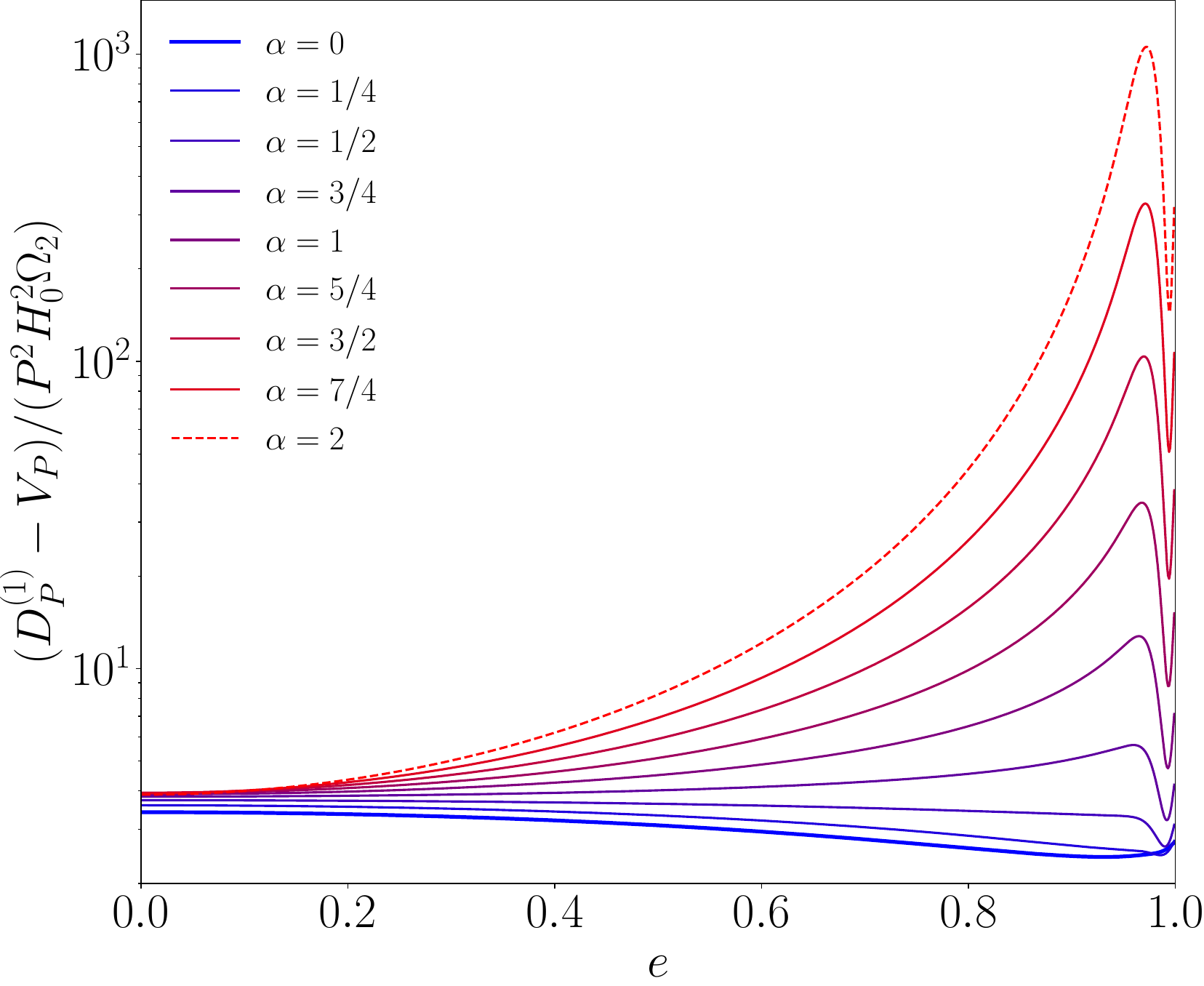}
            \includegraphics[width=0.497\textwidth]{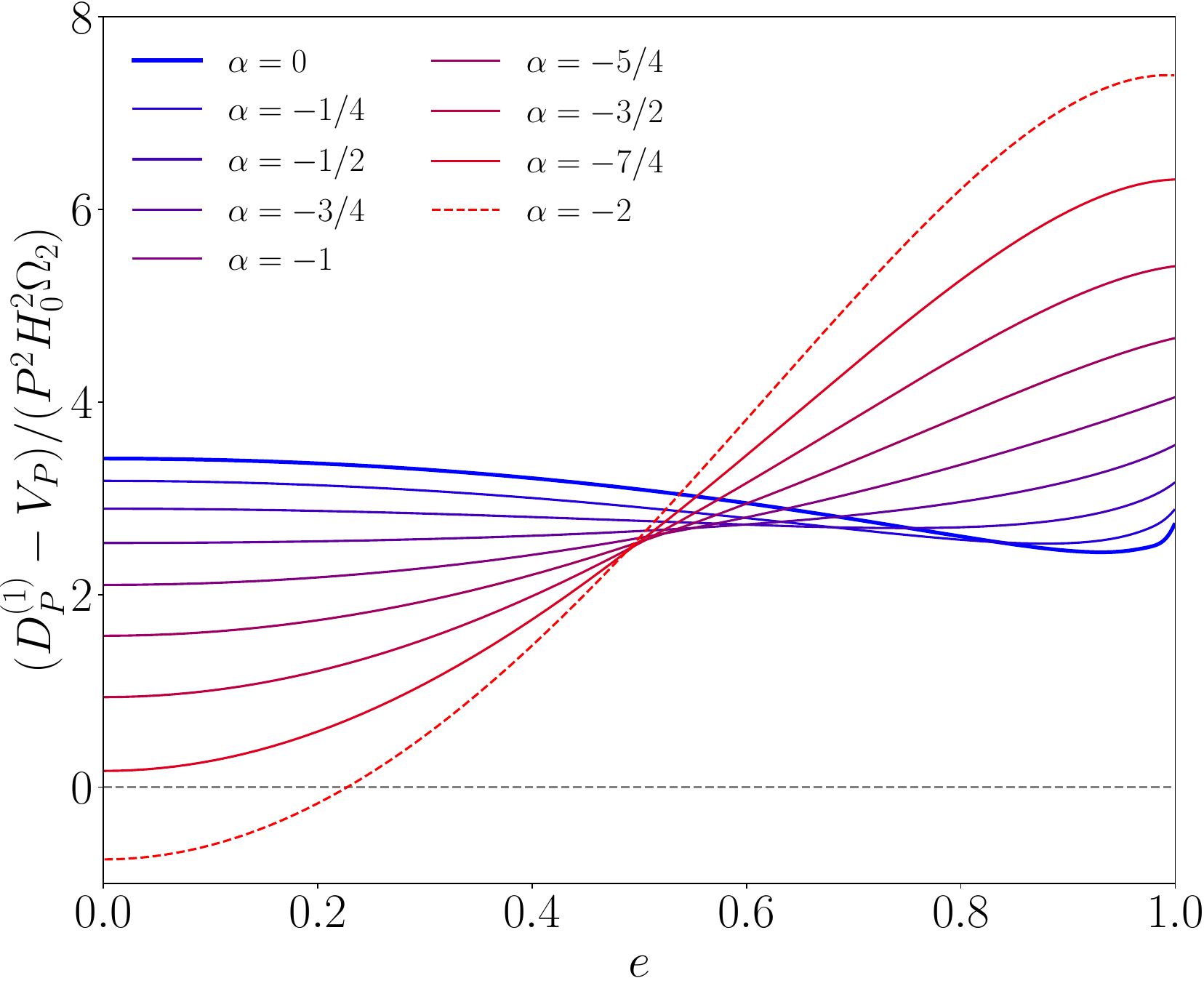}
            \includegraphics[width=0.497\textwidth]{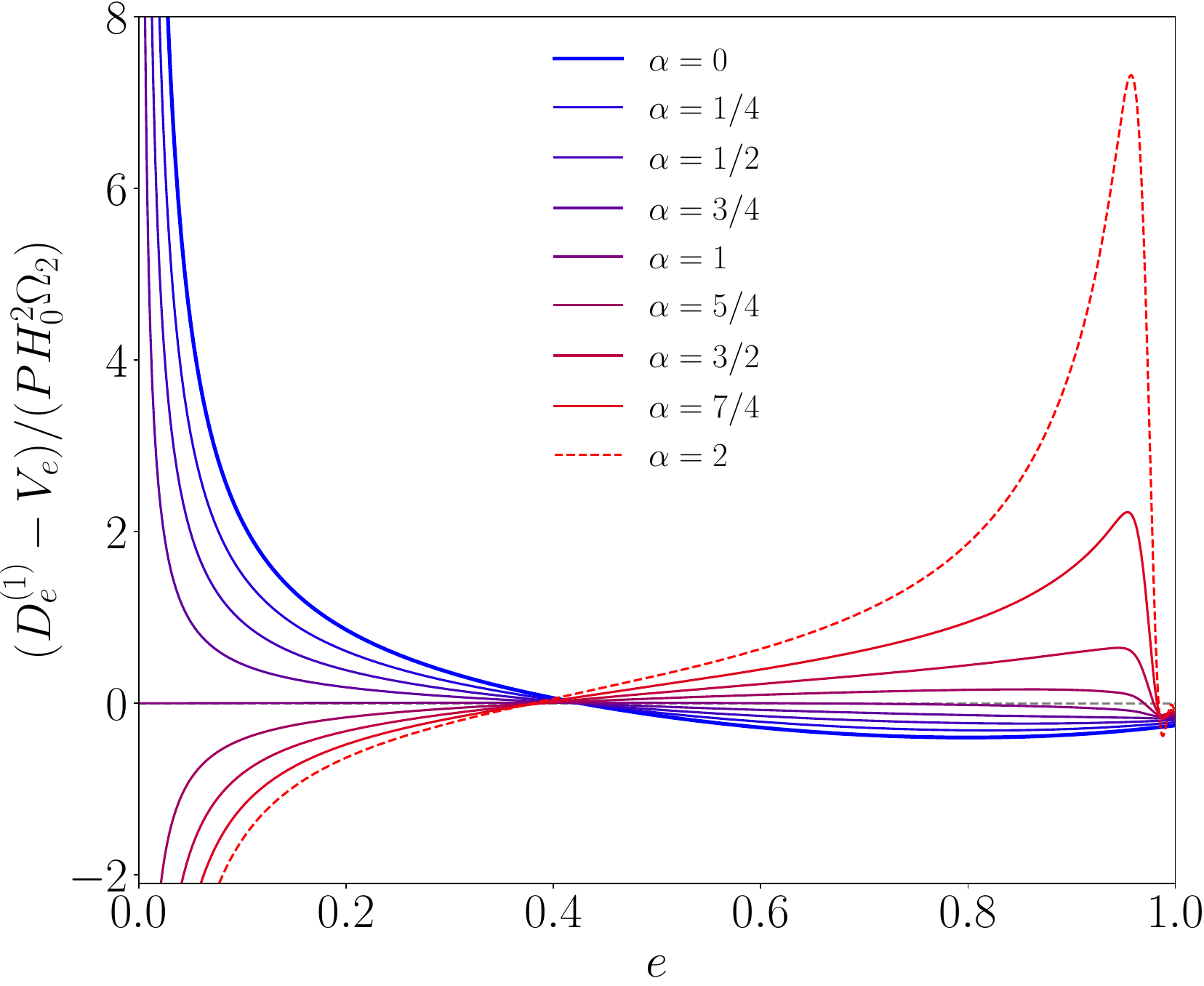}
            \includegraphics[width=0.497\textwidth]{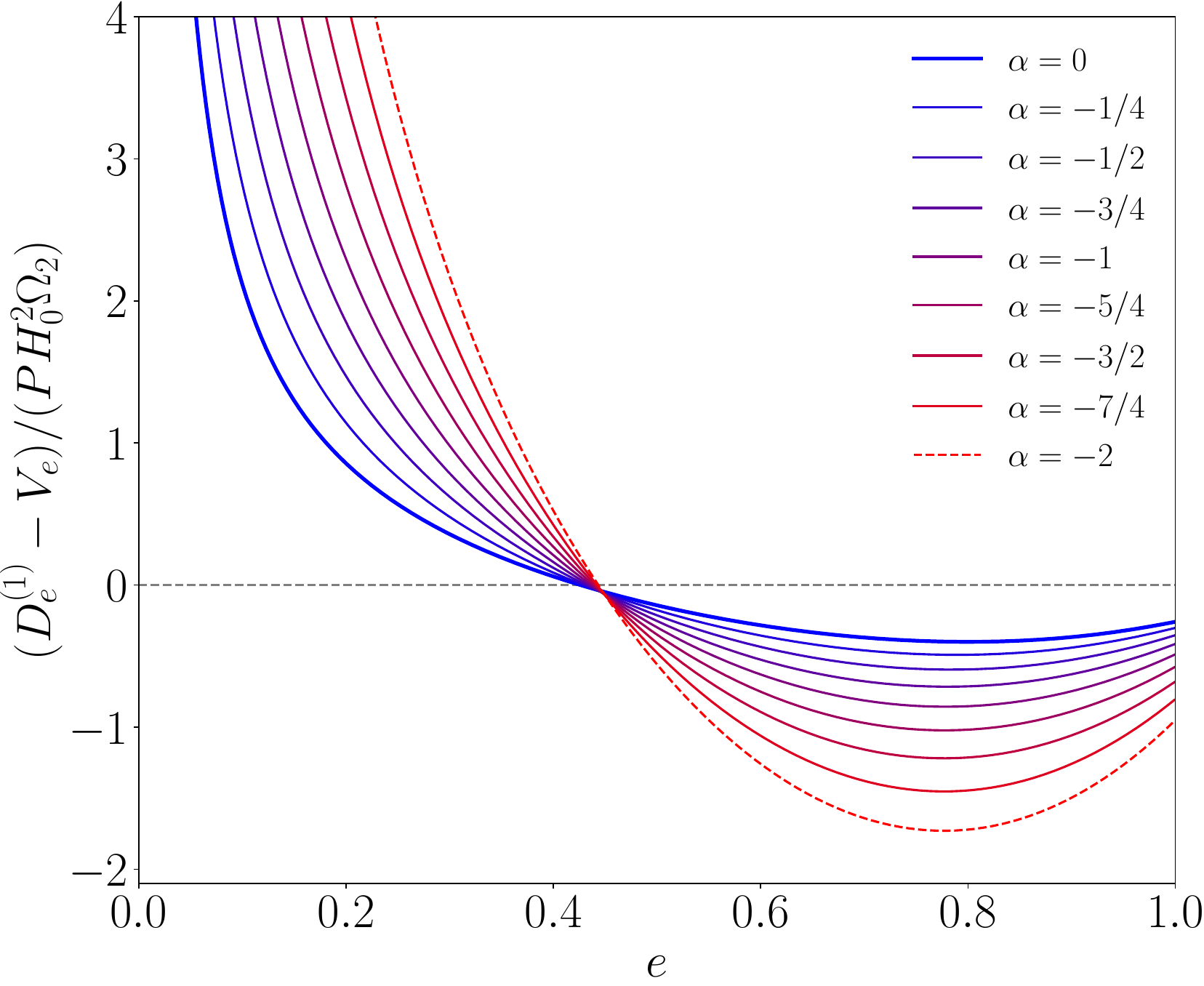}
            \caption{%
            Stochastic parts of the secular drift coefficients $D^{(1)}_P$ (first row) and $D^{(1)}_e$ (second row) as functions of eccentricity, for various power-law SGWB spectra, $\Omega(f)\sim f^\alpha$.
            The left column shows positive power-law indices, $\alpha=0,\dots,2$, while the right column shows negative indices, $\alpha=0,\dots,-2$.}
            \label{fig:km1-ecc}
        \end{figure*}
    \begin{align}
    \begin{split}
    \label{eq:ecc-diffusion}
        D^{(2)}_{PP}&=\frac{27P^3\gamma^2}{20}\sum_{n=1}^\infty nH_0^2\Omega_n\qty[\qty|E^{02}_n+\frac{e}{2}(E^{11}_n-E^{13}_n)|^2-\frac{(eS^{11}_n)^2}{3}],\\
        D^{(2)}_{Pe}&=\frac{\gamma^2D^{(2)}_{PP}}{3Pe}-\frac{9P^2\gamma^6}{40}\sum_{n=1}^\infty nH_0^2\Omega_nE^{22}_n\qty(\frac{2}{e}E^{02}_n+E^{11}_n-E^{13}_n)^*,\\
        D^{(2)}_{ee}&=\frac{3P\gamma^6}{20e^2}\sum_{n=1}^\infty nH_0^2\Omega_n\qty[\qty|E^{02}_n+\frac{e}{2}(E^{11}_n-E^{13}_n)-\gamma^2E^{22}_n|^2-\frac{(eS^{11}_n)^2}{3}],\\
        D^{(2)}_{II}&=\frac{3P\gamma^6}{80}\sum_{n=1}^\infty nH_0^2\Omega_n\qty[\qty|E^{20}_n|^2+\qty|E^{22}_n|^2+2\cos2\omega E^{20}_n(E^{22}_n)^*],\\
        D^{(2)}_{I\asc}&=\frac{3P\gamma^6}{40}\frac{\sin2\omega}{\sin I}\sum_{n=1}^\infty nH_0^2\Omega_nE^{20}_n(E^{22}_n)^*,\qquad D^{(2)}_{I\omega}=-\frac{3P\gamma^6}{40}\frac{\sin2\omega}{\tan I}\sum_{n=1}^\infty nH_0^2\Omega_nE^{20}_n(E^{22}_n)^*,\\
        D^{(2)}_{\asc\asc}&=\frac{3P\gamma^6}{80\sin^2I}\sum_{n=1}^\infty nH_0^2\Omega_n\qty[\qty|E^{20}_n|^2+\qty|E^{22}_n|^2-2\cos2\omega E^{20}_n(E^{22}_n)^*],\qquad D^{(2)}_{\asc\omega}=-\cos ID^{(2)}_{\asc\asc},\\
        D^{(2)}_{\omega\omega}&=\cos^2ID^{(2)}_{\asc\asc}+\frac{3P\gamma^6}{80e^2}\sum_{n=1}^\infty nH_0^2\Omega_n\qty[\qty|E^{11}_n+E^{13}_n+2E^{21}_n-2E^{23}_n+\frac{e}{2}\qty(E^{20}_n-E^{24}_n)|^2+\frac{4}{3}\qty(C^{11}_n)^2],\\
        D^{(2)}_{\omega\eps}&=-\frac{3P\gamma^7}{80e^2}\sum_{n=1}^\infty nH_0^2\Omega_n\qty[\qty|E^{11}_n+E^{13}_n+2(E^{21}_n-E^{23}_n)+\frac{e}{2}(E^{20}_n-4E^{22}_n-E^{24}_n)|^2+\frac{4}{3}C^{11}_n(C^{11}_n-2eC^{20}_n)-4|eE^{22}_n|^2],\\
        D^{(2)}_{\eps\eps}&=\frac{3P\gamma^8}{80e^2}\sum_{n=1}^\infty nH_0^2\Omega_n\qty[\qty|E^{11}_n+E^{13}_n+2(E^{21}_n-E^{23}_n)+\frac{e}{2}(E^{20}_n-8E^{22}_n-E^{24}_n)|^2+\frac{4}{3}(C^{11}_n-2eC^{20}_n)^2],\\
        D^{(2)}_{PI}&=D^{(2)}_{P\omega}=D^{(2)}_{P\asc}=D^{(2)}_{P\eps}=D^{(2)}_{eI}=D^{(2)}_{e\omega}=D^{(2)}_{e\asc}=D^{(2)}_{e\eps}=D^{(2)}_{I\eps}=D^{(2)}_{\asc\eps}=0.
    \end{split}
    \end{align}
\end{widetext}

The Hansen coefficients are defined by
    \begin{align}
    \begin{split}
    \label{eq:hansen-definition}
        C^{lm}_n(e)&\equiv\int_{t_0}^{t_0+P}\frac{\dd{t}}{P}\exp(\frac{2\uppi\rmi nt}{P})\frac{\cos m\psi}{(1+e\cos\psi)^l},\\
        S^{lm}_n(e)&\equiv\int_{t_0}^{t_0+P}\frac{\dd{t}}{P}\exp(\frac{2\uppi\rmi nt}{P})\frac{\sin m\psi}{(1+e\cos\psi)^l},\\
        E^{lm}_n(e)&\equiv C^{lm}_n(e)+S^{lm}_n(e),
    \end{split}
    \end{align}
    with primes indicating eccentricity derivatives, and the deterministic drift terms are those given in Sec.~\ref{sec:relativistic-drift},
    \begin{align}
    \begin{split}
        V_P&=-\frac{192\uppi\eta v_P^5}{5\gamma^7}\qty(1+\tfrac{73}{24}e^2+\tfrac{37}{96}e^4),\\
        V_e&=-\frac{608\uppi\eta v_P^5}{15P\gamma^5}\qty(e+\tfrac{121}{304}e^3),\\
        V_\omega&=\frac{6\uppi v_P^2}{P\gamma^2},\\
        V_\eps&=\frac{2\uppi v_P^2}{P}\qty(6-7\eta-\frac{15-9\eta}{\gamma}).
    \end{split}
    \end{align}

\begin{figure*}[t!]
    \includegraphics[width=0.497\textwidth]{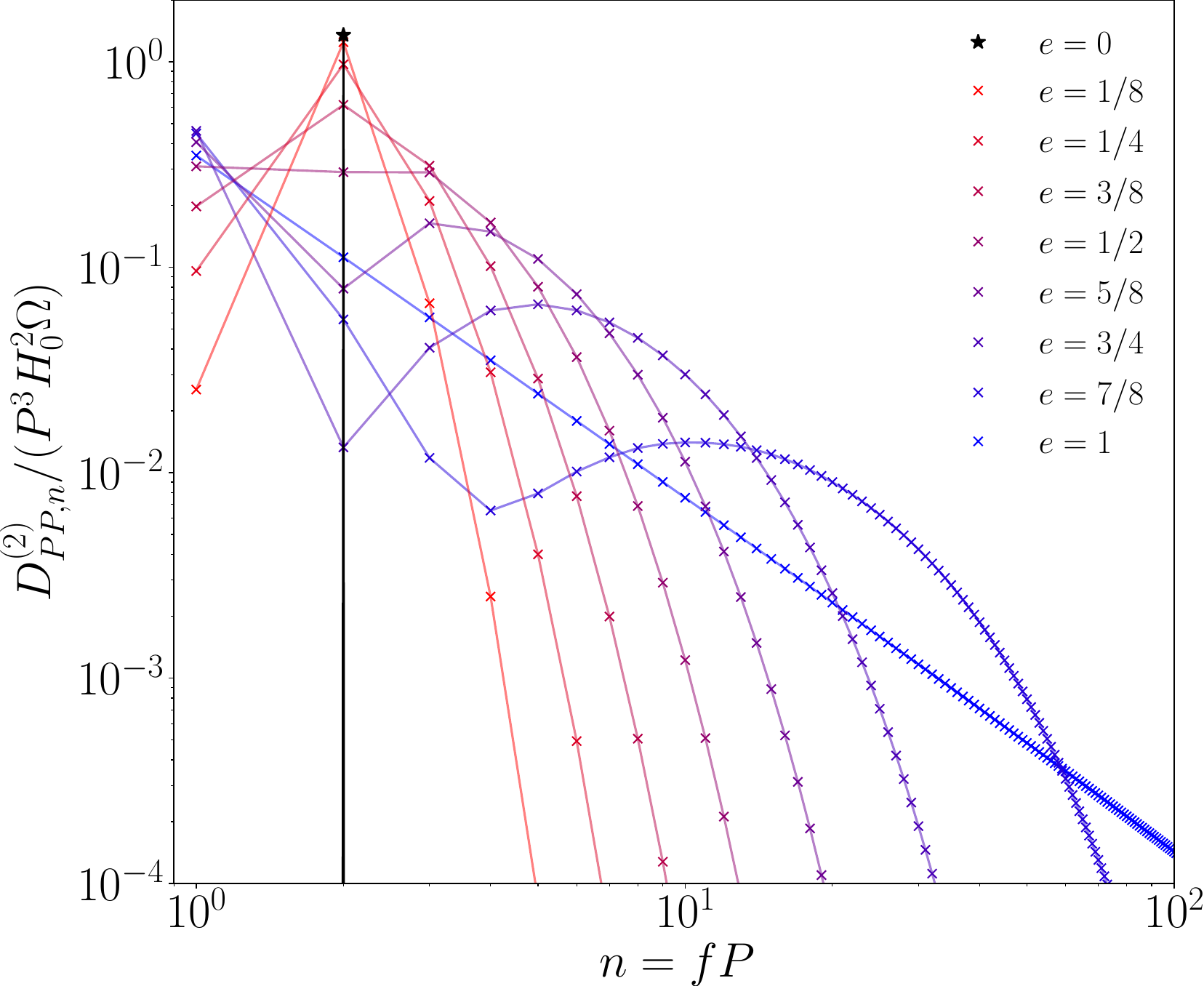}
    \includegraphics[width=0.497\textwidth]{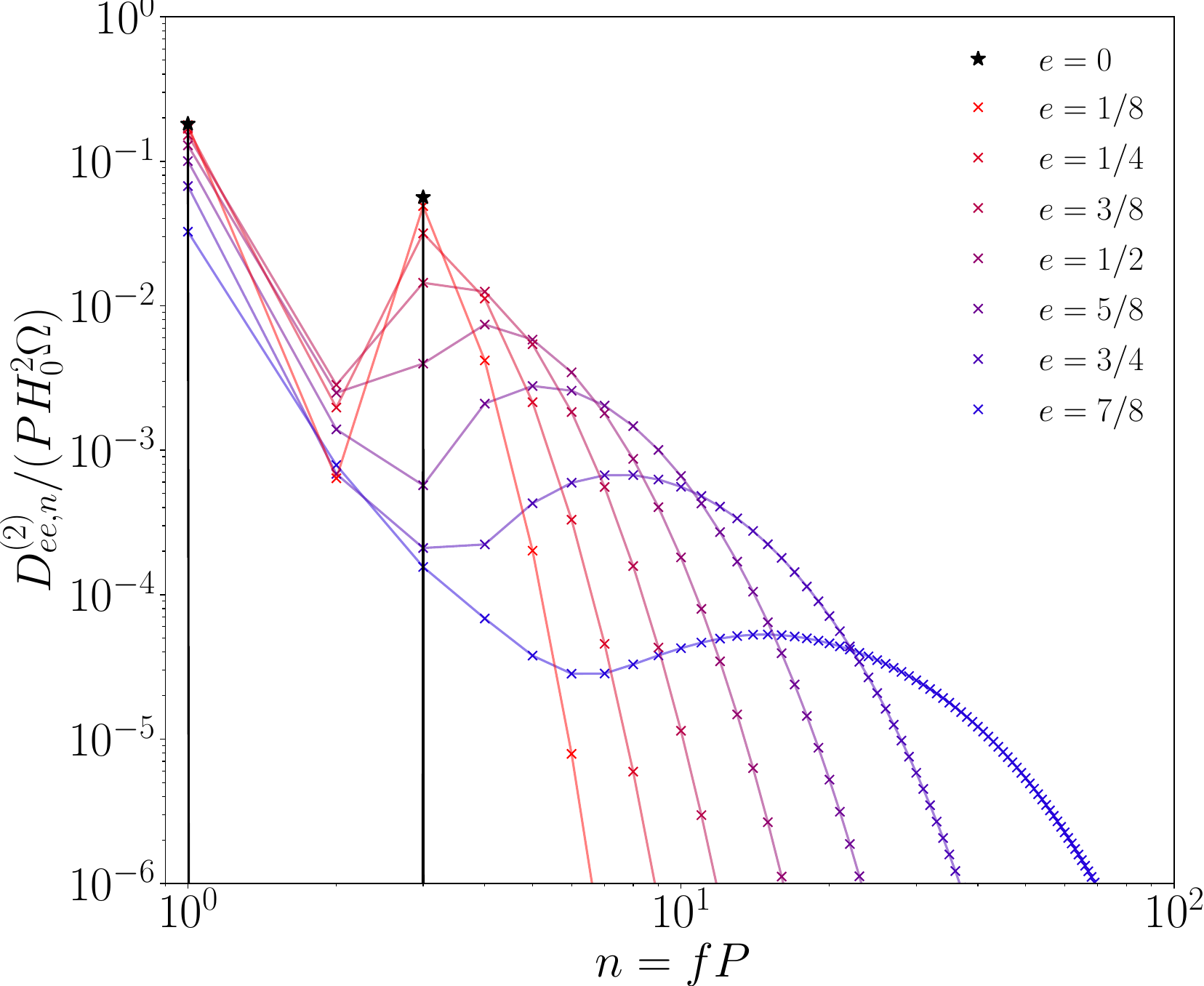}
    \caption{%
    Contributions to the secular diffusion coefficients $D^{(2)}_{PP}$ (left panel) and $D^{(2)}_{ee}$ (right panel) from different harmonic frequencies, for binaries with various eccentricities $e=0,\dots,1$.
    (The subscript ``$n$'' here indicates that we have extracted the contribution from the $n$th harmonic.)
    We set $\Omega(f)=\mathrm{constant}$ here, so that each harmonic receives equal weighting; alternative SGWB spectra will give different weighting to each harmonic.
    Note that $D^{(2)}_{ee}=0$ when $e\to1$.}
    \label{fig:km2-n}
\end{figure*}

\begin{figure*}[t!]
    \includegraphics[width=0.497\textwidth]{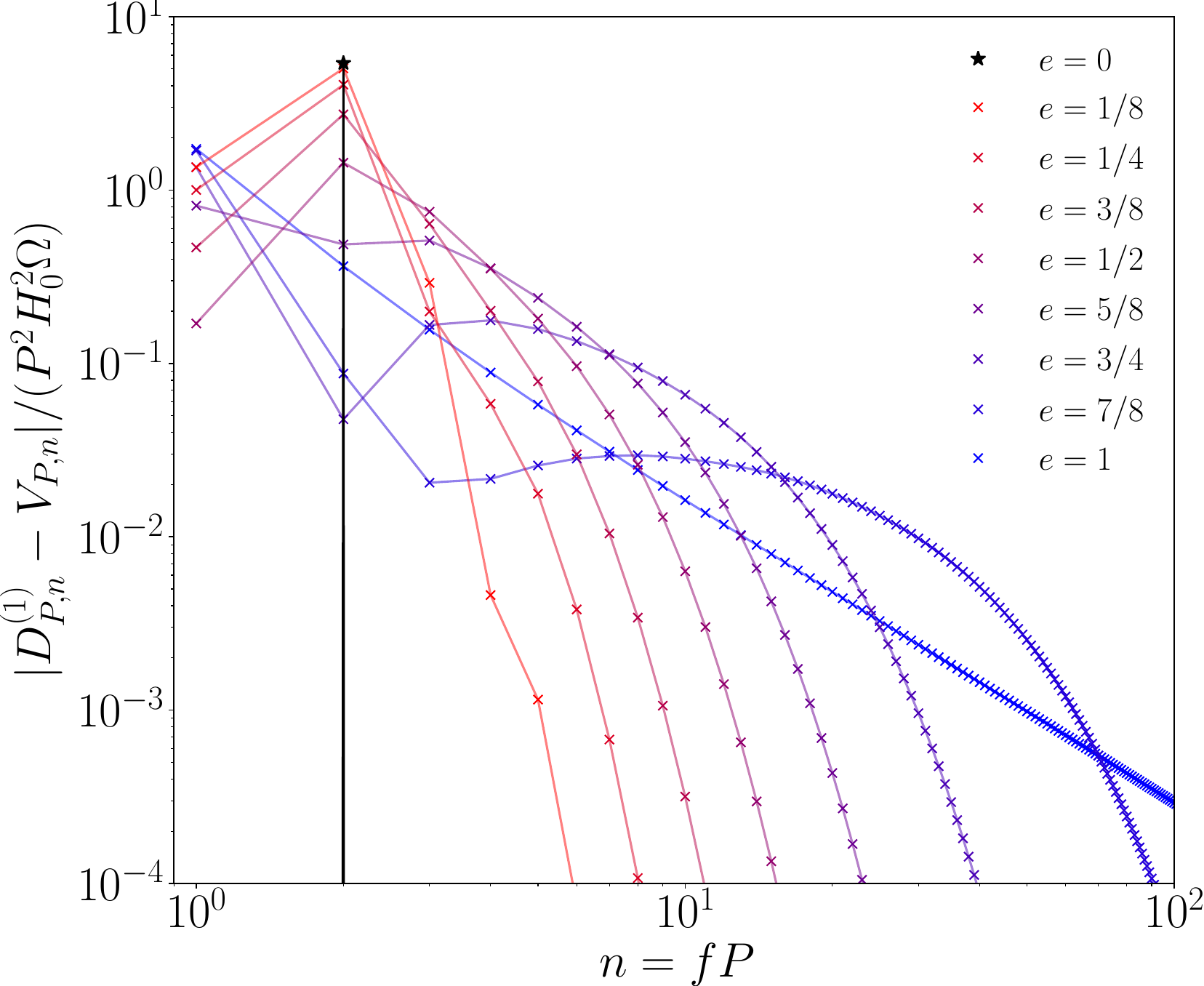}
    \includegraphics[width=0.497\textwidth]{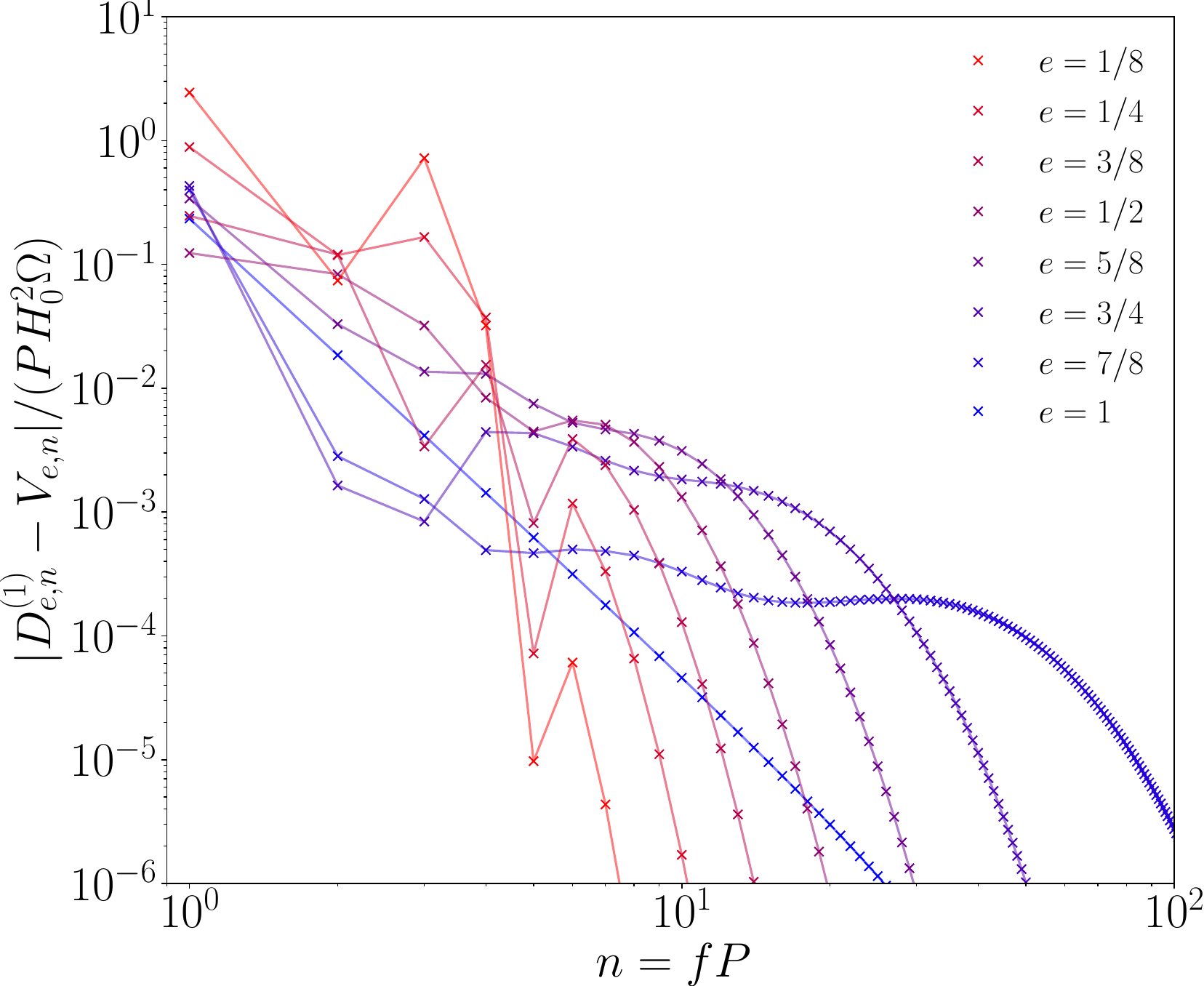}
    \caption{%
    Contributions to the stochastic parts of the secular drift coefficients $D^{(1)}_P$ (left panel) and $D^{(1)}_e$ (right panel) from different harmonic frequencies, for binaries with various eccentricities $e=0,\dots,1$.
    (The subscript ``$n$'' here indicates that we have extracted the contribution from the $n$th harmonic.)
    We show the absolute values, as the drift coefficients have both positive and negative contributions.
    Note that $D^{(1)}_e\to+\infty$ as $e\to0$.}
    \label{fig:km1-n}
\end{figure*}

The values of some of these drift and diffusion coefficients as functions of eccentricity and harmonic frequency are shown in Figs.~\ref{fig:km1-ecc}, \ref{fig:km2-ecc}, \ref{fig:km2-n}, and~\ref{fig:km1-n}; we focus on the coefficients pertaining to the period and eccentricity of the binary, as these are the most important for our envisaged observational applications.
These coefficients, describing the evolution of the size and shape of the orbit, are independent of the remaining four orbital elements $(I,\asc,\omega,\eps)$, which all describe the orientation of the orbit in space.
In Figs.~\ref{fig:km1-ecc} and~\ref{fig:km2-ecc} we plot the coefficients against eccentricity, assuming various power-law SGWB spectra $\Omega(f)\sim f^\alpha$, $\alpha\in[-2,+2]$.
There are two particularly striking features that are worth mentioning: the first is that $D^{(2)}_{Pe}$ and $D^{(2)}_{ee}$ both vanish as $e\to1$, regardless of the SGWB spectrum; and the second is that $D^{(1)}_e$ changes sign at an eccentricity $e\approx0.4$ that is approximately (though not exactly) independent of the SGWB spectrum; the significance of this near-universal crossover eccentricity is not immediately clear.

In Figs.~\ref{fig:km2-n} and~\ref{fig:km1-n} we plot the contributions from each of the binary's harmonic frequencies $f=n/P$.
We see that the evolution of the period is driven by the $n=2$ harmonic for near-circular binaries, with all other harmonics having zero contribution to $D^{(2)}_{PP}$ and $D^{(1)}_P$ in the circular limit $e\to0$.
This mirrors the frequency content of GW \emph{emission} from binaries, which is also dominated by the $n=2$ harmonic for small eccentricity.
The significance of $n=2$ here can be understood by recognising that advancing a circular orbit in time by $P/2$ (i.e. the inverse of the $n=2$ harmonic) is equivalent to exchanging the positions of the two bodies, resulting in a setup which has the same GW emission and absorption properties.
The eccentricity evolution, on the other hand, is dominated by the $n=1$ and $n=3$ harmonics when $e$ is small.
In all cases, the contribution from higher harmonics becomes stronger for larger eccentricities, as the Fourier spectrum of the binary's response to GWs becomes richer.
(Again, this is the same qualitative pattern that one finds in the case of GW \emph{emission} by the binary.)
In the limit $e\to1$, we see a very simple pattern emerge in which each coefficient approaches a power law in the harmonic number $n$ (except for $D^{(2)}_{ee}$, which vanishes in this limit).

The qualitative link between the GW emission and absorption spectra of binaries noted above is intriguing, and raises the question of whether this relationship can be established more formally.
This might lead to deeper insights into binary--GW interactions, for example by allowing us to prove something akin to a fluctuation-dissipation theorem for this system.
It would also be interesting to make contact with existing results on the scattering of GWs by binaries~\cite{Annulli:2018quj}.
However, on the face of it there are several important differences between the GW absorption and emission processes: for example, the masses of the orbiting bodies are crucial in determining the radiated GW flux, but have no influence at all on GW absorption, since the GW-induced oscillations in the orbital separation are independent of the masses.
(This last statement is a manifestation of the equivalence principle.)
We leave a more thorough exploration of these questions for future work.

Inserting the coefficients in Eqs.~\eqref{eq:ecc-drift} and~\eqref{eq:ecc-diffusion} into Eq.~\eqref{eq:fpe}, we obtain a FPE which completely describes the secular evolution of a general binary system under SGWB resonance; this is the main result of our analysis.
Sections~\ref{sec:circular}, \ref{sec:full-fpe}, and~\ref{sec:observations} explore various strategies for solving this equation, and for using these solutions to place constraints on the SGWB spectrum.
Before moving on, we derive some simplified expressions for the KM coefficients in the cases where the eccentricity and/or the inclination are small.

\subsection{Small-eccentricity and small-inclination orbits}
\label{sec:KM-small-e-I}

For binaries with very small eccentricity, we want to recast the results above in terms of the alternative orbital elements $(P,\zeta,\kappa,I,\asc,\xi)$, as defined in Eq.~\eqref{eq:small-e-I}.
We do this using the coordinate transformation laws for the KM coefficients (see, e.g., Sec.~4.9 of Ref.~\cite{Risken:1989fpe} for a derivation),
    \begin{align}
    \begin{split}
        D^{(1)}_i&=\pdv{X_i}{X_{i'}}D^{(1)}_{i'}+\pdv{X_i}{X_{i'}}{X_{j'}}D^{(2)}_{i'j'},\\
        D^{(2)}_{ij}&=\pdv{X_i}{X_{i'}}\pdv{X_j}{X_{j'}}D^{(2)}_{i'j'},
    \end{split}
    \end{align}
    where summation over the primed indices is implied.
Neglecting terms of order $e^2\sim\zeta^2\sim\zeta\kappa\sim\kappa^2$, we find the drift coefficients
    \begin{align}
    \begin{split}
    \label{eq:small-e-drift}
        D^{(1)}_P&=V_P+\frac{3P^2}{160}H_0^2(-79\Omega_1+288\Omega_2-27\Omega_3),\\
        D^{(1)}_\zeta&=V_\zeta+\frac{P}{160}\frac{\zeta}{\zeta^2+\kappa^2}H_0^2(25\Omega_1-27\Omega_3),\\
        D^{(1)}_\kappa&=V_\kappa+\frac{P}{160}\frac{\kappa}{\zeta^2+\kappa^2}H_0^2(25\Omega_1-27\Omega_3),\\
        D^{(1)}_I&=\frac{3P}{80}H_0^2\Omega_2\cot I,\\
        D^{(1)}_\asc&=0,\\
        D^{(1)}_\xi&=V_\xi,
    \end{split}
    \end{align}
    where the deterministic drift terms are given by
    \begin{align}
    \begin{split}
        V_\zeta&=\frac{6\uppi v_P^2}{P}\qty(\kappa-\tfrac{304}{45}\eta v_P^3\zeta),\\
        V_\kappa&=-\frac{6\uppi v_P^2}{P}\qty(\zeta+\tfrac{304}{45}\eta v_P^3\kappa),\\
        V_\xi&=-\frac{4\uppi v_P^2}{P}(3-\eta),
    \end{split}
    \end{align}
    and the diffusion coefficients are
    \begin{align}
    \begin{split}
    \label{eq:small-e-diffusion}
        D^{(2)}_{PP}&=\frac{27P^3}{20}H_0^2\Omega_2,\\
        D^{(2)}_{P\zeta}&=-\frac{3P\zeta}{160}H_0^2(25\Omega_1+12\Omega_2-27\Omega_3),\\
        D^{(2)}_{P\kappa}&=-\frac{3P\kappa}{160}H_0^2(25\Omega_1+12\Omega_2-27\Omega_3),\\
        D^{(2)}_{\zeta\zeta}&=D^{(2)}_{\kappa\kappa}=\frac{P}{160}H_0^2(29\Omega_1+9\Omega_3),\\
        D^{(2)}_{\zeta\asc}&=-\frac{3P\kappa}{80}H_0^2\Omega_2\frac{\cos I}{\sin^2I},\\
        D^{(2)}_{\zeta\xi}&=-\frac{P\kappa}{320}H_0^2\qty[203\Omega_1-12\Omega_2(20+\cot^2I)+63\Omega_3],\\
        D^{(2)}_{\kappa\asc}&=\frac{3P\zeta}{80}H_0^2\Omega_2\frac{\cos I}{\sin^2I},\\
        D^{(2)}_{\kappa\xi}&=\frac{P\zeta}{320}H_0^2\qty[203\Omega_1-12\Omega_2(20+\cot^2I)+63\Omega_3],\\
        D^{(2)}_{II}&=\frac{3P}{80}H_0^2\Omega_2,\\
        D^{(2)}_{\asc\asc}&=\frac{3P}{80}H_0^2\Omega_2\csc^2I,\\
        D^{(2)}_{\asc\xi}&=-\frac{3P}{80}H_0^2\Omega_2\frac{\cos I}{\sin^2I},\\
        D^{(2)}_{\xi\xi}&=\frac{3P}{80}H_0^2\Omega_2(16+\cot^2I),\\
        D^{(2)}_{PI}&=D^{(2)}_{P\asc}=D^{(2)}_{P\xi}=D^{(2)}_{\zeta\kappa}\\
        &=D^{(2)}_{\zeta I}=D^{(2)}_{\kappa I}=D^{(2)}_{I\asc}=D^{(2)}_{I\xi}=0.
    \end{split}
    \end{align}
(We have checked explicitly that the $D^{(2)}_{PP}$ coefficient here matches the corresponding result in Ref.~\cite{Hui:2012yp}.)

\begin{figure}[t!]
    \includegraphics[width=0.48\textwidth]{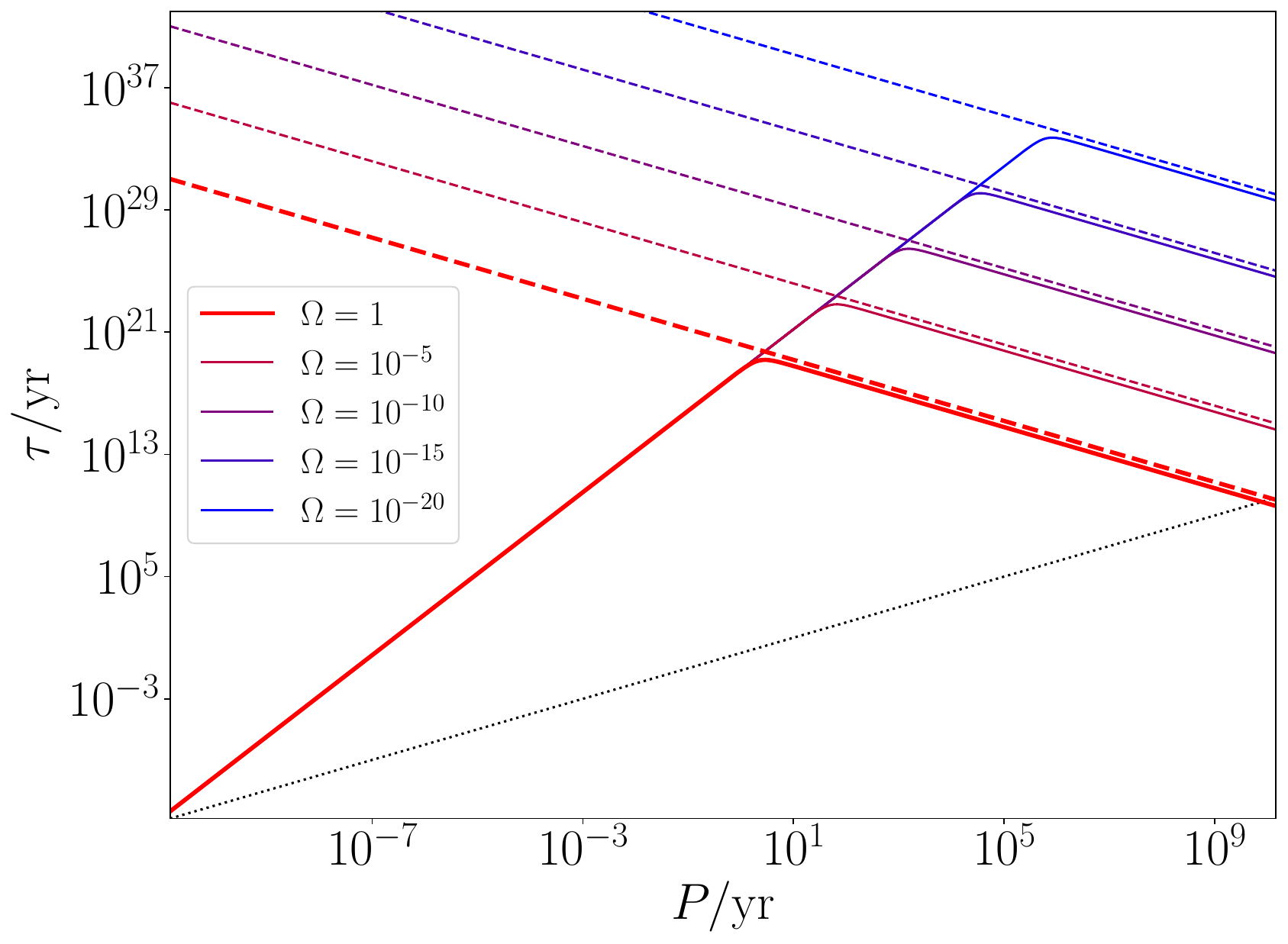}
    \caption{%
   Evolution timescales for the DF of a circular binary with total mass $M=M_\odot$ immersed in a scale-invariant SGWB $\Omega(f)=\mathrm{constant}$.
    Solid curves show the drift timescale~\eqref{eq:drift-timescale}, which is dominated by GW emission at ``short'' periods $P\lesssim P_\mathrm{crit}$ and by GW absorption from the SGWB at ``long'' periods $P\gtrsim P_\mathrm{crit}$.
    Dashed curves show the diffusion timescale~\eqref{eq:diffusion-timescale}.
    The dotted black curve shows $\tau=P$; the region below this curve corresponds to a timescale faster than the binary period, which breaks our secular averaging assumption and should thus be taken with a grain of salt.}
    \label{fig:timescales}
\end{figure}

We see from Eq.~\eqref{eq:small-e-drift} that the stochastic drift term for the period $P$ is usually positive, and can thus be interpreted physically as describing the softening of the binary due to the absorption of energy from the SGWB (the term can become negative if $\Omega_2$ is significantly smaller than $\Omega_1$ and/or $\Omega_3$, but SGWB spectra typically vary sufficiently slowly with frequency that this does not occur).
Interestingly, this implies that the net secular drift of the binary period (deterministic plus stochastic) generally changes sign at some critical value of $P$; e.g., for a scale-invariant SGWB $\Omega(f)=\mathrm{constant}$, this value is given by
    \begin{align}
    \begin{split}
    \label{eq:Pcrit}
        P_\mathrm{crit}&=\qty(\frac{1024\uppi\eta}{91H_0^2\Omega})^{3/11}(2\uppi GM)^{5/11}\\
        &\approx95\,\mathrm{yr}\times\qty(\frac{\Omega}{10^{-6}}\frac{1/4}{\eta})^{-3/11}\qty(\frac{M}{M_\odot})^{5/11},
    \end{split}
    \end{align}
    which corresponds to a semi-major axis of
    \begin{equation}
        a_\mathrm{crit}\approx21\,\mathrm{au}\times\qty(\frac{\Omega}{10^{-6}}\frac{1/4}{\eta})^{-2/11}\qty(\frac{M}{M_\odot})^{1/3}.
    \end{equation}
Binaries with $P<P_\mathrm{crit}$ will decay through GW emission, decreasing their period further, whereas binaries with $P>P_\mathrm{crit}$ undergo a net softening through SGWB absorption, leading to a further increase in their period.
The point $P=P_\mathrm{crit}$ is thus an unstable fixed point of the Langevin equation for $P$.\footnote{%
    We can understand this instability through a thermodynamic analogy.
    In the absence of the SGWB, a binary system radiates GW energy with increasing intensity as it inspirals; its dynamical ``temperature'' grows as it loses energy, meaning that it has a \emph{negative heat capacity}.
    Similarly, the SGWB acts as a heat reservoir and imparts energy to the binary (on average), therefore slowing the orbital motion and decreasing the system's temperature.
    A conventional thermodynamic system with positive heat capacity would equilibriate at a point where the heat loss from GW radiation balanced the heat gain from the SGWB.
    Instead, the binary undergoes a runaway increase or decrease in its temperature due to its negative heat capacity.}
Note however that random diffusion due to $D^{(2)}_{PP}$ acts on a similar timescale [see Fig.~\ref{fig:timescales} and Eqs.~\eqref{eq:drift-timescale} and~\eqref{eq:diffusion-timescale}], and can easily push the system either side of this critical point.

We find a similar phenomenon for the eccentricity.
Transforming back from the Laplace-Lagrange variables $\zeta,\kappa$ for now, Eqs.~\eqref{eq:small-e-drift} and~\eqref{eq:small-e-diffusion} give
    \begin{equation}
        D^{(1)}_e=V_e+\frac{9P}{80e}H_0^2(3\Omega_1-\Omega_3),
    \end{equation}
    so that the (usually positive) stochastic drift diverges as $e\to0$, while the (always negative) deterministic part vanishes as $e\to0$.
The net eccentricity drift thus changes sign at a critical value, just as it does for the period.
For example, assuming a scale-invariant SGWB spectrum, this critical value is
    \begin{align}
    \begin{split}
        &e_\mathrm{crit}=\sqrt{\frac{27H_0^2\Omega}{4864\uppi\eta v_P^5}}\\
        &\approx5.9\times10^{-5}\times\qty(\frac{\Omega}{10^{-6}}\frac{1/4}{\eta})^{1/2}\qty(\frac{M}{M_\odot})^{-5/6}\qty(\frac{P}{\mathrm{yr}})^{11/6}.
    \end{split}
    \end{align}
Since $\partial_eD^{(1)}_e<0$ at this point, $e_\mathrm{crit}$ is a stable fixed point of the corresponding Langevin equation: binaries with larger eccentricity will tend to circularise through GW emission until they reach $e_\mathrm{crit}$, while binaries with smaller eccentricity will on average have their eccentricity excited through SGWB resonance.
This is particularly interesting from an observational point of view, as it suggests that eccentricities smaller than $e_\mathrm{crit}$ might be less frequently observed in sufficiently old systems, though random diffusion due to $D^{(2)}_{ee}$ can still push systems below this point.

We also see from Eq.~\eqref{eq:small-e-drift} that the stochastic drift term for the inclination changes sign at $I=\uppi/2$ [this is also true in the general-eccentricity case, see Eq.~\eqref{eq:ecc-drift}].
Since $\partial_ID^{(1)}_I<0$ at $I=\uppi/2$, this is a stable fixed point of the corresponding Langevin equation; stochastic drift will, on average, drive binaries toward $I=\uppi/2$.
This effect is counteracted, however, by binaries diffusing away from $I=\uppi/2$.
As we show explicitly in Sec.~\ref{sec:circular}, on extremely long timescales these two effects balance each other, leaving an isotropic distribution for the inclination with mean $\ev{I}=\uppi/2$.

In the limit where the binary's inclination is also small, we rewrite the KM coefficients in terms of the orbital elements $(P,\zeta,\kappa,p,q,\lambda)$.
This gives the drift coefficients
    \begin{align}
    \begin{split}
    \label{eq:small-e-I-drift}
        D^{(1)}_p&=-\frac{Pp}{40}H_0^2\Omega_2,\\
        D^{(1)}_q&=-\frac{Pq}{40}H_0^2\Omega_2,
    \end{split}
    \end{align}
    and the diffusion coefficients,
    \begin{align}
    \begin{split}
    \label{eq:small-e-I-diffusion}
        D^{(2)}_{\zeta\lambda}&=-\frac{P\kappa}{320}H_0^2\qty(203\Omega_1-240\Omega_2+63\Omega_3),\\
        D^{(2)}_{\kappa\lambda}&=\frac{P\zeta}{320}H_0^2\qty(203\Omega_1-240\Omega_2+63\Omega_3),\\
        D^{(2)}_{pp}&=D^{(2)}_{qq}=\frac{3P}{80}H_0^2\Omega_2,\\
        D^{(2)}_{p\lambda}&=\frac{3Pq}{160}H_0^2\Omega_2\\
        D^{(2)}_{q\lambda}&=-\frac{3Pp}{160}H_0^2\Omega_2\\
        D^{(2)}_{\lambda\lambda}&=\frac{3P}{5}H_0^2\Omega_2,\\
        D^{(2)}_{Pp}&=D^{(2)}_{Pq}=D^{(2)}_{P\lambda}=D^{(2)}_{\zeta\kappa}=D^{(2)}_{\zeta p}\\
        &=D^{(2)}_{\zeta q}=D^{(2)}_{\kappa p}=D^{(2)}_{\kappa q}=D^{(2)}_{pq}=0,
    \end{split}
    \end{align}
    where the coefficients not listed are identical to those in Eqs.~\eqref{eq:small-e-drift} and~\eqref{eq:small-e-diffusion}.
Note that this implies
    \begin{equation}
        D^{(1)}_I=\frac{3P}{80I}H_0^2\Omega_2,
    \end{equation}
    which diverges as $I\to0$.
This means that binaries are quickly excited away from zero inclination, similar to what happens for the eccentricity as $e\to0$.

\section{Some exact results for circular binaries}
\label{sec:circular}

We have shown that the osculating orbital elements of a binary coupled to the SGWB evolve according to a nonlinear six-dimensional FPE, for which no analytical solution is generally available.
However, in the small-eccentricity limit $e\to0$, we found in Eqs.~\eqref{eq:small-e-drift} and~\eqref{eq:small-e-diffusion} that the drift and diffusion of the binary period $P$ are independent of all of the other orbital elements.
This allows us to treat $P$ separately by solving the one-dimensional FPE
    \begin{equation}
        \pdv{W}{t}=-\pdv{J}{P},
    \end{equation}
    where $W(P,t)$ is now the single-variable DF for $P$, marginalised over the other orbital elements, and where
    \begin{equation}
    \label{eq:prob-current}
        J(P,t)\equiv D^{(1)}W-\partial_P(D^{(2)}W)
    \end{equation}
    is the \emph{probability current}~\cite{Risken:1989fpe,Gardiner:2004hsm}.
(We have suppressed the $P$ subscripts on the drift and diffusion coefficients for this single-variable case.)
In this section, we derive some exact results for this simplified equation.
These results highlight the power of our Fokker-Planck formalism, which allows us to answer these questions about the full shape of the DF in a way that previous analyses are unable to.

\subsection{Quasi-stationary period distribution}

The simplest kind of solution to look for is a stationary (i.e., time-independent) distribution, corresponding to constant probability current throughout the parameter space.
Setting $J=\mathrm{constant}$ in Eq.~\eqref{eq:prob-current}, we can use the integrating factor (which is defined up to an arbitrary constant multiplicative factor)
    \begin{equation}
    \label{eq:integrating-factor}
        I(P)\equiv\exp(\int\dd{P}\frac{D^{(1)}}{D^{(2)}})
    \end{equation}
    to obtain
    \begin{equation}
    \label{eq:quasi-stationary}
        W=\frac{I(P)}{D^{(2)}}\qty(C-J\int\frac{\dd{P}}{I(P)}),
    \end{equation}
    with $C$ a constant which, for a given value of $J$, is fixed by the normalisation of the DF.
Clearly, the functional form of Eq.~\eqref{eq:quasi-stationary} depends on the SGWB energy spectrum $\Omega(f)$.
As a simple example, consider a scale-invariant SGWB, $\Omega(f)=\mathrm{constant}$, which has
    \begin{align}
    \begin{split}
        D^{(1)}&=V_P+\frac{273}{80}P^2H_0^2\Omega,\qquad D^{(2)}=\frac{27}{20}P^3H_0^2\Omega,\\
        I(P)&\propto P^{91/36}\exp[\frac{91}{132}\qty(\frac{P_\mathrm{crit}}{P})^{11/3}].
    \end{split}
    \end{align}
A full exploration of spectra beyond this simple scale-invariant case is beyond our scope here, but we expect the qualitative results of this section to be reasonably robust to this choice.

We can fix $J$ and $C$ by imposing the appropriate boundary conditions.
For some minimum value of $P$ the binary merges or is tidally disrupted, whereas for some maximum value the binary becomes gravitationally unbound, so at both extremes we require absorbing boundary conditions---i.e., the DF must go to zero at both boundaries.
Systems with absorbing boundary conditions do not admit nonzero stationary solutions~\cite{Gardiner:2004hsm}; formally, the conditional probability of the binary having period $P$ at time $t$, given that it initially had period $P_0$ at time zero, obeys
    \begin{equation}
        \lim_{t\to\infty}W(P,t|P_0,0)=0
    \end{equation}
    across the entire parameter space.
Intuitively, this is because all of the initial probability mass is eventually absorbed by one or other of the boundaries.
We can also understand this in terms of the instability discussed in Sec.~\ref{sec:KM-small-e-I}; binaries either side of the critical period $P_\mathrm{crit}$ undergo a runaway evolution away from this point, reaching one of the two boundaries within finite time, thus leaving an empty distribution in the limit $t\to\infty$.

In practice, however, the timescale over which the binary evolves is set by its period, and binaries with short periods will approach stationarity much faster than binaries with long periods.
More concretely, for a flat SGWB spectrum we have a drift timescale
    \begin{equation}
    \label{eq:drift-timescale}
        \tau_\mathrm{drift}\equiv\frac{P}{|D^{(1)}|}\simeq
        \begin{cases}
            \displaystyle\frac{80}{273PH_0^2\Omega}, & P\gg P_\mathrm{crit}, \\[4pt]
            \displaystyle\frac{5P^{8/3}}{192\uppi\eta(2\uppi GM)^{5/3}}, & P\ll P_\mathrm{crit},
        \end{cases}
    \end{equation}
    and a diffusion timescale
    \begin{equation}
    \label{eq:diffusion-timescale}
        \tau_\mathrm{diff}\equiv\frac{P^2}{|D^{(2)}|}=\frac{20}{27PH_0^2\Omega}.
    \end{equation}
As shown in Fig.~\ref{fig:timescales}, the fastest timescale is many orders of magnitude shorter near the lower boundary than it is near the upper boundary.
It may be reasonable, therefore, to look for ``quasi-stationary'' solutions, where equilibrium is established near the lower boundary, but where the boundary condition for long periods is neglected.
We achieve this by choosing $J$ and $C$ such that the DF goes to zero at the lower boundary, and such that the DF is normalised, but without enforcing any condition at the upper boundary.

\begin{figure}[t!]
    \includegraphics[width=0.48\textwidth]{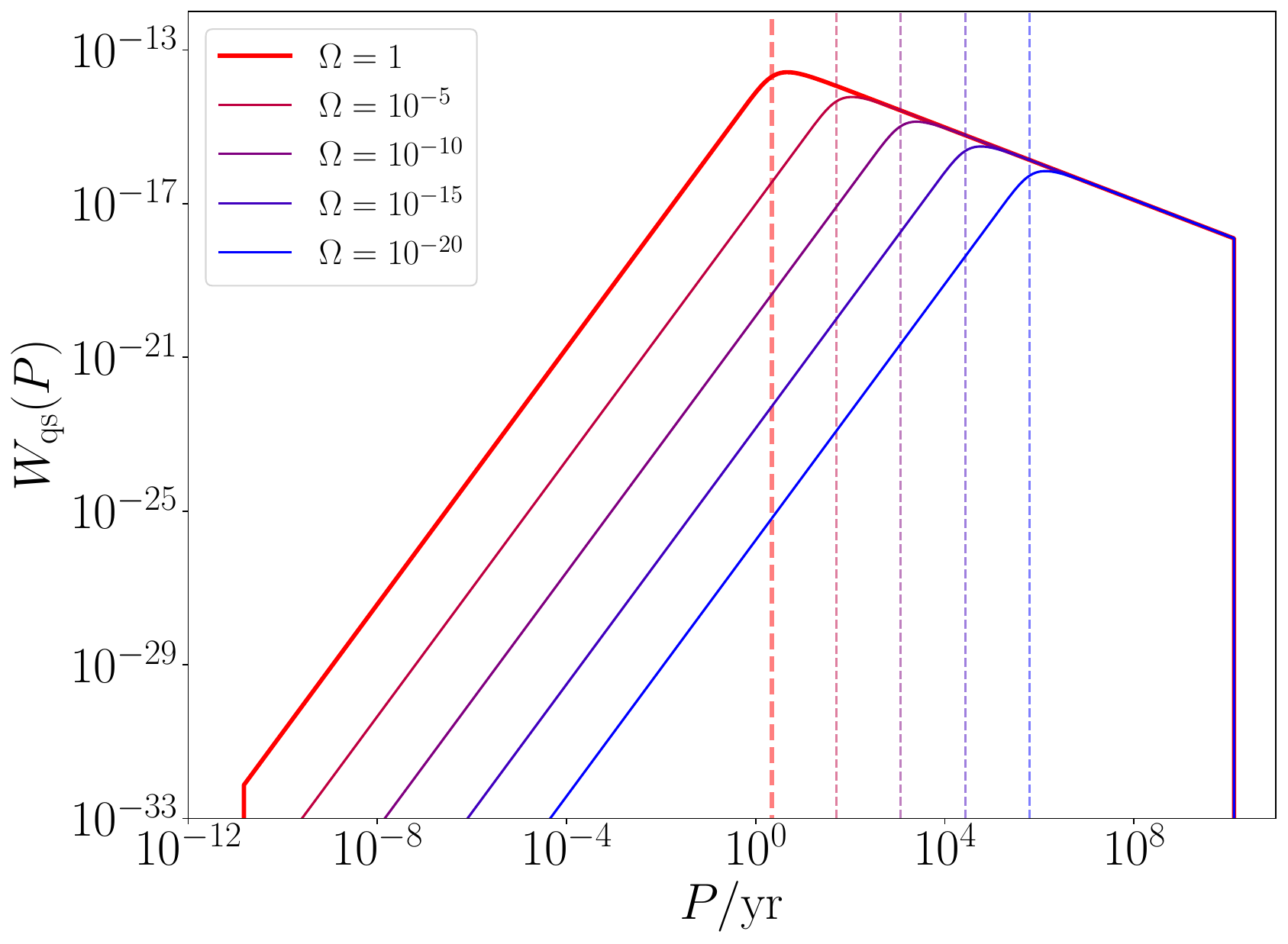}
    \caption{%
    The quasi-stationary distribution~\eqref{eq:quasi-stationary-solution} for the period $P$ of a binary with mass $M=M_\odot$ coupled to a scale-invariant SGWB, $\Omega(f)=\mathrm{constant}$.
    The upper and lower cutoffs are due to the age of the Universe and the binary's ISCO period~\eqref{eq:isco}, respectively.
    The dashed vertical lines indicate the value of $P_\mathrm{crit}$ for each $\Omega$, as defined by Eq.~\eqref{eq:Pcrit}---this is roughly the peak of the distribution in each case.
    Binaries with periods $P<P_\mathrm{crit}$ decay deterministically through GW emission and are removed from the distribution, whereas binaries with $P>P_\mathrm{crit}$ are supported against decay by resonant absorption of the SGWB.
    As we lower the SGWB intensity $\Omega$, the resonance becomes weaker, and the binaries must have longer periods to avoid decay.}
    \label{fig:quasi-stationary-solution}
\end{figure}

As a simple example, consider a scale-invariant SGWB spectrum $\Omega(f)=\mathrm{constant}$, and fix the DF to be zero at the period corresponding to the binary's innermost stable circular orbit (ISCO),
    \begin{equation}
    \label{eq:isco}
        P_\mathrm{ISCO}\equiv6^{3/2}\times2\uppi GM,
    \end{equation}
    with the binary assumed to merge at periods shorter than this.
The corresponding quasi-stationary distribution is then given in terms of the dimensionless variable $\varrho\equiv P/P_\mathrm{ISCO}$ by
    \begin{equation}
    \label{eq:quasi-stationary-solution}
        W_\mathrm{qs}\propto\exp(\frac{\lambda}{\varrho^{11/3}})\qty[\frac{E_{77/132}(\lambda\varrho^{-11/3})}{\varrho^2}-\frac{E_{77/132}(\lambda)}{\varrho^{17/36}}],
    \end{equation}
    where $E_n(z)\equiv\int_1^\infty\dd{t}\rme^{-zt}t^{-n}$ is the exponential integral function, and
    \begin{equation}
        \lambda\equiv\frac{91}{132}\qty(\frac{P_\mathrm{crit}}{P_\mathrm{ISCO}})^{11/3}=\frac{\sqrt{2}\eta(GMH_0)^{-2}}{8019\sqrt{3}\uppi\Omega}\gg1
    \end{equation}
    is a dimensionless constant which quantifies the strength of the deterministic drift $V_P$ relative to the secular diffusion $D^{(2)}$.
While the functional form of Eq.~\eqref{eq:quasi-stationary-solution} is somewhat opaque, one can show that it approximately interpolates between $W_\mathrm{qs}\sim P^{5/3}$ for $P_\mathrm{ISCO}\ll P\ll P_\mathrm{crit}$ and $W_\mathrm{qs}\sim P^{-17/36}$ for $P\gg P_\mathrm{crit}$ (this broken-power-law behaviour is clearly seen in Fig.~\ref{fig:quasi-stationary-solution}).
We normalise Eq.~\eqref{eq:quasi-stationary-solution} by integrating up to some maximum period, which we choose to be the age of the Universe.
By substituting Eq.~\eqref{eq:quasi-stationary-solution} into Eq.~\eqref{eq:prob-current} and evaluating at $\varrho=1$ where $W=0$ and $\pdv*{W}{P}>0$, we see that the probability current is strictly negative, $J<0$.
Since we specified that $J=\mathrm{constant}$, this means there is a uniform net flow towards shorter periods throughout the parameter space.

For hard binaries with $P<P_\mathrm{crit}$ it is obvious that we should have $J<0$, as the negative deterministic drift is the most important effect and quickly drives the binary toward merger.
For soft binaries with $P>P_\mathrm{crit}$ the negative probability current is less immediately obvious; it shows us that, while there is a net drift $D^{(1)}$ pushing these binaries towards longer periods, on average they are nonetheless expected to flow towards shorter periods.
We can understand this somewhat counter-intuitive behaviour by including diffusion as well as drift effects.
Indeed, note that by Eq.~\eqref{eq:prob-current}, the condition $J<0$ is equivalent to
    \begin{equation}
        \pdv{(\ln D^{(2)}W)}{(\ln P)}>\frac{PD^{(1)}}{D^{(2)}}=\frac{91}{36}\qty[1-(P/P_\mathrm{crit})^{-11/3}],
    \end{equation}
    where the RHS tends to a positive constant value in the $P\gg P_\mathrm{crit}$ region we are interested in.
What this is telling us is that, despite the positive drift coefficient, we can still have a negative probability current if and only if the diffusion coefficient $D^{(2)}$ grows sufficiently quickly with $P$, as this makes it sufficiently likely for the binary's random walk to wander below $P_\mathrm{crit}$ and then rapidly approach short periods through GW emission.
Interestingly, we find that for the quasi-stationary distribution we have $\pdv*{(\ln D^{(2)}W_\mathrm{qs})}{(\ln P)}\simeq91/36$, so this diffusive effect is \emph{only just} strong enough to cause a net negative probability current.

We can repeat the calculations above for any given SGWB spectrum to write down a corresponding quasi-stationary solution for the period of a circular binary.
However, the time taken to relax to this distribution is extremely long for typical SGWB spectra (see Fig.~\ref{fig:timescales}), so these solutions are physically uninteresting in most cases.
Besides, the assumption of a perfectly circular binary is overly simplistic, as we have shown in Sec.~\ref{sec:KM-small-e-I} that the eccentricity distribution relaxes away from zero on shorter timescales.
Nonetheless, the approach in this section is still useful for building intuitive understanding of the dynamics of the full DF, and may be useful for, e.g., studies of the orbital element distributions of old populations of binaries, where the full shape of the distribution is vitally important.
Finding quasi-stationary solutions for the full multivariate FPE is much more challenging, and for eccentric binaries there is no guarantee that such a solution with the appropriate lower boundary condition even exists.

\subsection{Mean coalescence time}

For any one-dimensional FPE, it is possible to write down an explicit formula for the \emph{mean first passage time} at either of its boundaries---i.e., the average time taken for an individual random trajectory to reach that boundary, as a function of the initial position~\cite{Gardiner:2004hsm}.
In our case, this is a useful tool for understanding how the presence of a particular SGWB spectrum impacts upon the eventual fate of a binary system.
Applying this to our lower absorbing boundary at the ISCO, we thus have the mean coalescence time of a binary coupled to the SGWB, as a function of its initial period $P_i$,
    \begin{equation}
        \ev{\tau(P_i)}=\int_{P_\mathrm{ISCO}}^{P_i}\dd{P}\int_P^{P_\mathrm{max}}\dd{P'}\frac{I(P')}{I(P)D^{(2)}(P')},
    \end{equation}
    where $I(P)$ is the integrating factor defined in Eq.~\eqref{eq:integrating-factor}.

Returning to the example of a scale-invariant SGWB, this becomes
    \begin{equation}
    \label{eq:coalescence-time}
        \ev{\tau(P_i)}=\int_1^{\varrho_i}\frac{\dd{\varrho}}{\varrho^{91/36}}\int_{\varrho}^{\varrho_\mathrm{max}}\frac{\dd{\varrho'}}{\varrho'^{17/36}}\frac{20\,\rme^{\lambda(\varrho'^{-11/3}-\varrho^{-11/3})}}{27P_\mathrm{ISCO}H_0^2\Omega}.
    \end{equation}
This double integral is challenging to evaluate in general.
However, we can easily verify Eq.~\eqref{eq:coalescence-time} by showing that it reproduces the standard expression for the coalescence time due to deterministic GW emission in cases where the SGWB resonance is weak.
Setting $P_i\ll P_\mathrm{crit}$, we can safely take the limit $\lambda\to\infty$ and extract the leading-order and next-to-leading-order terms,
    \begin{align}
    \begin{split}
        \ev{\tau(P_i)}&\simeq\frac{405GM}{16\eta}\qty(\frac{P_i}{P_\mathrm{ISCO}})^{8/3}\\
        &\times\qty[1-\frac{96}{91}\rme^{-\frac{91}{132}(P_\mathrm{crit}/P_i)^{11/3}}\frac{P_\mathrm{max}^{151/36}}{P_i^{19/36}P_\mathrm{crit}^{11/3}}].
    \end{split}
    \end{align}
The leading-order term agrees with the deterministic coalescence time one finds by integrating $V_P$, as expected.
There is a small negative contribution from the next-to-leading-order term, which indicates that SGWB resonance slightly speeds up the coalescence in this regime.
We can understand this by noticing that the term associated with $D^{(2)}$ in Eq.~\eqref{eq:prob-current} is always negative for the quasi-stationary distribution, so that diffusion always has a net negative contribution to the probability current and thus, on average, always help drive binaries towards merger.

\subsection{Including the remaining orbital elements}

Having found a quasi-stationary period distribution, it is now relatively easy to obtain stationary distributions for the remaining orbital elements $(I,\asc,\xi)$, so long as we hold $\zeta$ and $\kappa$ fixed at zero.
To do so, we write the full FPE as
    \begin{equation}
        \pdv{W}{t}=-\partial_iJ_i,\qquad J_i\equiv D^{(1)}_iW-\partial_j(D^{(2)}_{ij}W),
    \end{equation}
    where $W$ is now interpreted as the multivariate DF over the four orbital elements $(P,I,\asc,\xi)$, and the four corresponding probability currents are
    \begin{align}
    \begin{split}
    \label{eq:probability-currents}
        J_P&=D^{(1)}_PW-\partial_P(D^{(2)}_{PP}W),\\
        J_I&=D^{(1)}_IW-D^{(2)}_{II}\partial_IW,\\
        J_\asc&=-D^{(2)}_{\asc\asc}\partial_\asc W-D^{(2)}_{\asc\xi}\partial_\xi W,\\
        J_\xi&=V_\xi W-D^{(2)}_{\asc\xi}\partial_\asc W-D^{(2)}_{\xi\xi}\partial_\xi W.
    \end{split}
    \end{align}
We have used Eqs.~\eqref{eq:small-e-drift} and~\eqref{eq:small-e-diffusion} to simplify these expressions, in particular using the independence of the diffusion terms from most of the orbital elements to take them outside the partial derivatives.

We first consider the inclination $I$.
By definition, this is constrained to lie in the interval $[0,\uppi]$.
Unlike for the period $P$, a binary reaching one of the boundaries of this interval is not removed from the distribution; instead of absorbing boundary conditions, we have \emph{reflecting} boundary conditions, i.e., the probability current $J_I$ vanishes at both boundaries.
However, if the distribution is stationary then $J_I=\mathrm{constant}$, so the current must vanish everywhere on $[0,\uppi]$.
Setting $J_I=0$ in Eq.~\eqref{eq:probability-currents} gives
    \begin{equation}
        \partial_I\ln W=D^{(1)}_I/D^{(2)}_{II}=\cot I.
    \end{equation}
Integrating this, we find
    \begin{equation}
        W\propto\sin I,
    \end{equation}
    which corresponds to an isotropic distribution (since $\cos I$ is uniformly distributed).
This makes intuitive sense: on long timescales, SGWB resonance causes the binary to ``forget'' its initial orbital plane, such that the resulting stationary distribution is spherically symmetric.
This also agrees with our finding in Sec.~\ref{sec:KM-small-e-I} that $D^{(1)}_I=0$ at $I=\uppi/2$, which is the mean inclination of an isotropic distribution.

For $\asc$ and $\xi$ it is natural to impose periodic boundary conditions for the DF and for the probability currents,
    \begin{align}
    \begin{split}
        W(P,I,\asc,\xi,t)&=W(P,I,\asc+2\uppi,\xi,t)\\
        &=W(P,I,\asc,\xi+2\uppi,t),\\
        J_i(P,I,\asc,\xi,t)&=J_i(P,I,\asc+2\uppi,\xi,t)\\
        &=J_i(P,I,\asc,\xi+2\uppi,t).
    \end{split}
    \end{align}
Stationarity requires $\partial_\asc J_\asc=\partial_\xi J_\xi=0$.
By inspection, we see that this is achieved if $\partial_\asc W=\partial_\xi W=0$, so that the DF depends only on $P$ and $I$.
This corresponds to $\asc$ and $\xi$ being uniformly distributed, which also satisfies the periodic boundary conditions; $\asc$ then has zero probability current, while $\xi$ has a uniform negative current due to the deterministic drift $V_\xi$, which depends only on the period.

\section{Solving the full Fokker-Planck equation}
\label{sec:full-fpe}

We now consider the full FPE for all six orbital elements $(P,e,I,\asc,\omega,\eps)$.
Allowing nonzero eccentricity $e>0$ leads to much more complicated KM coefficients, and means that the results of Sec.~\ref{sec:circular} are no longer applicable.
Nonetheless, we can use the fact that diffusion of the orbital elements due to the SGWB takes place on very long timescales $\tau_\mathrm{diff}\sim1/(PH_0^2\Omega)\gg P$ (see Fig.~\ref{fig:timescales}).
This allows us to develop some useful approximate solution schemes for much shorter observational timescales.

\subsection{Perturbative short-time solution}

The FPE can be written as an operator equation
    \begin{equation}
        \pdv{W}{t}=L_\mathrm{FP}W,
    \end{equation}
    defined by the linear differential operator,
    \begin{equation}
        L_\mathrm{FP}(\vb*X)\equiv-\partial_iD^{(1)}_i+\partial_i\partial_jD^{(2)}_{ij}.
    \end{equation}
This has the formal solution\footnote{%
    Here we take advantage of the time-independence of the secular KM coefficients.
    For time-dependent coefficients, one would need to instead construct a time-ordered Dyson series~\cite{Risken:1989fpe}, though this gives the same result for short times.}
    \begin{equation}
        W(\vb*X,t)=\rme^{tL_\mathrm{FP}}W(\vb*X,0),
    \end{equation}
    which, for short times $t\ll\tau_\mathrm{diff}$, can be expanded as
    \begin{equation}
    \label{eq:short-time-expansion}
        W(\vb*X,t)=\qty[1+tL_\mathrm{FP}(\vb*X,0)+\order{t/\tau_\mathrm{diff}}^2]W(\vb*X,0),
    \end{equation}
    where the RHS depends only on data at time zero.

Suppose that at time zero the binary's orbital elements take on the ``sharp'' values $x_i$.
The initial condition for the DF is then
    \begin{equation}
        W(\vb*X,0)=\delta^{(6)}(\vb*X-\vb*x).
    \end{equation}
By using a Fourier representation of the delta function, we can evaluate Eq.~\eqref{eq:short-time-expansion} to find~\cite{Risken:1989fpe}
    \begin{align}
    \begin{split}
    \label{eq:short-time-solution}
        &W(\vb*X,t)=\frac{1}{\sqrt{\det4\uppi tD^{(2)}}}\\
        &\times\exp{-\frac{[D^{(2)}]^{-1}_{ij}}{4t}(X_i-x_i-D^{(1)}_it)(X_j-x_j-D^{(1)}_jt)}\\
        &+\order{t/\tau_\mathrm{diff}}^2,
    \end{split}
    \end{align}
    i.e., on short timescales, the DF is a multivariate Gaussian with mean $x_i+D^{(1)}_it$ and covariance matrix $2tD^{(2)}_{ij}$.
Here $[D^{(2)}]^{-1}_{ij}$ represents the elements of the inverse of the diffusion matrix, and both the drift vector and diffusion matrix are evaluated at $(\vb*X,t)=(\vb*x,0)$.

\subsection{Evolution of moments of the orbital elements}

The short-time expansion shows that on observational timescales the DF of the orbital elements is approximately Gaussian, and is therefore completely characterised by its first two moments: the mean and the covariance matrix, which we write as
    \begin{align}
    \begin{split}
        \bar{X}_i&\equiv\ev{X_i},\\
        C_{ij}&\equiv\mathrm{Cov}[X_i,X_j]=\ev{(X_i-\bar{X}_i)(X_j-\bar{X}_j)}.
    \end{split}
    \end{align}
It is therefore useful to take moments of the FPE to find the time evolution of these quantities, rather than attempting to calculate the full time-dependent DF.
In doing so, we can calculate the backreaction of perturbations on the evolution of the binary, obtaining corrections to the linear growth found in Eq.~\eqref{eq:short-time-solution}.

The first moment of the FPE gives
    \begin{align}
    \begin{split}
        \dot{\bar{X}}_i&=\pdv{}{t}\int\dd{\vb*X}X_iW=\int\dd{\vb*X}X_i\pdv{W}{t}\\
        &=-\int\dd{\vb*X}X_i\partial_j(D^{(1)}_jW)+\int\dd{\vb*X}X_i\partial_j\partial_k(D^{(2)}_{jk}W).
    \end{split}
    \end{align}
We integrate by parts, and assume that the DF falls off fast enough that all boundary terms vanish, leaving
    \begin{equation}
    \label{eq:mean-evolution}
        \dot{\bar{X}}_i=\int\dd{\vb*X}D^{(1)}_iW=\ev*{D^{(1)}_i}.
    \end{equation}
This fall-off assumption is very reasonable here, as the diffusion rate is extremely small for realistic binaries, so the DF will only have support very near to the mean value.
Doing the same for the second moment gives
    \begin{align}
    \begin{split}
        \dv{}{t}&\ev{X_iX_j}=\int\dd{\vb*X}X_iX_j\pdv{W}{t}\\
        &=-\int\dd{\vb*X}X_iX_j\partial_k(D^{(1)}_kW)\\
        &\qquad\qquad+\int\dd{\vb*X}X_iX_j\partial_k\partial_\ell(D^{(2)}_{k\ell}W)\\
        &=\int\dd{\vb*X}(X_iD^{(1)}_j+X_jD^{(1)}_i)W+2\int\dd{\vb*X}D^{(2)}_{ij}W\\
        &=\ev*{X_iD^{(1)}_j}+\ev*{X_jD^{(1)}_i}+2\ev*{D^{(2)}_{ij}}.
    \end{split}
    \end{align}
We can combine these to give the evolution equation for the covariance matrix,
    \begin{align}
    \begin{split}
    \label{eq:covariance-evolution}
        \dot{C}_{ij}&=\dv{t}(\ev{X_iX_j}-\bar{X}_i\bar{X}_j)\\
        &=\ev*{X_iD^{(1)}_j}+\ev*{X_jD^{(1)}_i}+2\ev*{D^{(2)}_{ij}}\\
        &\qquad\qquad-\ev{X_i}\ev*{D^{(1)}_j}-\ev{X_j}\ev*{D^{(1)}_i}\\
        &=\mathrm{Cov}[X_i,D^{(1)}_j]+\mathrm{Cov}[X_j,D^{(1)}_i]+2\ev*{D^{(2)}_{ij}}.
    \end{split}
    \end{align}

\subsection{The slow-diffusion approximation}

Eqs.~\eqref{eq:mean-evolution} and~\eqref{eq:covariance-evolution} fully describe the evolution of the mean and covariance of the orbital elements.
However, they are given in terms of ensemble averages over nonlinear functions of the orbital elements, which we cannot perform without knowing the full DF.
Even if we were to evaluate them approximately by assuming a Gaussian distribution, the resulting expressions would be very cumbersome.

In the case where the variance is small and any given orbital element $X_i$ is ``close'' to its mean value $\ev{X_i}$ (in a probabilistic sense), one can instead Taylor expand an arbitrary function of the elements around the mean,
    \begin{align}
    \begin{split}
        f(\vb*X)&=f(\bar{\vb*X})+(X_i-\bar{X_i})\partial_if(\bar{\vb*X})\\
        &\qquad+\frac{1}{2}(X_i-\bar{X_i})(X_j-\bar{X_j})\partial_i\partial_jf(\bar{\vb*X})+\cdots,
    \end{split}
    \end{align}
    so that the mean of the function is approximated by
    \begin{equation}
    \label{eq:small-var-approx}
        \ev{f(\vb*X)}\simeq f(\bar{\vb*X})+\frac{1}{2}C_{ij}\partial_i\partial_jf(\bar{\vb*X}).
    \end{equation}
(Note that the first-order term in the expansion vanishes when taking the mean, as the first central moment is identically zero.)

We can justify using Eq.~\eqref{eq:small-var-approx} by noting that the diffusion matrix calculated in Sec.~\ref{sec:KM} is very small in most physical situations.
To keep track of how this smallness propagates into the evolution equations, we introduce a formal small parameter $\epsilon$ (which we will later set to unity), writing
    \begin{equation}
        D^{(2)}_{ij}\to\epsilon D^{(2)}_{ij}.
    \end{equation}
For sharp initial conditions $\bar{X}_i=x_i$, $C_{ij}=0$, we see from Eq.\eqref{eq:covariance-evolution} that $C_{ij}=\order{\epsilon}$, as the $\mathrm{Cov}[\vb*X,D^{(1)}]$ terms are initially zero.
We therefore also write
    \begin{equation}
        C_{ij}\to\epsilon C_{ij}.
    \end{equation}
We thus see that Eq.~\eqref{eq:small-var-approx} is justified if we neglect terms of order $\epsilon^2$.
We call this the \emph{slow-diffusion} approximation, as it relies on the fact that the timescale $\tau_\mathrm{diff}$ over which the covariance grows is long compared to the observation time.
Note that the stochastic contribution to the drift vector is generally of the same order as the diffusion matrix, so we also write
    \begin{equation}
        D^{(1)}_i=V_i+\epsilon\updelta D^{(1)}_i,
    \end{equation}
    where the stochastic term $\updelta D^{(1)}_i$ is suppressed by a factor of $\epsilon$.

\begin{figure*}[t!]
    \includegraphics[width=\textwidth]{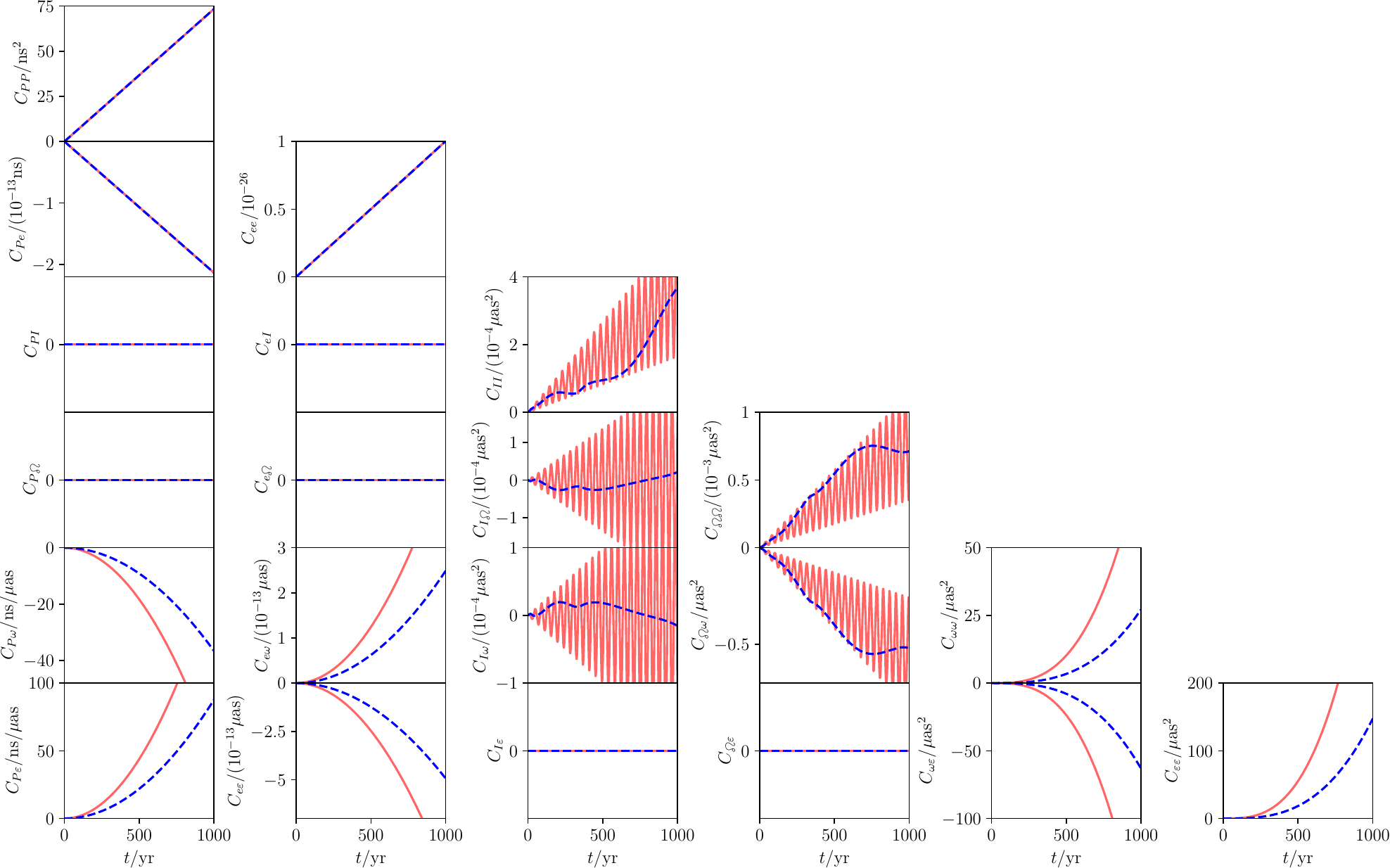}
    \caption{%
    The covariance matrix $C_{ij}(t)$ of the orbital elements of the binary pulsar B1913+16 (the Hulse-Taylor system) over a $1\,\mathrm{kyr}$ interval, assuming a scale-invariant SGWB spectrum $\Omega(f)=10^{-5}$, and including the first 400 harmonics.
    The blue dashed curves show numerical solutions of the evolution equations~\eqref{eq:slow-diffusion-evolution-equations}, while the pale red curves show the naive solution $t\times\dot{C}_{ij}$, which is exact only when $\dot{C}_{ij}=\mathrm{constant}$.}
    \label{fig:HT_cov_t}
\end{figure*}

Applying this approximation to Eqs.~\eqref{eq:mean-evolution} and~\eqref{eq:covariance-evolution}, we obtain the moment evolution equations to first order in $\epsilon$,
    \begin{align}
    \begin{split}
    \label{eq:moment-evolution-slow-diffusion}
        \dot{\bar{X}}_i&=V_i+\epsilon\updelta D^{(1)}_i+\frac{1}{2}\epsilon C_{jk}\partial_j\partial_kV_i+\order{\epsilon^2},\\
        \epsilon\dot{C}_{ij}&=2\epsilon D^{(2)}_{ij}+\epsilon C_{ik}\partial_kV_j+\epsilon C_{jk}\partial_kV_i+\order{\epsilon^2},
    \end{split}
    \end{align}
    where the drift vector and diffusion matrix are both evaluated at $\bar{\vb*X}$.

We see that the evolution equation~\eqref{eq:moment-evolution-slow-diffusion} for the mean orbital elements includes both $\order{\epsilon^0}$ and $\order{\epsilon^1}$ terms.
For numerical reasons, it is convenient to separate these.
We therefore write the mean orbital elements as
    \begin{equation}
        \bar{\vb*X}(t)=\bar{\vb*X}_0(t)+\epsilon\updelta\bar{\vb*X}(t),
    \end{equation}
    where $\bar{\vb*X}_0$ represents the values the elements would take in the absence of the SGWB, which obey the deterministic evolution equation
    \begin{equation}
        \dot{\bar{X}}_{0,i}=V_i(\bar{\vb*X}_0).
    \end{equation}
(Note that this separation of deterministic and stochastic parts of the drift is always possible, regardless of the size of the deterministic term.)
Meanwhile $\updelta\bar{\vb*X}$ represents the deviation in the mean due to SGWB resonance, and evolves according to
    \begin{align}
    \begin{split}
        \epsilon\updelta\dot{\bar{X}}_i=V_i(\bar{\vb*X})&-V_i(\bar{\vb*X}_0)+\epsilon\updelta D^{(1)}_i(\bar{\vb*X})\\
        &+\frac{1}{2}\epsilon C_{jk}\partial_j\partial_kV_i(\bar{\vb*X})+\order{\epsilon^2}.
    \end{split}
    \end{align}
We can Taylor-expand the terms that are evaluated at $\bar{\vb*X}=\bar{\vb*X}_0+\epsilon\updelta\bar{\vb*X}$ to give
    \begin{align}
    \begin{split}
        \epsilon\updelta\dot{\bar{X}}_i=\epsilon\updelta\bar{X}_j\partial_j&V_i(\bar{\vb*X}_0)+\epsilon\updelta D^{(1)}_i(\bar{\vb*X}_0)\\
        &+\frac{1}{2}\epsilon C_{jk}\partial_j\partial_kV_i(\bar{\vb*X}_0)+\order{\epsilon^2}.
    \end{split}
    \end{align}

With the appropriate terms identified, we can send $\epsilon\to1$.
Our full set of evolution equations, to leading order in the slow-diffusion approximation, reads
    \begin{align}
    \begin{split}
    \label{eq:slow-diffusion-evolution-equations}
        \dot{\bar{X}}_{0,i}&=V_i,\\
        \updelta\dot{\bar{X}}_i&\simeq\updelta D^{(1)}_i+\updelta\bar{X}_j\partial_jV_i+\frac{1}{2}C_{jk}\partial_j\partial_kV_i,\\
        \dot{C}_{ij}&\simeq2D^{(2)}_{ij}+C_{ik}\partial_kV_j+C_{jk}\partial_kV_i,
    \end{split}
    \end{align}
    with all terms evaluated at the deterministic mean elements $\bar{\vb*X}_0$.

\begin{figure*}[t!]
    \includegraphics[width=0.85\textwidth]{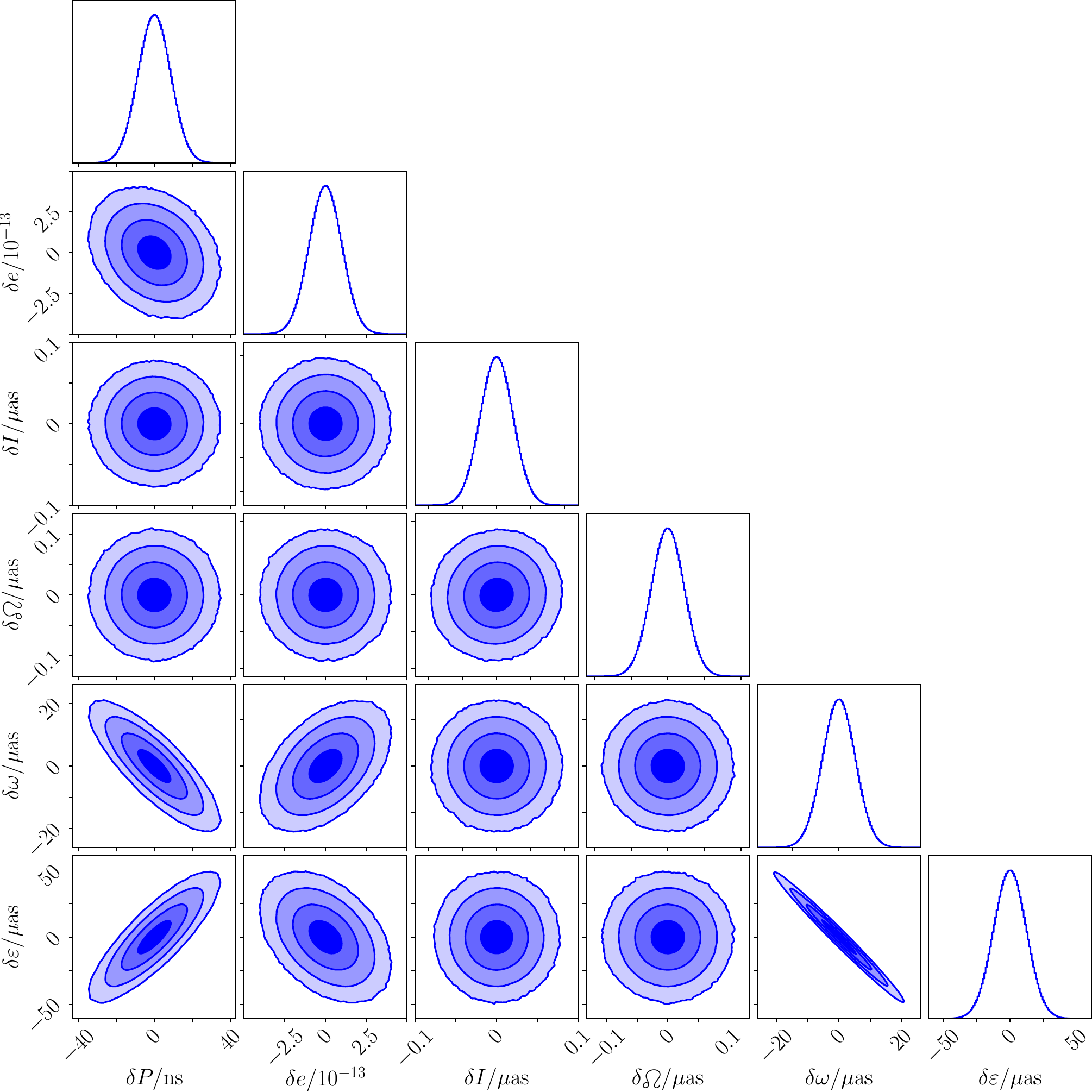}
    \caption{%
    Corner plot showing the distribution of the orbital elements of B1913+16 at the end of the $1\,\mathrm{kyr}$ numerical integration shown in Fig.~\ref{fig:HT_cov_t}.}
    \label{fig:HT_corner}
\end{figure*}

It is interesting to note that, starting from sharp initial conditions, some elements of the covariance matrix remain fixed at zero under Eq.~\eqref{eq:slow-diffusion-evolution-equations}:
    \begin{equation}
    \label{eq:zero-cov}
        C_{PI}=C_{P\asc}=C_{eI}=C_{e\asc}=C_{I\eps}=C_{\asc\eps}=0,
    \end{equation}
    i.e., stochastic perturbations to these pairs of orbital elements remain statistically uncorrelated at first order in the slow-diffusion approximation.
This is due to the vanishing of the corresponding elements of the diffusion matrix $D^{(2)}_{ij}$ in Eq.~\eqref{eq:ecc-diffusion}, as well as the fact that $V_I=V_\asc=0$.
However, all other elements of the covariance matrix generically grow over time.

In Fig.~\ref{fig:HT_cov_t}, we show an example of an integration of Eq.~\eqref{eq:slow-diffusion-evolution-equations} for the Hulse-Taylor binary pulsar B1913+16~\cite{Hulse:1974eb}.
We see that, in this example, the period-eccentricity sector of the covariance matrix grows linearly with time,
    \begin{equation}
        C_{PP}\sim C_{Pe}\sim C_{ee}\sim t,
    \end{equation}
    indicating that the right-hand side of Eq.~\eqref{eq:slow-diffusion-evolution-equations} is dominated by the $D^{(2)}$ term for these components.
In contrast, many of the components involving the argument of pericentre and compensated mean anomaly (specifically $C_{P\omega}$, $C_{P\eps}$, $C_{e\omega}$, $C_{e\eps}$, $C_{\omega\omega}$, $C_{\omega\eps}$, and $C_{\eps\eps}$) have relatively smaller values for the diffusion matrix, and are instead driven by the $C\partial V$ terms, leading them to grow like $\sim t^2$ (since the components of the covariance matrix sourcing them grow like $C\sim t$).
The components $C_{II}$, $C_{I\asc}$, $C_{I\omega}$, $C_{\asc\asc}$, and $C_{\asc\omega}$ evolve more erratically; this is due to the presence of $\sin2\omega$, $\cos2\omega$ terms in the corresponding components of the diffusion matrix [see Eq.~\eqref{eq:ecc-diffusion}], which oscillate on a timescale $\uppi/\dot{\omega}_\mathrm{sec}\approx43\,\mathrm{yr}$ due to the perihelion precession of the Hulse-Taylor system~\cite{Weisberg:2010zz,Weisberg:2016jye}.
Finally, the remaining six components of the covariance matrix are zero throughout the integration time, as expected from Eq.~\eqref{eq:zero-cov}.

In Fig.~\ref{fig:HT_corner} we show the distribution of the orbital elements at the end of the integration in Fig.~\ref{fig:HT_cov_t}.
We see that the orbital elements are, on the whole, weakly correlated with each other, with the notable exceptions of the pairs $(P,\omega)$, $(P,e)$, and particularly $(\omega,\eps)$, which are highly covariant.
Note that these distributions include the overall offset due to stochastic drift, but that this is less important than diffusion in this case, and is therefore harder to distinguish by eye.

\subsection{Growth rate of non-Gaussianity}

By using Eq.~\eqref{eq:slow-diffusion-evolution-equations} we have neglected all higher-order moments of the distribution, and thus fail to capture any non-Gaussianities in the DF.
We can measure the departure from Gaussianity by tracking the evolution of the third central moment,
    \begin{equation}
        S_{ijk}\equiv\ev{(X_i-\bar{X}_i)(X_j-\bar{X}_j)(X_k-\bar{X}_k)},
    \end{equation}
    as this vanishes identically for a Gaussian distribution.
In particular, for each individual orbital element $X_i$ we have, to leading order in the slow-diffusion approximation,
    \begin{equation}
    \label{eq:skewness-evolution}
        \dot{S}_{iii}\simeq-3\bar{X}_iC_{ij}\partial_jV_i\, ,
    \end{equation}
    (with no summation over $i$).
Unlike the equations for the first two moments, the RHS of Eq.~\eqref{eq:skewness-evolution} is initially zero---non-Gaussianity is only sourced once the covariance has had a chance to grow.
We also see that non-Gaussianity can only be sourced at leading order for orbital elements which have a nonzero deterministic drift $V_i$; the inclination $I$ and longitude of ascending node $\asc$ remain Gaussian at leading order.
This justifies our assumption that the distribution is Gaussian on observational timescales.
However, studies of, e.g., the properties of old binary populations will require us to drop this assumption and solve for the full DF.
The results of Sec.~\ref{sec:circular} are an important first step in studying cases like these.

\section{Observational applications}
\label{sec:observations}

In this section we describe how to estimate the sensitivity of a given binary to a generic SGWB spectrum using two different high-precision probes of binary dynamics: timing of millisecond pulsars, and laser ranging experiments.
This will allow us to compute forecasts for the upper limits that each observational probe will be able to place on various SGWB spectra.

\subsection{Distribution of the observed orbital elements}

Consider a set of observations of a binary system, which are broken up into intervals much longer than the orbital period, but much shorter than the orbital diffusion timescale.
These observations give us a discrete series of measurements of the orbital elements $\vb*X(t)$, one for each interval, with the intervals being labelled by $t$.
(It is necessary to split up the data into intervals, since each high-precision measurement of the orbital elements requires a large number of individual data points.)
We assume these measurements are made with Gaussian noise and zero bias.
The measured values $\vu*X(t)$ are thus drawn from a multivariate Gaussian distribution centred on the true values $\vb*X(t)$, with log-likelihood $\mathcal{L}=\ln p$ given by
    \begin{equation}
        -2\mathcal{L}(\vu*X|\vb*X)=\sum_t\ln\det2\uppi\mathsf{M}+(\vu*X-\vb*X)^\mathsf{T}\mathsf{M}^{-1}(\vu*X-\vb*X),
    \end{equation}
    where the measurement noise is described by the covariance matrix $\mathsf{M}$.
The form of the covariance matrix depends on how sensitive the timing residuals are to each orbital element---we discuss this further below.

The true orbital elements $\vb*X(t)$ are themselves random due to the uncertainty caused by SGWB resonance.
As discussed in Sec.~\ref{sec:full-fpe}, we can usually approximate the distribution of the elements as Gaussian (at least on observational timescales), so that the log-likelihood reads
    \begin{equation}
        -2\mathcal{L}(\vb*X|\Omega)=\sum_t\ln\det2\uppi\mathsf{C}+(\vb*X-\bar{\vb*X})^\mathsf{T}\mathsf{C}^{-1}(\vb*X-\bar{\vb*X}),
    \end{equation}
    where the mean values $\bar{\vb*X}(t)$ and covariance matrix $\mathsf{C}(t)$ both depend on the SGWB spectrum $\Omega(f)$, and are computed as described in Sec.~\ref{sec:full-fpe}.

We can marginalise over the unknown ``true'' elements $\vb*X(t)$ to obtain a likelihood function for the measured elements $\vu*X(t)$ for a given SGWB spectrum,
    \begin{align}
    \begin{split}
    \label{eq:likelihood}
        -2\mathcal{L}(\vu*X|\Omega)&\equiv-2\ln\qty[\int\dd{\vb*X}p(\vu*X|\vb*X)p(\vb*X|\Omega)]\\
        &=\sum_t\ln\det2\uppi\mathsf{N}+(\vu*X-\bar{\vb*X})^\mathsf{T}\mathsf{N}^{-1}(\vu*X-\bar{\vb*X}),
    \end{split}
    \end{align}
    where
    \begin{equation}
        \mathsf{N}\equiv\mathsf{M}+\mathsf{C}
    \end{equation}
    is the combined covariance matrix, incorporating the measurement uncertainty as well as the stochasticity of the orbital elements.

\subsection{Likelihood-ratio test}

Given a set of measured orbital elements $\vu*X$, we can phrase the SGWB detection problem as a hypothesis test, where
    \begin{itemize}
        \item $\mathcal{H}_0$ is the \emph{null hypothesis}, that there is no SGWB present, $\Omega=0$.
        \item $\mathcal{H}_\Omega$ is the \emph{alternative hypothesis}, that there is a SGWB present, $\Omega\ne0$.
    \end{itemize}
The simplest version of this problem is when we are searching for a SGWB with a fixed spectral shape (e.g., a power law), in which case the only unknown is a single parameter setting the overall amplitude, which we denote $\Omega$.
(Concretely, this parameter $\Omega$ should then refer to the amplitude of the SGWB at some fixed reference frequency.)

A very natural way of carrying out such a hypothesis test is by using the log-likelihood-ratio statistic
    \begin{equation}
        \Lambda(\vu*X)\equiv2\max_{\Omega>0}\qty[\mathcal{L}(\vu*X|\Omega)-\mathcal{L}(\vu*X|0)],
    \end{equation}
    i.e., we compare the likelihood~\eqref{eq:likelihood} in the case of the null hypothesis $\Omega=0$ with the maximum value of the likelihood as a function of $\Omega$ in the case of the alternative hypothesis.
Since the maximum value of the likelihood over the entire range of values of $\Omega$ is always greater than or equal to its value at $\Omega=0$, we see that $\Lambda\ge0$.

Using Eq.~\eqref{eq:likelihood}, along with the fact that $\mathsf{N}=\mathsf{M}$ when $\Omega=0$, we find
    \begin{align}
    \begin{split}
    \label{eq:detection-statistic}
        \Lambda(\vu*X)&=\max_{\Omega>0}\sum_t(\vu*X-\bar{\vb*X}_0)^\mathsf{T}\mathsf{M}^{-1}(\vu*X-\bar{\vb*X}_0)\\
        &-(\vu*X-\bar{\vb*X})^\mathsf{T}\mathsf{N}^{-1}(\vu*X-\bar{\vb*X})-\ln\det(\mathsf{I}+\mathsf{M}^{-1}\mathsf{C}),
    \end{split}
    \end{align}
    where $\bar{\vb*X}_0$ is the value of $\bar{\vb*X}$ when $\Omega=0$, and where we have used
    \begin{equation}
        \frac{\det\mathsf{N}}{\det\mathsf{M}}=\det\mathsf{M}^{-1}\mathsf{N}=\det(\mathsf{I}+\mathsf{M}^{-1}\mathsf{C}).
    \end{equation}

Given a set of measurements $\vu*X(t)$, we can therefore compute the likelihood ratio statistic using Eq.~\eqref{eq:detection-statistic} by solving the FPE to find $\bar{\vb*X}(t)$ and $\mathsf{C}(t)$ for a large number of possible of values of $\Omega$ in order to maximise the likelihood.
If the resulting value of $\Lambda$ is large enough, then we reject the null hypothesis and claim a detection of the SGWB.

An obvious question is:
What value of $\Lambda$ is ``large enough''?
To define the detection threshold, we need to know the distribution of $\Lambda$ in the case where $\mathcal{H}_0$ is true.
Comparing the observed value of $\Lambda$ with this null distribution then allows us to directly infer the statistical significance of the results.
Since $\Lambda$ is a complicated function of the data $\vu*X$, it is difficult to find an exact distribution, even though we have fully specified how the data are distributed.
However, in the limit where we have a large number of measurements (i.e., the data cover a large number of time intervals $t$), Wilks' theorem tells us that $\Lambda$ follows a chi-squared distribution~\cite{Casella:2002stat},
    \begin{equation}
        \lim_{n_t\to\infty}\Lambda\sim\chi^2_1,
    \end{equation}
    where $n_t$ is the number of time segments.
(In this case the chi-squared distribution has one degree of freedom, as $\mathcal{H}_\Omega$ has one free parameter compared to $\mathcal{H}_0$; we could imagine using a more complicated model for the SGWB with $k$ parameters, in which case the appropriate distribution would be $\chi^2_k$.)
In this limit, we can therefore set the threshold for detecting the SGWB at a given confidence level according to the corresponding $p$-value of $\chi^2_1$; e.g., a detection with 95\% confidence would require $\Lambda\ge3.841$.

\subsection{Sensitivity forecasts}

We can also use the likelihood-ratio statistic discussed above to estimate the sensitivity of future observing campaigns to different SGWB spectra.
To do so, we simply compute the expected value of $\Lambda$ under the SGWB hypothesis $\mathcal{H}_\Omega$, and find the smallest value of $\Omega$ for which this expectation surpasses the detection threshold---this tells us the weakest SGWB that we can expect to detect with a given set of observations.

Let us use $\ev{\cdots}_\Omega$ to denote an expectation value under $\mathcal{H}_\Omega$.
By definition, we have
    \begin{align}
    \begin{split}
        \ev*{\hat{X}_i}_\Omega&=\bar{X}_i,\\
        \ev*{(\hat{X}_i-\bar{X}_i)(\hat{X}_j-\bar{X}_j)}_\Omega&=\mathsf{N}_{ij},
    \end{split}
    \end{align}
    where the mean vector $\bar{\vb*X}$ and covariance matrix $\mathsf{C}$ here are computed using the true underlying value of $\Omega$.
In principle, this true $\Omega$ is different from the value $\hat{\Omega}$ that maximises the likelihood, and it is the latter which determines the values for $\bar{\vb*X}$ and $\mathsf{C}$ that appear in the statistic $\Lambda$.
However, in the $n_t\to\infty$ limit discussed above we have $\hat{\Omega}\to\Omega$ (in statistics parlance, the maximum-likelihood estimator is efficient~\cite{Casella:2002stat}), so the two are interchangeable.
It is thus straightforward to show that, in this limit, the expected value of $\Lambda$ under $\mathcal{H}_\Omega$ is
    \begin{align}
    \begin{split}
    \label{eq:expected-lambda}
        \ev{\Lambda}_\Omega&=\sum_t\updelta\bar{\vb*X}^\mathsf{T}\mathsf{M}^{-1}\updelta\bar{\vb*X}+\tr\mathsf{M}^{-1}\mathsf{C}-\ln\det(\mathsf{I}+\mathsf{M}^{-1}\mathsf{C}).
    \end{split}
    \end{align}
We see that the stochastic drift in the orbital elements appears quadratically here, which means that it typically contributes less to the detectability of the SGWB than diffusion, which enters linearly through the covariance matrix.

Once we have specified the covariance matrix $\mathsf{M}$ for our observations, we can compute Eq.~\eqref{eq:expected-lambda} and find the smallest value of $\Omega$ that we can expect to detect.

\subsection{Application to pulsar timing}

While the formalism we have developed above is applicable to a very broad class of astrophysical binary systems, one of the main applications is in the case where one member of the binary is a millisecond pulsar (MSP)~\cite{Lorimer:2008se}.
Analysis of the timing data from this MSP then allows us to determine its orbit with incredible precision, with uncertainties as small as a few parts per billion in some cases.
These precision measurements allow us to search for the very small stochastic perturbations to the orbit described above.
(The same principle has been used to set novel constraints on ultralight dark matter, due to its resonant effects on binary pulsar orbits~\cite{Blas:2016ddr,LopezNacir:2018epg,Blas:2019hxz,Armaleo:2019gil,Desjacques:2020fdi}.)
Here we follow the approach of Refs.~\cite{Blandford:1976pt,Epstein:1977pn,Damour:1986pn} to estimate the covariance matrix $\mathsf{M}$ for the orbital elements of a binary pulsar.

For each of the $n_t$ observation intervals used to construct the likelihood ratio statistic above, one must observe some number $n_\obs$ of pulse arrival times, compare these arrival times with a timing formula for the binary, and thereby infer the binary's orbital elements at that time from the timing residuals.
In the limit $n_\obs\to\infty$, the resulting covariance matrix $\mathsf{M}$ describing the uncertainty in the orbital elements is given by the inverse of the \emph{Fisher matrix},
    \begin{equation}
        \mathsf{M}=\mathsf{F}^{-1},\qquad\mathsf{F}_{ij}\equiv-\ev{\partial_i\partial_j\mathcal{L}(\vb*t|\vb*X)},
    \end{equation}
    where $\mathcal{L}$ is the log-likelihood describing the distribution of arrival times $\vb*t=(t_1,t_2,\ldots,t_{n_\obs})$ as a function of the orbital elements $\vb*X$, and the angle brackets here denote an expectation value under that distribution.
We make the standard assumptions that the pulse arrival times form a set of uncorrelated Gaussian random variables with constant variance $\sigma^2$ (i.e., the timing noise is independent of the binary's orbit), and with mean values given by the timing formula,
    \begin{equation}
        \ev{t_a}\equiv\mathcal{T}_a(\vb*X),\qquad\mathrm{Cov}[t_a,t_b]\equiv\delta_{ab}\sigma^2,
    \end{equation}
    so that the log-likelihood is given by
    \begin{equation}
        -2\mathcal{L}(\vb*t|\vb*X)=\sum_{a=1}^{n_\obs}\ln2\uppi\sigma^2+\frac{1}{\sigma^2}\qty(t_a-\mathcal{T}_a)^2.
    \end{equation}
The Fisher matrix for a likelihood of this form (as derived in, e.g., Ref.~\cite{Tegmark:1996bz}) is
    \begin{equation}
    \label{eq:fisher-sum}
        \mathsf{F}_{ij}\equiv\frac{1}{\sigma^2}\sum_{a=1}^{n_\obs}\pdv{\mathcal{T}_a}{X_i}\pdv{\mathcal{T}_a}{X_j}.
    \end{equation}

To evaluate the Fisher matrix, we therefore need the derivatives of the timing formula $\mathcal{T}$ with respect to each of the orbital elements.
Using the standard Blandford-Teukolsky timing formula~\cite{Blandford:1976pt}, we write these as
    \begin{align}
    \begin{split}
    \label{eq:blandford-teukolsky}
        \pdv{\mathcal{T}}{P}&=\frac{v_P\sin I}{1+m_1/m_2}\\
        &\times\frac{E-e\sin E}{2\uppi(1-e\cos E)}(\sin\omega\sin E-\gamma\cos\omega\cos E),\\
        \pdv{\mathcal{T}}{e}&=-\frac{P}{2\uppi}\frac{v_P\sin I}{1+m_1/m_2}\\
        &\times\qty[\sin\omega(1+\sin^2E)+\frac{\cos\omega\sin E}{\gamma}(e-\gamma^2\cos E)],\\
        \pdv{\mathcal{T}}{I}&=\frac{P}{2\uppi}\frac{v_P\cos I}{1+m_1/m_2}\qty[\sin\omega(\cos E-e)+\gamma\cos\omega\sin E],\\
        \pdv{\mathcal{T}}{\asc}&=0,\\
        \pdv{\mathcal{T}}{\omega}&=\frac{P}{2\uppi}\frac{v_P\sin I}{1+m_1/m_2}\qty[\cos\omega(\cos E-e)-\gamma\sin\omega\sin E],\\
        \pdv{\mathcal{T}}{\eps}&=-\frac{P}{2\uppi}\frac{v_P\sin I}{1+m_1/m_2}\frac{\sin\omega\sin E-\gamma\cos\omega\cos E}{1-e\cos E},
    \end{split}
    \end{align}
    where the \emph{eccentric anomaly} $E(t)$ is defined by
    \begin{equation}
        r=a(1-e\cos E),\qquad\cos\psi=\frac{\cos E-e}{1-e\cos E},
    \end{equation}
    and acts as an alternative to the true anomaly $\psi$ as a way of parameterising the orbital ellipse.
We see that the timing formula does not depend on the longitude of ascending node, $\pdv*{\mathcal{T}}{\asc}=0$, which means that $\asc$ cannot be determined with pulsar timing (physically this is because $\asc$ corresponds to a rotation around the line-of-sight axis, and thus does not affect the observable motion parallel to the line of sight).

In the limit of many observed pulses, $n_\obs\to\infty$, and assuming that these observations are distributed uniformly in time, we can replace the sum in Eq.~\eqref{eq:fisher-sum} with an integral to obtain
    \begin{align}
    \begin{split}
    \label{eq:fisher-integral}
        \mathsf{F}_{ij}&\simeq\frac{n_\obs}{T_\obs\sigma^2}\int_0^{T_\obs}\dd{t}\pdv{\mathcal{T}}{X_i}\pdv{\mathcal{T}}{X_j}\\
        &\simeq\frac{n_\obs}{T_\obs\sigma^2}\frac{P}{2\uppi}\int_0^{2\uppi T_\obs/P}\dd{E}(1-e\cos E)\pdv{\mathcal{T}}{X_i}\pdv{\mathcal{T}}{X_j},
    \end{split}
    \end{align}
    where $T_\obs$ is the time interval over which the $n_\obs$ pulse measurements are made, and where we have used Kepler's equation,
    \begin{equation}
        \dv{E}{t}=\frac{2\uppi/P}{1-e\cos E}.
    \end{equation}
For simplicity, we assume that $T_\obs$ is an integer multiple of the binary period $P$, and begins when the binary is at pericentre; if this is not the case, then the following formulae contain additional phase factors which do not affect the order of magnitude of the results.

We find that it is convenient to define
    \begin{equation}
        \mathsf{F}_{ij}=\frac{n_\obs P^2}{16\uppi^2\sigma^2}\qty(\frac{v_P\sin I}{1+m_1/m_2})^2\;\tilde{\mathsf{F}}_{ij},
    \end{equation}
    pulling out some factors which appear in all of the Fisher matrix elements.
Inserting the derivatives~\eqref{eq:blandford-teukolsky} into Eq.~\eqref{eq:fisher-integral}, we find that this factorised form of the Fisher matrix is given by
    \begin{align}
    \begin{split}
        \tilde{\mathsf{F}}_{PP}&=\frac{8e}{P^2}(1+\tfrac{1}{4}e^2)+\frac{\cos2\omega}{P^2}f_1(e)\\
        &+\frac{2\uppi T_\obs}{P^3}\sin2\omega f_2(e)+\frac{8\uppi^2T_\obs^2}{3P^4}\qty[1-\tfrac{1}{4}\cos2\omega f_3(e)],\\
        \tilde{\mathsf{F}}_{Pe}&=\frac{4}{P}\qty(1+\tfrac{3}{8}e)\qty(1-\tfrac{1}{6}e^2)+\frac{\uppi T_\obs e}{P^2\gamma}\sin2\omega\\
        &-\frac{8\cos2\omega}{3P}\qty(1+\tfrac{3}{4}e+\tfrac{1}{6}e^2+\tfrac{1}{16}e^3),\\
        \tilde{\mathsf{F}}_{PI}&=\frac{\cot I}{P}\qty[2e\qty(1+\tfrac{1}{4}e)+\cos2\omega\qty(1-2e-\tfrac{3}{2}e^2)],\\
        \tilde{\mathsf{F}}_{P\omega}&=-\frac{\sin2\omega}{P}(1-2e-\tfrac{3}{2}e^2)-\frac{2\uppi T_\obs\gamma}{P^2},\\
        \tilde{\mathsf{F}}_{P\eps}&=-\frac{\sin2\omega}{P}f_2(e)-\frac{2\uppi T_\obs}{P^2}\qty[1-\tfrac{1}{4}\cos2\omega f_3(e)],\\
        \tilde{\mathsf{F}}_{ee}&=\frac{5}{\gamma^2}\qty[1-\tfrac{3}{4}e^2-\tfrac{1}{20}e^4-\tfrac{9}{10}\cos2\omega\qty(1-\tfrac{23}{18}e^2+\tfrac{1}{18}e^4)],\\
        \tilde{\mathsf{F}}_{eI}&=3e\cot I\qty[1+\tfrac{1}{12}e^2-\tfrac{11}{6}\cos2\omega\qty(1-\tfrac{1}{22}e^2)],\\
        \tilde{\mathsf{F}}_{e\omega}&=\tfrac{11}{2}e\sin2\omega\qty(1-\tfrac{1}{22}e^2),\\
        \tilde{\mathsf{F}}_{e\eps}&=\frac{e}{\gamma}\sin2\omega,\\
        \tilde{\mathsf{F}}_{II}&=2\cot^2I\qty(1+\tfrac{3}{2}e^2-\tfrac{5}{2}e^2\cos2\omega),\\
        \tilde{\mathsf{F}}_{I\omega}&=5e^2\cot I\sin2\omega,\\
        \tilde{\mathsf{F}}_{I\eps}&=0,\\
        \tilde{\mathsf{F}}_{\omega\omega}&=2+3e^2+5e^2\cos2\omega,\\
        \tilde{\mathsf{F}}_{\omega\eps}&=2\gamma,\\
        \tilde{\mathsf{F}}_{\eps\eps}&=4\frac{\sin^2\omega+\gamma\cos^2\omega}{1+\gamma},
    \end{split}
    \end{align}
    where the $f_i(e)$ are functions of eccentricity, which are given to $\order{e^{14}}$ by
    \begin{align}
    \begin{split}
        f_1(e)&\simeq1+\tfrac{8}{9}e-\tfrac{3}{16}e^2-\tfrac{448}{225}e^3-\tfrac{175}{288}e^4-\tfrac{11584}{11025}e^5\\
        &\qquad-\tfrac{2975}{9216}e^6-\tfrac{67264}{99225}e^7-\tfrac{96733}{460800}e^8-\tfrac{5818432}{12006225}e^9\\
        &\qquad-\tfrac{278579}{1843200}e^{10}-\tfrac{149726912}{405810405}e^{11}-\tfrac{20910823}{180633600}e^{12},\\
        f_2(e)&\simeq1+\tfrac{4}{3}e+\tfrac{5}{8}e^2+\tfrac{2}{15}e^3+\tfrac{1}{48}e^4+\tfrac{1}{210}e^5\\
        &\qquad-\tfrac{29}{768}e^6-\tfrac{31}{1260}e^7-\tfrac{359}{7680}e^8-\tfrac{3559}{110880}e^9\\
        &\qquad-\tfrac{469}{10240}e^{10}-\tfrac{19087}{576576}e^{11}-\tfrac{6099}{143360}e^{12},\\
        f_3(e)&\simeq e^2+\tfrac{1}{2}e^4+\tfrac{5}{16}e^6+\tfrac{7}{32}e^8+\tfrac{21}{128}e^{10}+\tfrac{33}{256}e^{12}.
    \end{split}
    \end{align}
We have checked numerically that these expressions are accurate to within $\approx2\%$ even at large eccentricity, $e=0.95$; for smaller eccentricities, the accuracy is even better.

For binaries with small eccentricity, we can change to the alternative orbital elements $(P,\zeta,\kappa,I,\asc,\xi)$ to find
    \begin{align}
    \begin{split}
        \tilde{\mathsf{F}}_{PP}&=\frac{1}{P^2}\qty[-\frac{\zeta^2-\kappa^2}{\zeta^2+\kappa^2}+\frac{4\uppi T_\obs}{P}\frac{\zeta\kappa}{\zeta^2+\kappa^2}+\frac{8\uppi^2T_\obs^2}{3P^2}],\\
        \tilde{\mathsf{F}}_{P\zeta}&=\frac{4\zeta}{3P}\frac{5\zeta^2+6\kappa^2}{(\zeta^2+\kappa^2)^{3/2}},\\
        \tilde{\mathsf{F}}_{P\kappa}&=\frac{4\kappa}{3P}\frac{\kappa^2}{(\zeta^2+\kappa^2)^{3/2}},\\
        \tilde{\mathsf{F}}_{PI}&=-\frac{\cot I}{P}\frac{\zeta^2-\kappa^2}{\zeta^2+\kappa^2},\\
        \tilde{\mathsf{F}}_{P\xi}&=-\frac{2}{P}\qty(\frac{\zeta\kappa}{\zeta^2+\kappa^2}+\frac{\uppi T_\obs}{P}),\\
        \tilde{\mathsf{F}}_{\zeta\zeta}&=\frac{19}{2},\\
        \tilde{\mathsf{F}}_{\zeta\kappa}&=\frac{3}{4}\zeta\kappa,\\
        \tilde{\mathsf{F}}_{\zeta I}&=\frac{1}{2}\zeta\cot I\qty(15+\frac{2\zeta^2}{\zeta^2+\kappa^2}),\\
        \tilde{\mathsf{F}}_{\zeta\xi}&=2\kappa\frac{\zeta^2-\kappa^2}{\zeta^2+\kappa^2},\\
        \tilde{\mathsf{F}}_{\kappa\kappa}&=\frac{1}{2},\\
        \tilde{\mathsf{F}}_{\kappa I}&=-\frac{1}{2}\kappa\cot I\qty(5-\frac{2\zeta^2}{\zeta^2+\kappa^2}),\\
        \tilde{\mathsf{F}}_{\kappa\xi}&=\zeta\qty(\frac{5}{2}-\frac{\zeta^2}{\zeta^2+\kappa^2}),\\
        \tilde{\mathsf{F}}_{II}&=2\cot^2I,\\
        \tilde{\mathsf{F}}_{I\xi}&=0,\\
        \tilde{\mathsf{F}}_{\xi\xi}&=2.
    \end{split}
    \end{align}

Assuming some values for the observation interval $T_\obs$, number of observations $n_\obs$, and rms timing noise $\sigma$, we can thus use the Fisher matrix elements given above to calculate the expected likelihood ratio~\eqref{eq:expected-lambda} for a given SGWB spectrum, and therefore compute upper limit forecasts.

\subsection{Application to laser ranging}

Another extremely precise observational probe of binary dynamics is \emph{laser ranging} (LR), in which laser pulses are fired at retroreflectors on bodies orbiting the Earth; typically the Moon (Lunar Laser Ranging, LLR)~\cite{Murphy:2013qya} or artificial satellites (Satellite Laser Ranging, SLR)~\cite{Ciufolini:2016ntr}.
By measuring the round-trip times of these pulses, the size of the orbit can be tracked over time with millimetre precision.

We estimate the sensitivity of a generic LR experiment in a very similar way to our treatment of pulsar timing above.
Individual range measurements are assumed to be unbiased, with uncorrelated Gaussian noise of variance $\sigma^2$.
We write their mean value at time $t$ in terms of the eccentric anomaly,
    \begin{equation}
        r(t)=a(1-e\cos E(t)),
    \end{equation}
    and define the Fisher matrix in the limit of many uniformly-spaced observations analogously to Eq.~\eqref{eq:fisher-integral},
    \begin{equation}
    \label{eq:fisher-lr}
        \mathsf{F}_{ij}\simeq\frac{n_\mathrm{obs}}{T_\mathrm{obs}\sigma^2}\frac{P}{2\uppi}\int_0^{2\uppi T_\mathrm{obs}/P}\dd{E}(1-e\cos E)\pdv{r}{X_i}\pdv{r}{X_j}.
    \end{equation}
The necessary partial derivatives are given by
    \begin{align}
    \begin{split}
        \label{eq:derivatives-lr}
        \pdv{r}{P}&=\frac{2a}{3P}(1-e\cos E)-\frac{ae\sin E}{P}\frac{E-e\sin E}{1-e\cos E},\\
        \pdv{r}{e}&=a\frac{e-\cos E}{1-e\cos E},\\
        \pdv{r}{\eps}&=\frac{ae\sin E}{1-e\cos E},
    \end{split}
    \end{align}
    where we have used Kepler's equation,
    \begin{equation}
        E-e\sin E=\frac{2\uppi t}{P}+\eps.
    \end{equation}
Note that we have neglected the three angles describing the orientation of the orbital plane in space, $(I,\asc,\omega)$, as these are not directly measured by the round-trip times of the laser pulses.

Inserting Eq.~\eqref{eq:derivatives-lr} into Eq.~\eqref{eq:fisher-lr}, we obtain the Fisher matrix,
    \begin{align}
    \begin{split}
        \mathsf{F}_{PP}&\simeq\frac{4n_\mathrm{obs}a^2}{9P^2\sigma^2}\bigg[\frac{3\uppi^2T_\mathrm{obs}^2}{2P^2}(e^2+\tfrac{1}{4}e^4)\\
        &\qquad\qquad\qquad+1+3e+\tfrac{27}{16}e^2+4e^3+\tfrac{315}{256}e^4\bigg],\\
        \mathsf{F}_{Pe}&\simeq\frac{3n_\mathrm{obs}a^2}{4P\sigma^2}\qty(e+\tfrac{8}{9}e^2+\tfrac{5}{8}e^3+\tfrac{8}{45}e^4),\\
        \mathsf{F}_{P\eps}&\simeq-\frac{\uppi n_\mathrm{obs}T_\mathrm{obs}a^2}{2P^2\sigma^2}\qty(e^2+\tfrac{1}{4}e^4),\\
        \mathsf{F}_{ee}&=\frac{n_\mathrm{obs}a^2}{\sigma^2}\qty(2-\gamma-\frac{1-\gamma}{e^2}),\\
        \mathsf{F}_{\eps\eps}&=\frac{n_\mathrm{obs}a^2}{\sigma^2}\qty(1-\gamma).
    \end{split}
    \end{align}
We have neglected $\order{e^6}$ terms for the first three entries here, since the Moon and the artificial satellites that are typically tracked with SLR have small eccentricities (e.g., the Lunar eccentricity is $e_{\leftmoon}\approx0.055$).

\section{Summary and outlook}
\label{sec:summary}

In this paper we have developed a powerful new formalism for calculating the statistical evolution of binary systems coupled to the SGWB, deriving a secular FPE which captures the full probability distribution of all six orbital elements on timescales much longer than the binary period.
The KM coefficients describing this FPE, given in Eqs.~\eqref{eq:ecc-drift} and~\eqref{eq:ecc-diffusion}, and illustrated in Figs.~\ref{fig:km1-ecc}, \ref{fig:km2-ecc}, \ref{fig:km2-n}, and~\ref{fig:km1-n}, encapsulate the rich dynamical structure that arises from the interactions between tensorial GW perturbations and elliptical orbits.

The full FPE is a six-dimensional nonlinear PDE, and it is therefore challenging to find exact solutions.
Nonetheless, we have extracted some qualitative features of the late-time behaviour in Sec.~\ref{sec:circular} by fixing the eccentricity to zero.
This analysis illustrates one of the key advantages of our formalism over previous approaches: its ability to capture the full shape of the DF.
We find that, while the stochastic drift effect due to the SGWB tends to increase the energy of the binary and counteract its orbital decay through GW emission, diffusion tends to have the opposite effect, so that the net influence of SGWB resonance is to drive the binary towards merger.
At the same time, the SGWB perturbations act to erase any memory of the binary's initial configuration in space, driving the orbit towards an isotropic distribution in the inclination $I$, and uniform distributions in the other angular variables.
We also find, however, that there is a strong drive towards larger eccentricities in the $e\to0$ limit, so these results with $e=0$ must be taken with a pinch of salt.

We have also developed, in Sec.~\ref{sec:full-fpe}, a practical approach for numerically integrating the full FPE with general eccentricity $e\in(0,1)$ on observational timescales, taking advantage of the fact that these are typically much shorter than the diffusion timescale.
The resulting set of equations~\eqref{eq:slow-diffusion-evolution-equations} are illustrated in Figs.~\ref{fig:HT_cov_t} and~\ref{fig:HT_corner}, using the binary pulsar B1913+16 as an example.
Combined with the statistical tools developed in Sec.~\ref{sec:observations}, this allows us to calculate SGWB sensitivity curves for probes of binary dynamics such as pulsar timing and Lunar/satellite laser ranging, which we present in a companion paper~\cite{Blas:2021mqw}, showing that these can improve upon existing bounds in the $\upmu$Hz frequency band by several orders of magnitude.

Our results motivate further work to develop binary resonance into a precision tool for GW astronomy.
The most important task will be to develop the necessary data analysis pipelines to conduct GW searches with pulsar-timing and laser-ranging data, in a way that fully utilises the theoretical developments in this work.
However, there also many possible avenues for developing our theoretical formalism further.
One important problem is to develop practical numerical integration schemes that go beyond the approach in Sec.~\ref{sec:full-fpe} and capture non-Gaussian features in the DF, thereby taking full advantage of the Fokker-Planck approach (this will be particularly important for studies of populations of binaries~\cite{Barr:2016vxv,Hui:2018mkc}).
It would also be interesting to relax some of our assumptions, for example abandoning the secular-averaging approach and attempting to capture the evolution of binaries within a single orbital period, or perhaps relaxing some of the usual assumptions about the GW strain statistics to develop searches for SGWBs that are non-Gaussian, anisotropic, or have nonstandard polarisation content.
One could also explore the sensitivity of binaries to narrowband sources, using a similar approach to Ref.~\cite{Blas:2019hxz} to consider GW frequencies between the binary's resonant frequencies.
There is also no reason to restrict ourselves to just binaries; in future work, we plan to consider the SGWB-driven evolution of other gravitationally-bound systems such as hierarchical triples or many-body systems such as galaxies and globular clusters.
Finally, our work could be extended even further by considering other stochastic perturbing fields which may exist in the Universe, such as ultralight scalars \cite{Foster:2017hbq,Kalaydzhyan:2018zsx,Centers:2019dyn,Dror:2021nyr}.

\begin{acknowledgments}
    A.C.J. thanks Andrew Pontzen for interesting discussions related to this work.
    We acknowledge the use of \href{https://numpy.org/}{\texttt{NumPy}}~\cite{Harris:2020arr} and \href{https://www.scipy.org/}{\texttt{SciPy}}~\cite{Virtanen:2020sci} in producing the numerical solutions in Figs.~\ref{fig:HT_cov_t} and~\ref{fig:HT_corner}.
    Figs.~\ref{fig:km1-ecc}--\ref{fig:HT_cov_t} were produced using \href{https://matplotlib.org/}{\texttt{Matplotlib}}~\cite{Hunter:2007}, while Fig.~\ref{fig:HT_corner} was produced using \href{https://corner.readthedocs.io/en/latest/}{\texttt{corner.py}}~\cite{ForemanMackey:2016cor}.
    A.C.J. was supported by King's College London through a Graduate Teaching Scholarship.
    D.B. is supported by a ``Ayuda Beatriz Galindo Senior'' from the Spanish ``Ministerio de Universidades'', grant BG20/00228.
    D.B. acknowledges support from the Fundaci\'on Jesus Serra and the Instituto de Astrof\'isica de Canarias under the Visiting Researcher Programme 2021 agreed between both institutions.
\end{acknowledgments}

\appendix
\section{Polarisation tensors in the binary's coordinate frame}
\label{sec:polarisation-tensor-projections}

In order to describe the GW polarisation tensors $e^A_{ij}$, we introduce the orthonormal frame $(\vu*u,\vu*v,\vu*n)$, where $\vu*n$ is the GW propagation direction (see Fig.~\ref{fig:orbital-elements}).
We want to find the components of these basis vectors in the frame of the binary, $(\vu*r,\vu*\theta,\vu*\ell)$, as this determines the binary's response to the GW.
First, we transform from the GW frame to the fixed reference frame $(\vu*x,\vu*y,\vu*z)$ by applying the standard rotations with respect to the zenith $\vartheta$ and azimuth $\phi$,
    \begin{equation}
        \begin{pmatrix}
            x \\ y \\ z
        \end{pmatrix}
        =\mathsf{R}_\phi\mathsf{R}_\vartheta
        \begin{pmatrix}
            u \\ v \\ n
        \end{pmatrix},
    \end{equation}
    where
    \begin{equation}
        \mathsf{R}_\vartheta=
        \begin{pmatrix}
            \cos\vartheta & 0 & \sin\vartheta \\
            0 & 1 & 0 \\
            -\sin\vartheta & 0 & \cos\vartheta
        \end{pmatrix},\quad
        \mathsf{R}_\phi=
        \begin{pmatrix}
            \cos\phi & -\sin\phi & 0 \\
            \sin\phi & \cos\phi & 0 \\
            0 & 0 & 1
        \end{pmatrix}.
    \end{equation}
This reference frame is transformed to the binary frame $(\vu*r,\vu*\theta,\vu*\ell)$ with three further rotations, which specify the inclination $I$, the longitude of ascending node $\asc$, and the argument of the binary in the orbital plane $\theta=\psi+\omega$~\cite{Murray:2000ssd},
    \begin{equation}
        \begin{pmatrix}
            r \\ \theta \\ \ell
        \end{pmatrix}
        =\mathsf{R}_\theta\mathsf{R}_I\mathsf{R}_\asc
        \begin{pmatrix}
            x \\ y \\ z
        \end{pmatrix},
    \end{equation}
    where
    \begin{align}
    \begin{split}
        \mathsf{R}_\asc&=
        \begin{pmatrix}
            \cos\asc & \sin\asc & 0 \\
            -\sin\asc & \cos\asc & 0 \\
            0 & 0 & 1
        \end{pmatrix},\quad
        \mathsf{R}_I=
        \begin{pmatrix}
            1 & 0 & 0 \\
            0 & \cos I & \sin I \\
            0 & -\sin I & \cos I
        \end{pmatrix},\\
        &\qquad\qquad\quad\mathsf{R}_\theta=
        \begin{pmatrix}
            \cos\theta & \sin\theta & 0 \\
            -\sin\theta & \cos\theta & 0 \\
            0 & 0 & 1
        \end{pmatrix}.
    \end{split}
    \end{align}
We thus obtain the desired relationship between the binary frame and the GW frame by applying all five rotations,
    \begin{equation}
        \begin{pmatrix}
            r \\ \theta \\ \ell
        \end{pmatrix}
        =\mathsf{R}_\theta\mathsf{R}_I\mathsf{R}_\asc\mathsf{R}_\phi\mathsf{R}_\vartheta
        \begin{pmatrix}
            u \\ v \\ n
        \end{pmatrix}.
    \end{equation}

The various contractions with the polarisation tensors are then given by
    \begin{align}
    \begin{split}
    \label{eq:polarisation-tensor-contractions}
        &e_{ij}^+\hat{r}^i\hat{r}^j=-\qty[\cos\varphi\cos I\sin\theta-\sin\varphi\cos\theta]^2\\
        &+\qty[\sin\vartheta\sin I\sin\theta-\cos\vartheta(\cos\varphi\cos\theta+\sin\varphi\cos I\sin\theta)]^2,\\
        &e_{ij}^\times\hat{r}^i\hat{r}^j=2(\cos\varphi\cos I\sin\theta-\sin\varphi\cos\theta)\\
        &\times[\cos\vartheta(\cos\varphi\cos\theta+\sin\varphi\cos I\sin\theta)-\sin\vartheta\sin I\sin\theta],\\
        &e_{ij}^+\hat{r}^i\hat{\theta}^j=(\sin\varphi\cos\theta-\cos\varphi\cos I\sin\theta)\\
        &\qquad\qquad\times(\cos\varphi\cos I\cos\theta+\sin\varphi\sin\theta)\\
        &-[\cos\vartheta\cos\varphi\sin\theta+(\sin\vartheta\sin I-\cos\vartheta\sin\varphi\cos I)\cos\theta]\\
        &\times[\cos\vartheta(\cos\varphi\cos\theta+\sin\varphi\cos I\sin\theta)-\sin\vartheta\sin I\sin\theta],\\
        &e_{ij}^\times\hat{r}^i\hat{\theta}^j=\cos\vartheta\cos2\varphi\cos I\cos2\theta\\
        &\qquad\qquad+\sin\vartheta\sin I(\sin\varphi\cos2\theta-\cos\varphi\cos I\sin2\theta)\\
        &\qquad\qquad+\frac{1}{2}\cos\vartheta\sin2\varphi(1+\cos^2I)\sin2\theta,\\
        &e_{ij}^+\hat{r}^i\hat{\ell}^j=\cos\varphi\sin I(\cos\varphi\cos I\sin\theta-\sin\varphi\cos\theta)\\
        &-(\sin\vartheta\cos I+\cos\vartheta\sin\varphi\sin I)\\
        &\times\qty[\cos\vartheta(\cos\varphi\cos\theta+\sin\varphi\cos I\sin\theta)-\sin\vartheta\sin I\sin\theta],\\
        &e_{ij}^\times\hat{r}^i\hat{\ell}^j=\sin\vartheta(\sin\varphi\cos I\cos\theta-\cos\varphi\cos2I\sin\theta)\\
        &\qquad\qquad-\cos\vartheta\sin I(\cos2\varphi\cos\theta+\sin2\varphi\cos I\sin\theta),
    \end{split}
    \end{align}
    where we define $\varphi\equiv\phi-\asc$.
We recall that $\vartheta,\phi$ are the spherical coordinates of the incoming plane GW, $\asc$ is the longitude of ascending node, and $\theta\equiv\psi+\omega$ is the orbital argument with respect to the ascending node, with $\psi$ the true anomaly and $\omega$ the argument of pericentre; these are all illustrated in Fig.~\ref{fig:orbital-elements}.

\section{Transfer functions}
\label{sec:transfer}

The Fourier components of the transfer functions are defined in terms of the polarisation tensor contractions discussed in Appendix~\ref{sec:polarisation-tensor-projections} by
    \begin{align}
    \begin{split}
    \label{eq:transfer-functions}
        T^A_{P,n}&=\frac{3P^2\gamma}{4\uppi}\ev{\frac{e\sin\psi}{1+e\cos\psi}e_{ij}^A\hat{r}_i\hat{r}_j+e_{ij}^A\hat{r}_i\hat{\theta}_j}_n,\\
        T^A_{e,n}&=\frac{\gamma^2T^A_{P,n}}{3Pe}-\frac{P\gamma^5}{4\uppi e}\ev{\frac{e_{ij}^A\hat{r}_i\hat{\theta}_j}{(1+e\cos\psi)^2}}_n,\\
        T^A_{I,n}&=\frac{P\gamma^3}{4\uppi}\ev{\frac{\cos\theta}{(1+e\cos\psi)^2}e_{ij}^A\hat{r}_i\hat{\ell}_j}_n,\\
        T^A_{\asc,n}&=\frac{P\gamma^3}{4\uppi\sin I}\ev{\frac{\sin\theta}{(1+e\cos\psi)^2}e_{ij}^A\hat{r}_i\hat{\ell}_j}_n,\\
        T^A_{\omega,n}&=\frac{P\gamma^3}{4\uppi e}\ev{\frac{\sin\psi(2+e\cos\psi)}{(1+e\cos\psi)^2}e_{ij}^A\hat{r}_i\hat{\theta}_j-\frac{\cos\psi e_{ij}^A\hat{r}_i\hat{r}_j}{1+e\cos\psi}}_n\\
        &-T^A_{\asc,n}\cos I,\\
        T^A_{\eps,n}&=-\frac{P\gamma^4}{2\uppi}\ev{\frac{e^A_{ij}\hat{r}_i\hat{r}_j}{(1+e\cos\psi)^2}}_n-\gamma\cos IT^A_{\asc,n}-\gamma T^A_{\omega,n},
    \end{split}
    \end{align}
    where we have introduced the secular averaging operation
    \begin{equation}
    \label{eq:secular-average}
        \ev{\cdots}_n\equiv\int_0^P\frac{\dd{t}}{P}\exp(\frac{2\uppi\rmi nt}{P})(\cdots)
    \end{equation}
    which extracts the $n^\mathrm{th}$-order Fourier coefficient of a given function of the true anomaly $\psi(t)$, holding the orbital elements fixed (as they vary over much longer timescales).
The subscript $n$ distinguishes this from the ensemble average $\ev{\cdots}$.

The Fourier components arising in the eccentric transfer functions can be expressed in terms of \emph{Hansen coefficients}, $C^{lm}_n$, $S^{lm}_n$~\cite{Brumberg:1995atcm}.
These are functions of eccentricity that have been used for centuries in celestial mechanics to describe Keplerian motion.
We define them here by
    \begin{align}
    \begin{split}
    \label{eq:hansen-definition-appendix}
        C^{lm}_n(e)&=\ev{\frac{\cos m\psi}{(1+e\cos\psi)^l}}_n,\\
        S^{lm}_n(e)&=\ev{\frac{\sin m\psi}{(1+e\cos\psi)^l}}_n,
    \end{split}
    \end{align}
    with explicit expressions for particular sets of $(l,m)$ given in Appendix~\ref{sec:hansen}.

Inserting Eq.~\eqref{eq:polarisation-tensor-contractions} into Eq.~\eqref{eq:transfer-functions} and using Eq.~\eqref{eq:hansen-definition-appendix}, we can thus write the transfer functions as linear combinations of the Hansen coefficients.
These are then the input to computing the KM coefficients.

\section{Transforming the reference frame}
\label{sec:reference-frame}

The polarisation tensor contractions in Eq.~\eqref{eq:polarisation-tensor-contractions} are completely general and apply to any choice of reference frame.
However, the corresponding expressions for the transfer functions are very lengthy, making it prohibitively difficult to calculate the KM coefficients.
We circumvent this difficulty by choosing a particular reference frame in which the transfer functions are much simpler, before transforming back to a general reference frame.

Given two reference frames, $(\vu*x,\vu*y,\vu*z)$ and $(\vu*{x}{}',\vu*{y}{}',\vu*{z}{}')$, we have two corresponding sets of orbital elements for the binary, $X=(P,e,I,\asc,\omega,\eps)$ and $X'=(P,e,I',\asc',\omega',\eps)$ (the period, eccentricity, and compensated mean anomaly are the same in both frames, as they do not depend on the orientation of the binary in space).
The KM coefficients for the unprimed elements are given in terms of those for the primed ones by (see, e.g., Sec.~4.9 of Ref.~\cite{Risken:1989fpe} for a derivation)
    \begin{align}
    \begin{split}
        D^{(1)}_i&=\pdv{X_i}{X_{i'}}D^{(1)}_{i'}+\pdv{X_i}{X_{i'}}{X_{j'}}D^{(2)}_{i'j'},\\
        D^{(2)}_{ij}&=\pdv{X_i}{X_{i'}}\pdv{X_j}{X_{j'}}D^{(2)}_{i'j'},
    \end{split}
    \end{align}
    where primed indices run over the primed orbital elements.
We therefore require the first and second partial derivatives of the unprimed elements with respect to the primed ones.
These can be deduced from the following relations, which are derived from Sec.~2.8 of Ref.~\cite{Murray:2000ssd},
    \begin{align}
    \begin{split}
    \label{eq:orbit-angles-transformation}
        \cos I&=s_{I'}\sin I'\sin\asc'\vu*{x}{}'\vdot\vu*z\\
        &-s_{I'}\sin I'\cos\asc'\vu*{y}{}'\vdot\vu*z+\cos I'\vu*{z}{}'\vdot\vu*z,\\
        \sin I\sin\asc&=s_{I'}\sin I'\sin\asc'\vu*{x}{}'\vdot\vu*x\\
        &-s_{I'}\sin I'\cos\asc'\vu*{y}{}'\vdot\vu*x+\cos I'\vu*{z}{}'\vdot\vu*x,\\
        \sin I\sin\omega&=(\cos\asc'\cos\omega'-\cos I'\sin\asc'\sin\omega')\vu*{x}{}'\vdot\vu*z\\
        &+(\sin\asc'\cos\omega'+\cos I'\cos\asc'\sin\omega')\vu*{y}{}'\vdot\vu*z\\
        &+\sin I'\sin\omega\vu*{z}{}'\vdot\vu*z,
    \end{split}
    \end{align}
    where
    \begin{equation}
        s_{I'}\equiv
        \begin{cases}
            +1 & \text{if }\cos I'>0,\\
            -1 & \text{if }\cos I'<0.
        \end{cases}
    \end{equation}

We are now free to specify the primed frame such that the KM coefficients are easier to compute.
One particularly simple choice is to choose $(\vu*{x}{}',\vu*{y}{}',\vu*{z}{}')$ such that $I'=\uppi/4$ and $\asc'=\omega'=0$.
(It may seem at first that $I'=0$ is a simpler choice, as the reference frame then coincides with the binary's frame.
However, there is a coordinate singularity associated with $I'=0$ which makes some of the associated coefficients poorly-behaved.
Taking $I'=\uppi/4$ is much easier, particularly since we then have $\sin I'=\cos I'$, which simplifies many of the resulting expressions.)

\begin{widetext}
Having specified the primed frame, we require the first and second derivatives to transform back to the unprimed frame, which is relevant for a general observer.
Using Eq.~\eqref{eq:orbit-angles-transformation}, we find that the nonzero derivatives, evaluated at $(I',\asc',\omega')=(\tfrac{\uppi}{4},0,0)$, are given by
    \begin{align}
    \begin{split}
        \pdv{I}{I'}&=\cos\omega,\qquad\pdv{I}{\asc'}=-\frac{\sin\omega}{\sqrt{2}},\qquad\pdv{\asc}{I'}=\frac{\sin\omega}{\sin I},\qquad\pdv{\asc}{\asc'}=\frac{\cos\omega}{\sqrt{2}\sin I},\\
        \pdv{\omega}{I'}&=-\frac{\sin\omega}{\tan I},\qquad\pdv{\omega}{\asc'}=\frac{1}{\sqrt{2}}\qty(1-\frac{\cos\omega}{\tan I}),\qquad\pdv{\omega}{\omega'}=1,\\
        \pdv[2]{I}{{I'}}&=\frac{\sin^2\omega}{\tan I},\qquad\pdv{I}{I'}{\asc'}=\frac{\sin\omega}{\sqrt{2}}\qty(\frac{\cos\omega}{\tan I}-1),\qquad\pdv[2]{I}{{\asc'}}=\frac{\cos\omega}{2}\qty(\frac{\cos\omega}{\tan I}-1),\\
        \pdv[2]{\asc}{{I'}}&=-\frac{\sin2\omega}{\sin I\tan I},\qquad\pdv{\asc}{I'}{\asc'}=\frac{\cos\omega-\cos2\omega\cot I}{\sqrt{2}\sin I},\qquad\pdv[2]{\asc}{{\asc'}}=\frac{\sin\omega}{\sin I}\qty(\frac{\cos\omega}{\tan I}-\frac{1}{2}),\\
        \pdv[2]{\omega}{{I'}}&=\sin2\omega\frac{2-\sin^2I}{2\sin^2I},\quad\pdv{\omega}{I'}{\asc'}=\frac{1}{2\sqrt{2}}\qty(\cos2\omega\frac{2-\sin^2I}{\sin^2I}-\frac{2\cos\omega}{\tan I}-1),\quad\pdv[2]{\omega}{{\asc'}}=\frac{\sin\omega}{\tan I}\qty(\frac{1}{2}-\cos\omega\frac{2-\sin^2I}{\sin2I}).
    \end{split}
    \end{align}
It is straightforward to confirm that these are well-behaved and take on the appropriate values when $(I,\asc,\omega)\to(I',\asc',\omega')$.

We thus find that the unprimed drift coefficients are given in terms of those in the primed frame by
    \begin{align}
    \begin{split}
    \label{eq:drift-transformation}
        D^{(1)}_I&=\cos\omega\qty[D^{(1)}_{I'}+\qty(\frac{\cos\omega}{\tan I}-1)\frac{D^{(2)}_{\asc'\asc'}}{2}]+\frac{\sin^2\omega}{\tan I}D^{(2)}_{I'I'},\\
        D^{(1)}_\asc&=\frac{1}{\sin I}\qty[\sin\omega D^{(1)}_{I'}-\frac{\sin2\omega}{\tan I}D^{(2)}_{I'I'}+\sin\omega\qty(\frac{\cos\omega}{\tan I}-\frac{1}{2})D^{(2)}_{\asc'\asc'}],\\
        D^{(1)}_\omega&=\frac{\sin\omega}{\sin I}\qty[-\cos ID^{(1)}_{I'}+\cos\omega\frac{2-\sin^2I}{\sin I}D^{(2)}_{I'I'}+\qty(\cos I-\cos\omega\frac{2-\sin^2I}{\sin I})\frac{D^{(2)}_{\asc'\asc'}}{2}],
    \end{split}
    \end{align}
    with the diffusion coefficients given by
    \begin{align}
    \begin{split}
    \label{eq:diffusion-transformation}
        D^{(2)}_{XI}&=D^{(2)}_{X\asc}=D^{(2)}_{X\omega}=D^{(2)}_{X\eps}=D^{(2)}_{I\eps}=D^{(2)}_{\asc\eps}=0,\\
        D^{(2)}_{II}&=\cos^2\omega D^{(2)}_{I'I'}+\frac{\sin^2\omega}{2}D^{(2)}_{\asc'\asc'},\\
        D^{(2)}_{I\asc}&=\frac{\sin2\omega}{2\sin I}\qty(D^{(2)}_{I'I'}-\frac{1}{2}D^{(2)}_{\asc'\asc'}),\\
        D^{(2)}_{I\omega}&=\sin\omega\qty[-\frac{\cos\omega}{\tan I}D^{(2)}_{I'I'}+\frac{1}{2}\qty(\frac{\cos\omega}{\tan I}-1)D^{(2)}_{\asc'\asc'}-\frac{1}{\sqrt{2}}D^{(2)}_{\asc'\omega'}],\\
        D^{(2)}_{\asc\asc}&=\frac{1}{\sin^2I}\qty(\sin^2\omega D^{(2)}_{I'I'}+\frac{\cos^2\omega}{2}D^{(2)}_{\asc'\asc'}),\\
        D^{(2)}_{\asc\omega}&=\frac{1}{\sin I}\qty[-\frac{\sin^2\omega}{\tan I}D^{(2)}_{I'I'}+\frac{\cos\omega}{2}\qty(1-\frac{\cos\omega}{\tan I})D^{(2)}_{\asc'\asc'}+\frac{\cos\omega}{\sqrt{2}}D^{(2)}_{\asc'\omega'}],\\
        D^{(2)}_{\omega\omega}&=\frac{\sin^2\omega}{\tan^2I}D^{(2)}_{I'I'}+\frac{1}{2}\qty(1-\frac{\cos\omega}{\tan I})^2D^{(2)}_{\asc'\asc'}+\sqrt{2}\qty(1-\frac{\cos\omega}{\tan I})D^{(2)}_{\asc'\omega'}+D^{(2)}_{\omega'\omega'},\\
        D^{(2)}_{\omega\eps}&=D^{(2)}_{\omega'\eps},
    \end{split}
    \end{align}
    where $X$ here stands for $P$ or $e$.

\section{Kramers-Moyal coefficients in the primed frame}
\label{sec:KM-primed-frame}

Here we give the KM coefficients evaluated in the primed coordinate frame $(I',\asc',\omega')=(\tfrac{\uppi}{4},0,0)$.
To calculate these, we first evaluate the polarisation tensor contractions from Appendix~\ref{sec:polarisation-tensor-projections} in the primed frame, and insert these into Eq.~\eqref{eq:transfer-functions} to find the appropriate GW transfer functions, expressing the secular averages in terms of Hansen coefficients that are listed below in Appendix~\ref{sec:hansen}.
These secular transfer functions are then inserted into Eq.~\eqref{eq:km-final}, integrating over the GW propagation direction $\vu*n=(\vartheta,\phi)$ to obtain the KM coefficients.
The resulting expressions for the diffusion matrix are
    \begin{align}
    \begin{split}
        D^{(2)}_{PP}&=\frac{27P^3\gamma^2}{20}\sum_{n=1}^\infty nH_0^2\Omega_n\qty[\qty|E^{02}_n+\frac{e}{2}(E^{11}_n-E^{13}_n)|^2-\frac{(eS^{11}_n)^2}{3}],\\
        D^{(2)}_{Pe}&=\frac{\gamma^2D^{(2)}_{PP}}{3Pe}-\frac{9P^2\gamma^6}{40}\sum_{n=1}^\infty nH_0^2\Omega_nE^{22}_n\qty(\frac{2}{e}E^{02}_n+E^{11}_n-E^{13}_n)^*,\\
        D^{(2)}_{ee}&=\frac{3P\gamma^6}{20e^2}\sum_{n=1}^\infty nH_0^2\Omega_n\qty[\qty|E^{02}_n+\frac{e}{2}(E^{11}_n-E^{13}_n)-\gamma^2E^{22}_n|^2-\frac{(eS^{11}_n)^2}{3}],\\
        D^{(2)}_{I'I'}&=\frac{3P\gamma^6}{80}\sum_{n=1}^\infty nH_0^2\Omega_n\qty|E^{20}_n+E^{22}_n|^2,\\
        D^{(2)}_{\asc'\asc'}&=\frac{3P\gamma^6}{40}\sum_{n=1}^\infty nH_0^2\Omega_n\qty|E^{20}_n-E^{22}_n|^2,\\
        D^{(2)}_{\asc'\omega'}&=-\frac{1}{\sqrt{2}}D^{(2)}_{\asc'\asc'},\\
        D^{(2)}_{\omega'\omega'}&=\frac{1}{2}D^{(2)}_{\asc'\asc'}+\frac{3P\gamma^6}{80e^2}\sum_{n=1}^\infty nH_0^2\Omega_n\qty[\qty|E^{11}_n+E^{13}_n+2E^{21}_n-2E^{23}_n+\frac{e}{2}\qty(E^{20}_n-E^{24}_n)|^2+\frac{4}{3}\qty(C^{11}_n)^2],\\
        D^{(2)}_{\omega'\eps}&=-\frac{3P\gamma^7}{80e^2}\sum_{n=1}^\infty nH_0^2\Omega_n\qty[\qty|E^{11}_n+E^{13}_n+2(E^{21}_n-E^{23}_n)+\frac{e}{2}(E^{20}_n-4E^{22}_n-E^{24}_n)|^2+\frac{4}{3}C^{11}_n(C^{11}_n-2eC^{20}_n)-4e^2|E^{22}_n|^2],\\
        D^{(2)}_{\eps\eps}&=\frac{3P\gamma^8}{80e^2}\sum_{n=1}^\infty nH_0^2\Omega_n\qty[\qty|E^{11}_n+E^{13}_n+2(E^{21}_n-E^{23}_n)+\frac{e}{2}(E^{20}_n-8E^{22}_n-E^{24}_n)|^2+\frac{4}{3}(C^{11}_n-2eC^{20}_n)^2],\\
        D^{(2)}_{PI'}&=D^{(2)}_{P\asc'}=D^{(2)}_{P\omega'}=D^{(2)}_{P\eps}=D^{(2)}_{eI'}=D^{(2)}_{e\asc'}=D^{(2)}_{e\omega'}=D^{(2)}_{e\eps}=D^{(2)}_{I'\asc'}=D^{(2)}_{I'\omega'}=D^{(2)}_{I'\eps}=D^{(2)}_{\asc'\eps}=0,
    \end{split}
    \end{align}
    while the drift vector is given by
    \begin{align}
    \begin{split}
        D^{(1)}_P&=V_P+\frac{9P^2\gamma^2}{80}\sum_{n=1}^\infty nH_0^2\Omega_n\bigg[2E^{02}_n\qty(10eE^{02}_n+(3+7e^2)E^{11}_n+(1-11e^2)E^{13}_n-2e\gamma^2E^{22}_n)^*\\
        &\qquad\qquad\quad+E^{11}_n\qty(e(3+2e^2)E^{11}_n-2e(1+4e^2)E^{13}_n-2\gamma^2E^{22}_n)^*-e(1-6e^2)\qty|E^{13}_n|^2+2\gamma^2E^{13}_n\qty(E^{22}_n)^*\\
        &\qquad\qquad\quad+8\gamma^2\qty(E^{02}_n+\frac{e}{2}(E^{11}_n-E^{13}_n))\qty(E^{21}_n-E^{23}_n+\frac{e}{4}(E^{20}_n+6E^{22}_n-E^{24}_n))^*\\
        &\qquad\qquad\quad+4\gamma^2\qty(E^{02}_n+\frac{e}{2}(E^{11}_n-E^{13}_n)-\gamma^2E^{22}_n)\qty({E^{02}_n}'+\frac{e}{2}(E^{11}_n-E^{13}_n)')^*-\frac{4e}{3}S^{11}_n\qty((1+4e^2)S^{11}_n+e\gamma^2{S^{11}_n}')\bigg],\\
        D^{(1)}_e&=V_e+\frac{3P\gamma^6}{20e^2}\sum_{n=1}^\infty nH_0^2\Omega_n\bigg[\frac{1}{e}\qty|E^{02}_n+\frac{e}{2}(E^{11}_n-E^{13}_n)|^2+\frac{\gamma^4}{e}(S^{22}_n)^2-\frac{e^2}{3}{S^{11}_n}'S^{11}_n+\qty(E^{02}_n+\frac{e}{2}(E^{11}_n-E^{13}_n)-\gamma^2E^{22}_n)\times\\
        &\times\qty(-\frac{2}{e}E^{02}_n+\frac{1}{2}(E^{11}_n+3E^{13}_n)+2(E^{21}_n-E^{23}_n)+{E^{02}_n}'-\gamma^2{E^{22}_n}'+\frac{e}{2}(E^{20}_n+12E^{22}_n-E^{24}_n+(E^{11}_n-E^{13}_n)'))^*\bigg],\\
        D^{(1)}_{I'}&=\frac{1}{2}D^{(2)}_{\asc'\asc'},\\
        D^{(1)}_{\asc'}&=0,\\
        D^{(1)}_{\omega'}&=V_{\omega'},\\
        D^{(1)}_{\eps}&=V_\eps,
    \end{split}
    \end{align}
    where primes on the Hansen coefficients denote derivatives with respect to eccentricity.
These KM coefficients can be transformed back to a general reference frame using Eqs.~\eqref{eq:drift-transformation} and~\eqref{eq:diffusion-transformation}, resulting in the expressions given in Sec.~\ref{sec:KM}.

\section{Hansen coefficients}
\label{sec:hansen}

Using various formulae given in Ref.~\cite{Brumberg:1995atcm}, we write the Hansen coefficients as
    \begin{align}
    \begin{split}
        \label{eq:hansen-general}
        C^{lm}_n&=\frac{{}_2F_1(-l-m-1,-l+m-1;1;\beta^2)}{2\gamma^{2l}(1+\beta^2)^{l+1}}[J_{n-m}(ne)+J_{n+m}(ne)]\\
        &+\sum_{k=1}^\infty\bigg\{\frac{(-l+m-1)_k\beta^k}{2k!\gamma^{2l}(1+\beta^2)^{l+1}}{}_2F_1(-l-m-1,-l+m-1+k;1+k;\beta^2)[J_{n-m-k}(ne)+J_{n+m+k}(ne)]\\
        &\qquad\quad+\frac{(-l-m-1)_k\beta^k}{2k!\gamma^{2l}(1+\beta^2)^{l+1}}{}_2F_1(-l-m-1+k,-l+m-1;1+k;\beta^2)[J_{n-m+k}(ne)+J_{n+m-k}(ne)]\bigg\},\\
        S^{lm}_n&=\frac{\rmi\,{}_2F_1(-l-m-1,-l+m-1;1;\beta^2)}{2\gamma^{2l}(1+\beta^2)^{l+1}}[J_{n-m}(ne)-J_{n+m}(ne)]\\
        &+\sum_{k=1}^\infty\bigg\{\frac{\rmi(-l+m-1)_k\beta^k}{2k!\gamma^{2l}(1+\beta^2)^{l+1}}{}_2F_1(-l-m-1,-l+m-1+k;1+k;\beta^2)[J_{n-m-k}(ne)-J_{n+m+k}(ne)]\\
        &\qquad\quad+\frac{\rmi(-l-m-1)_k\beta^k}{2k!\gamma^{2l}(1+\beta^2)^{l+1}}{}_2F_1(-l-m-1+k,-l+m-1;1+k;\beta^2)[J_{n-m+k}(ne)-J_{n+m-k}(ne)]\bigg\},
    \end{split}
    \end{align}
    where we define the expansion variable
    \begin{equation}
        \beta\equiv\frac{e}{1+\gamma},
    \end{equation}
    and where $(\cdots)_k$ is a rising Pochhammer symbol, defined by
    \begin{equation}
        \label{eq:pochhammer}
        (n)_k\equiv n(n+1)(n+2)\cdots(n+k-1),
    \end{equation}
    while ${}_2F_1(a,b;c;z)$ is a hypergeometric function, and $J_n(z)$ is a Bessel function of the first kind.
We see that in the circular case $e=0$ we have $\beta=0$, ${}_2F_1(a,b;c;0)=1$, and $J_n(0)=\delta_{n,0}$, and these expressions in Eq.~\eqref{eq:hansen-general} simplify to
    \begin{align}
    \begin{split}
        C^{lm}_n&=\frac{1}{2}(\delta_{n,m}+\delta_{n,-m}),\\
        S^{lm}_n&=\frac{\rmi}{2}(\delta_{n,m}-\delta_{n,-m}),
    \end{split}
    \end{align}
    which can be confirmed by directly integrating Eq.~\eqref{eq:hansen-definition-appendix}.
Note also that for general eccentricity $e\in(0,1)$, from the definition of the Pochhammer symbol Eq.~\eqref{eq:pochhammer}, the sums over $k$ in Eq.~\eqref{eq:hansen-general} terminate if and only if $m\le l+1$, otherwise the corresponding Hansen coefficients will have an infinite number of terms.

Using Eq.~\eqref{eq:hansen-general}, we can directly compute all the Hansen coefficients that appear in Eqs.~\eqref{eq:ecc-drift} and~\eqref{eq:ecc-diffusion}, obtaining the cosine coefficients
    \begin{align}
    \begin{split}
        \label{eq:hansen-cosine-explicit}
        C^{02}_n&=-\qty[\frac{1+\gamma-\frac{2}{3}e^2(1+\frac{1}{4}\gamma)}{(1+\gamma)^4}]12\gamma^2J_n+\sum_{k=0}^\infty\frac{\beta^k(1-\beta^2)^3}{2(1+\beta^2)}(J_{n-k-2}+J_{n+k+2}),\\
        C^{11}_n&=\frac{1}{2n\gamma^2}(J_{n-1}-J_{n+1}),\\
        C^{13}_n&=-\qty[\frac{1+\gamma-\frac{9}{8}e^2(1+\frac{5}{9}\gamma)+\frac{1}{5}e^4}{\beta\gamma^2(1+\gamma)^6}]20e^4J_n-\qty[\frac{1-\gamma-\frac{9}{4}e^2(1-\frac{7}{9}\gamma)+\frac{3}{4}e^4(1-\frac{5}{3}\gamma)}{e^3\beta\gamma^2}]2(J_{n-1}+J_{n+1})\\
        &-\qty[\frac{1-\gamma-3e^2(1-\frac{5}{6}\gamma)+3e^4(1-\frac{5}{8}\gamma)-e^6(1-\frac{15}{32}\gamma)}{e^5\gamma^2}]8(J_{n-2}+J_{n+2})+\sum_{k=0}^\infty\frac{\beta^k(1-\beta^2)^5}{2(1+\beta^2)^2\gamma^2}(J_{n-k-3}+J_{n+k+3}),\\
        C^{20}_n&=-\qty[\frac{n^2(1-\frac{1}{2}e^2)-3n+2}{n^4e^2\gamma^4}]8J_{n-2}+\qty(\frac{n-1}{n^3e\gamma^4})4J_{n-3},\\
        C^{21}_n&=\qty[\frac{n^3(1-\tfrac{7}{4}e^2+\tfrac{3}{4}e^4)-n^2(1-\tfrac{7}{2}e^2+e^4)-4n(1+\tfrac{1}{2}e^2)+4}{n^4e^3\gamma^4}]4J_{n-2}-\qty[\frac{n^2(1-\tfrac{3}{2}e^2+\tfrac{1}{2}e^4)+n(1+e^2)-2}{n^3e^2\gamma^4}]2J_{n-3},\\
        C^{22}_n&=-\qty[\frac{n^3(1-\frac{7}{4}e^2+\frac{3}{4}e^4)-2n^2(1-\frac{13}{8}e^2+\frac{7}{16}e^4)-n(1+\frac{1}{2}e^2)+2-e^2}{n^4e^4\gamma^4}]16J_{n-2}\\
        &+\qty[\frac{n^2(1-\frac{3}{2}e^2+\frac{1}{2}e^4)+\frac{1}{2}ne^2-1+\frac{1}{2}e^2}{n^3e^3\gamma^4}]8J_{n-3},\\
        C^{23}_n&=\qty[\frac{n^2\gamma^6+3n(1-\frac{5}{4}e^2+\frac{1}{4}e^4)+2-\frac{3}{2}e^2}{n^2e^3\gamma^4}]4J_n-\qty(\frac{1-\frac{5}{4}e^2+\frac{1}{4}e^4}{n^2e^2\gamma^4})12J_{n-1},\\
        C^{24}_n&=\frac{8\gamma^3}{(1+\gamma)^4}(J_{n-4}+J_{n+4})\\
        &+\sum_{k=0}^\infty\frac{\beta^k(1+\gamma)^3}{16\gamma^4}\qty[(1-\beta^2)^7(J_{n-k-4}+J_{n+k+4})+\frac{(-7)_k}{k!}{}_2F_1(1,k-7;k+1;\beta^2)(J_{n+k-4}+J_{n-k+4})],
    \end{split}
    \end{align}
    and the sine coefficients,
    \begin{align}
    \begin{split}
        \label{eq:hansen-sine-explicit}
        S^{02}_n&=-\qty[\frac{1-\gamma-2e^2(1-\frac{3}{4}\gamma)+e^4}{e^3}]2\rmi(J_{n-1}-J_{n+1})+\sum_{k=0}^\infty\frac{\rmi\beta^k(1-\beta^2)^3}{2(1+\beta^2)}(J_{n-k-2}-J_{n+k+2}),\\
        S^{11}_n&=\frac{\rmi J_n}{ne\gamma},\\
        S^{13}_n&=-\qty[\frac{1-3e^2(1-\frac{5}{6}\gamma)+3e^4(1-\frac{5}{8}\gamma)-e^6}{e^4\gamma^2}]4\rmi(J_{n-1}-J_{n+1})\\
        &-\qty[\frac{1-\gamma-3e^2(1-\frac{5}{6}\gamma)+3e^4(1-\frac{5}{8}\gamma)-e^6(1-\frac{15}{32}\gamma)}{e^5\gamma^2}]8\rmi(J_{n-2}-J_{n+2})+\sum_{k=0}^\infty\frac{\rmi\beta^k(1-\beta^2)^5}{2(1+\beta^2)^2\gamma^2}(J_{n-k-3}-J_{n+k+3}),\\
        S^{21}_n&=\frac{-\rmi}{n^2e\gamma^3}[(n\gamma^2+2)J_n-2eJ_{n-1}],\\
        S^{22}_n&=\qty[\frac{n^3(1-\frac{5}{4}e^2+\frac{1}{4}e^4)-2n^2(1-\frac{9}{8}e^2)-n(1+e^2)+2}{n^4e^4\gamma^3}]16\rmi J_{n-2}-\qty[\frac{n^3\gamma^2+n^2e^2-n}{n^4e^3\gamma^3}]8\rmi J_{n-3},\\
        S^{23}_n&=\frac{15e^2}{8\gamma^3}\rmi(J_{n-1}-J_{n+1})-\frac{3e}{2\gamma^3}\rmi(J_{n-2}-J_{n+2})+\frac{1-\frac{1}{4}e^2}{2\gamma^3}\rmi(J_{n-3}-J_{n+3}),\\
        S^{24}_n&=\frac{8\rmi\gamma^3}{(1+\gamma)^4}(J_{n-4}-J_{n+4})\\
        &+\sum_{k=0}^\infty\frac{\beta^k(1+\gamma)^3}{16\gamma^4}\qty[(1-\beta^2)^7(J_{n-k-4}-J_{n+k+4})+\frac{(-7)_k}{k!}{}_2F_1(1,k-7;k+1;\beta^2)(J_{n+k-4}-J_{n-k+4})].
    \end{split}
    \end{align}
Note that all the Bessel functions appearing in Eqs.~\eqref{eq:hansen-cosine-explicit} and~\eqref{eq:hansen-sine-explicit} have their argument equal to $ne$, even if they are not of order $n$; we suppress this argument for brevity.
\end{widetext}

\bibliography{binary-resonance}
\end{document}